\documentclass[useAMS,usenatbib]{mn2e}

\usepackage{amssymb}
\usepackage{graphicx}
\usepackage[usenames,dvipsnames]{color}
\usepackage{verbatim}

\usepackage{times}
\usepackage{mathptmx}

\newcommand{\um}[1]{\ \mathrm{#1}}
\newcommand{\mic}[1]{\ \umu\mathrm{#1}}
\newcommand{\HAng}[3]{#1\!:\!#2\!:\!#3}
\newcommand{\DAng}[3]{#1^\circ\,#2'\,#3''}

\title[High-resolution VLA observations of MBs]{High-resolution Very Large Array observations of 18 MIPSGAL bubbles}
\author[A. Ingallinera et al.]{{\small A. Ingallinera$^1$\thanks{E-mail:ingallinera@oact.inaf.it}, C. Trigilio$^1$, P. Leto$^1$, G. Umana$^1$, C. Buemi$^1$, F. Bufano$^1$, C. Agliozzo$^{2,3}$, S. Riggi$^1$, N. Flagey$^4$, K. Silva$^{4,5}$, L. Cerrigone$^6$, F. Cavallaro$^{1,7,8}$}\\
$^1$INAF-Osservatorio Astrofisico di Catania, Via Santa Sofia 78, 95123 Catania, Italy\\
$^2$Millennium Institute of Astrophysics, 7591245 Santiago, Chile\\
$^3$Universidad Andr\'es Bello, Avda. Republica 252, 8320000 Santiago, Chile\\
$^4$Canada-France-Hawaii Telescope Corporation, 65-1238 Mamalahoa Hwy, Kamuela, HI 96743, USA\\
$^5$Physics and Astronomy Department, University of Hawaii Hilo, Hilo, HI 96720, USA\\
$^6$ASTRON, the Netherlands Institute for Radioastronomy, PO Box 2, 7990 AA Dwingeloo, The Netherlands\\
$^7$Universit\`a di Catania, Dipartimento di Fisica e Astronomia, Via Santa Sofia, 64, 95123 Catania, Italy\\
$^8$CSIRO Astronomy and Space Science, PO Box 76, Epping, NSW 1710, Australia
}

\begin{document}

\date{\textbf{Draft}}

\pagerange{\pageref{firstpage}--\pageref{lastpage}} \pubyear{2002}

\maketitle

\label{firstpage}

\begin{abstract}
We present radio observations of 18 MIPSGAL bubbles performed at $5\um{GHz}$ ($6\um{cm}$) with the Karl G. Jansky Very Large Array in configuration B and BnA. The observations were aimed at understanding what kind of information high-resolution and high-sensitivity radio maps can supply on the circumstellar envelopes of different kinds of evolved stars and what their comparison with infrared images with similar resolution can tell us. We found that the 18 bubbles can be grouped into five categories according to their radio morphology. The three bubbles presenting a central point source in the radio images all correspond to luminous blue variable star candidates. Eleven bubbles show an elliptical shape and the total lack of a central object in the radio, and are likely associated with planetary nebulae. Under this assumption we derive their distance, their ionized mass and their distribution on the Galactic plane. We discuss the possibility that the MIPSGAL bubbles catalogue (428 objects) may contain a large fraction of all Galactic planetary nebulae located at a distance between $1.4\um{kpc}$ and $6.9\um{kpc}$ and lying in the MIPSGAL field of view. Among the remaining bubbles we identify also a H \textsc{ii} region and a proto-planetary nebula candidate.

\end{abstract}


\section{Introduction}
The \textit{Spitzer Space Telescope} is permitting a great improvement of the knowledge on the celestial objects populating our Galaxy. Different surveys and targeted observations have allowed astronomers to observe through the Galactic dust, which prevents observations at shorter wavelengths, and many unknown sources have been unveiled. Thanks to the MIPSGAL Legacy Survey\footnote{http://mipsgal.ipac.caltech.edu/} \citep{Carey2009}, conducted with MIPS \citep[The Multiband Imaging Photometer for \textit{Spitzer};][]{Rieke2004}, \citet{Mizuno2010} identified at $24\mic{m}$ 428 roundish objects presenting diffuse emission. These small ($\lesssim1'$) rings, disks or shells (hereafter denoted as `bubbles') are pervasive throughout the entire Galactic plane in the mid-infrared (IR). The main hypothesis about the nature of the bubbles is that they are different types of evolved stars. In fact, follow up studies have found among them planetary nebulae (PNe), supernova remnants, Wolf--Rayet (WR) stars, luminous blue-variable (LBV) stars, asymptotic giant branch (AGB) stars, and so on (see \citealt{Nowak2014} for a review). However, currently, only about 30 per cent of the bubbles are classified. The discovery potential and the implications on the Galactic physics are huge both for high- and low-mass evolved stars.

Massive stars play a pivotal role in the evolution of their host galaxies. They are among the major contributors to the interstellar ultraviolet radiation and, via their strong stellar winds and final explosion, provide enrichment of processed material (gas and dust) and mechanical energy to the interstellar medium, strongly influencing subsequent local star formation. The details of post-main sequence (MS) evolution of massive stars are still poorly understood. Increasing the sample of these stars is fundamental to better constrain evolutionary model parameters. On the other hand the bubbles catalogue is the right place where to find new PNe. Currently only about 15 per cent of the expected Galactic PNe have been observed \citep{Sabin2014}. This difference is usually explained in terms of strong interstellar extinction in the Galactic plane.

In general, many kinds of evolved stars are known to be radio emitters. In the last phases of their life the stars undergo several periods of enhanced mass-loss, with their external layers ejected to form a circumstellar envelope (CSE). The now-exposed hot inner part of the star can ionize the CSE and continuum radio emission originates from it. The radio emission is therefore a helpful probe to complement IR observations in classifying the bubbles and deriving their physical properties, being a powerful tool to investigate circumstellar ionized material. Between 2010 and 2012 we carried out radio continuum observations with the Karl G.~Jansky Very Large Array (VLA) at $5\um{GHz}$ (configuration D) and at $1.4\um{GHz}$ (configuration C and CnB) of a subset of 55 bubbles (\citealt{Ingallinera2014a}; `Paper I' hereafter). We were able to calculate the radio spectral index ($\alpha$ defined as $S\propto\nu^\alpha$, where $S$ is the flux density and $\nu$ the frequency) for 18 of them and to determine an upper limit for 13. We found that at least 70 per cent of them have spectral indices compatible with a thermal emission. Since one of the main goal of Paper I was to accurately calculate the radio flux densities, we chose VLA compact configurations to avoid issues from the lack of short-spacing information. The drawback was a poorer resolution with respect to IR images, and morphological comparisons proved impossible. To overcome this problem, in this paper we present new VLA observations at $5\um{GHz}$ of 18 bubbles, using the more extended B and BnA configurations. The main goal of this work is to show how high-resolution and high-sensitivity radio images are an extremely powerful tool to discriminate low- and high-mass evolved stars, derive physical properties of PNe, disclose the mutual interaction of massive star ejecta and, in synergy with 8-$\umu$m images, immediately recognize possible H \textsc{ii} regions among the bubbles.

In Section \ref{sec:obs} we supply technical details of observations and data reduction, grouping the observed bubbles by their radio morphology. In Sections \ref{sec:bub_cs}, \ref{sec:ell}, \ref{sec:fill} and \ref{sec:others} we discuss each morphological category separately. In Section \ref{sec:PNe} we comment the result obtained on PNe. Summary and conclusions are reported in Section \ref{sec:sumcon}.

\section{Observations and data reduction}
\label{sec:obs}
\subsection{Sample selection and observations}
In Paper I we had found that 44 out of 55 bubbles were unambiguously detected at $5\um{GHz}$, with sensitivity limits around $\sim\!100\mic{Jy\ beam}^{-1}$. For 18 of them we were able to measure their flux density both at $1.4\um{GHz}$ and at $5\um{GHz}$. This was the starting sample for this follow-up work. However two of them (MGE 040.3704-00.4750 and MGE 031.9075-00.3087) resulted too much extended to be properly imaged with VLA (being more extended than the VLA largest angular scale) and were excluded. Other two bubbles (MGE 034.8961+00.3018 and MGE 027.3839-00.3031), non-detected at $1.4\um{GHz}$, were included in the sample. We selected these 18 bubbles for higher-resolution 6-cm images. The sample was split in two groups, each one containing 9 objects, depending on the source declination. For bubbles with declination above $-20^\circ$ we used the VLA in configuration B (baselines from $210\um{m}$ to $11.1\um{km}$), while for the others we opted for configuration BnA. Our choice was driven by the better imaging performance of VLA hybrid configurations when observing low-declination sources.

The observations were carried out in March 2015 (configuration B) and in May 2015 (configuration BnA). The flux calibrator was, for both groups, 3C286, while different gain calibrators were chosen from the standard VLA calibrators list, selected to be as close as possible to targeted sources. The time on source for the bubbles varied from 8 to 20 minutes, as reported in Table \ref{tab:obs}.

\begin{table*}
\caption{Observation summary. Flux densities are taken from Paper I. ($^a$\citealt{Pottasch1988}; $^b$\citealt{Anderson2011}; $^c$\citealt{Flagey2014}; $^d$\citealt{Gvaramadze2010}; $^e$\citealt{Clark2005})}
\begin{tabular}{ccccccc}\hline
[MGE] & $\alpha$ & $\delta$ & Config. & Obs. time & Flux density & Classification\\
& (J2000) & (J2000) & & (min) & at $6\um{cm}$ (mJy)\\\hline
002.2128-01.6131 & $\HAng{17}{57}{03.9}$ & $-\DAng{27}{51}{30}$ & BnA & 20 & $\phantom{0}1.6\pm0.1$\\ 
005.2641+00.3775 & $\HAng{17}{56}{13.4}$ & $-\DAng{24}{13}{13}$ & BnA & 10 & $\phantom{0}9.0\pm0.2$\\ 
005.6102-01.1516 & $\HAng{18}{02}{48.4}$ & $-\DAng{24}{40}{54}$ & BnA & 10 & $38.1\pm0.8$\\ 
008.9409+00.2532 & $\HAng{18}{04}{36.3}$ & $-\DAng{21}{05}{26}$ & BnA & 10 & $\phantom{0}8.2\pm0.4$\\ 
009.3523+00.4733 & $\HAng{18}{04}{38.9}$ & $-\DAng{20}{37}{27}$ & BnA & 10 & $17.7\pm0.7$ & PN$^a$\\ 
016.2280-00.3680 & $\HAng{18}{21}{36.9}$ & $-\DAng{14}{59}{41}$ & B & 8 & $10.3\pm0.6$ & PN?$^b$\\ 
027.3839-00.3031 & $\HAng{18}{42}{22.5}$ & $-\DAng{05}{04}{29}$ & B & 8.5 & $18.1\pm0.5$\\ 
028.7440+00.7076 & $\HAng{18}{41}{16.0}$ & $-\DAng{03}{24}{11}$ & B & 9 & $\phantom{0}6.4\pm0.3$\\ 
030.1503+00.1237 & $\HAng{18}{45}{55.2}$ & $-\DAng{02}{25}{08}$ & B & 8 & $22.9\pm1.5$ & LBV?$^c$\\ 
031.7290+00.6993 & $\HAng{18}{46}{45.2}$ & $-\DAng{00}{45}{06}$ & B & 8 & $12.3\pm0.1$ & H \textsc{ii} region?$^b$\\ 
032.4982+00.1615 & $\HAng{18}{50}{04.3}$ & $-\DAng{00}{18}{45}$ & B & 8 & $\phantom{0}4.7\pm0.9$ & O5-6V?$^c$\\ 
034.8961+00.3018 & $\HAng{18}{53}{56.8}$ & $\phantom{-}\DAng{01}{53}{08}$ & B & 9 & $17.8\pm0.4$\\ 
042.0787+00.5084 & $\HAng{19}{06}{24.6}$ & $\phantom{-}\DAng{08}{22}{02}$ & B & 8 & $10.5\pm0.1$ & LBV?$^c$\\ 
042.7665+00.8222 & $\HAng{19}{06}{33.6}$ & $\phantom{-}\DAng{09}{07}{20}$ & B & 8 & $12.7\pm0.4$ & WC5-6$^c$; [WC]$^d$; PN?$^e$\\ 
352.3117-00.9711 & $\HAng{17}{29}{58.3}$ & $-\DAng{35}{56}{56}$ & BnA & 10 & $15.5\pm0.9$\\ 
356.1447+00.0550 & $\HAng{17}{35}{54.5}$ & $-\DAng{32}{10}{35}$ & BnA & 10 & $15.0\pm0.7$\\ 
356.7168-01.7246 & $\HAng{17}{44}{29.6}$ & $-\DAng{32}{38}{11}$ & BnA & 12 & $\phantom{0}5.8\pm0.1$\\ 
356.8155-00.3843 & $\HAng{17}{39}{21.3}$ & $-\DAng{31}{50}{44}$ & BnA & 10 & $15.1\pm0.4$\\ 
\hline
\end{tabular}
\label{tab:obs}
\end{table*}

We used the $C$-band receiver with a total bandwidth of $2\um{GHz}$, namely from 4 to $6\um{GHz}$, in order to have approximately the same central frequency as for the D-configuration observations described in Paper I. We used the pseudo-continuum observing mode with 1024 2-MHz-wide channels, divided in 16 spectral windows, to easily identify and flag possible narrow-spectrum radio-frequency interferences (RFI).

\subsection{Data reduction}
\label{sec:data_red}
A preliminary editing and calibration was applied by the NRAO staff with the CASA calibration pipeline version 1.3.1. After a careful check of the pipeline output, we found that the calibration process was performed perfectly, while a further bad data flagging was needed. RFI only slightly affected observations and more than 80 per cent of the data were useful.

A first imaging test was performed for all the bubbles that had not been resolved in Paper I to determine their angular size. The $uv$ data of bubbles with size greater than the VLA B-configuration largest angular scale ($\sim\!15\um{arcsec}$) were concatenated to the respective visibilities in D-configuration from our previous observations, in order to mitigate flux-loss issues and imaging artefacts. For the remaining bubbles such a concatenation was not needed and was not performed, since we found that, in general, it increases the background noise. The noise increase is probably due to the higher noise of the D-configuration data and to the fact that we use the `natural' weighting (inner part of the $uv$ plane weighted more than the outer part). The final imaging process was performed with CASA using the task \textsc{clean} and the multi-scale algorithm \citep{Cornwell2008}. To take into account wide-band effects we used the multi-frequency synthesis algorithm considering the first three terms of the Taylor expansion \citep{Rau2011}. The achieved angular resolution was about $2\um{arcsec}$. This algorithm also returned a spectral index map. In Table \ref{tab:data} we report the angular size of each bubble and the rms of the final map.

\begin{table}
\caption{Image data. Radio morphology: `C' bubbles with a central object; `E' elliptical; `F' filled elliptical.}
\begin{tabular}{cccc}\hline 
[MGE] & Diameter & Map rms & Radio\\
& (arcsec) & ($\umu\mathrm{Jy\ beam}^{-1}$) & morph.\\\hline
002.2128-01.6131 & 13 & $\phantom{0}9.0$ & F\\ 
005.2641+00.3775 & 14 & 12.4 & E\\ 
005.6102-01.1516 & 18 & 15.8 & E\\ 
008.9409+00.2532 & 16 & 11.8 & E\\ 
009.3523+00.4733 & 14 & 13.7 & E\\ 
016.2280-00.3680 & 17 & 13.7 & E\\ 
027.3839-00.3031 & 68 & 11.0 & C\\ 
028.7440+00.7076 & 33 & 12.2 & O\\ 
030.1503+00.1237 & 13 & 23.7 & C\\ 
031.7290+00.6993 & 31 & 10.1 & O\\ 
032.4982+00.1615 & $\phantom{0}8$ & 20.0 & O\\ 
034.8961+00.3018 & 24 & 12.5 & E\\ 
042.0787+00.5084 & 20 & 19.5 & C\\ 
042.7665+00.8222 & 13 & 11.8 & F\\ 
352.3117-00.9711 & 22 & 13.6 & E\\ 
356.1447+00.0550 & 30 & 13.4 & F\\ 
356.7168-01.7246 & 17 & 11.3 & B\\ 
356.8155-00.3843 & 20 & 12.1 & F\\ 
\hline
\end{tabular}
\label{tab:data}
\end{table}

\subsection{Radio morphology}
\begin{figure*}
\includegraphics[height=5cm]{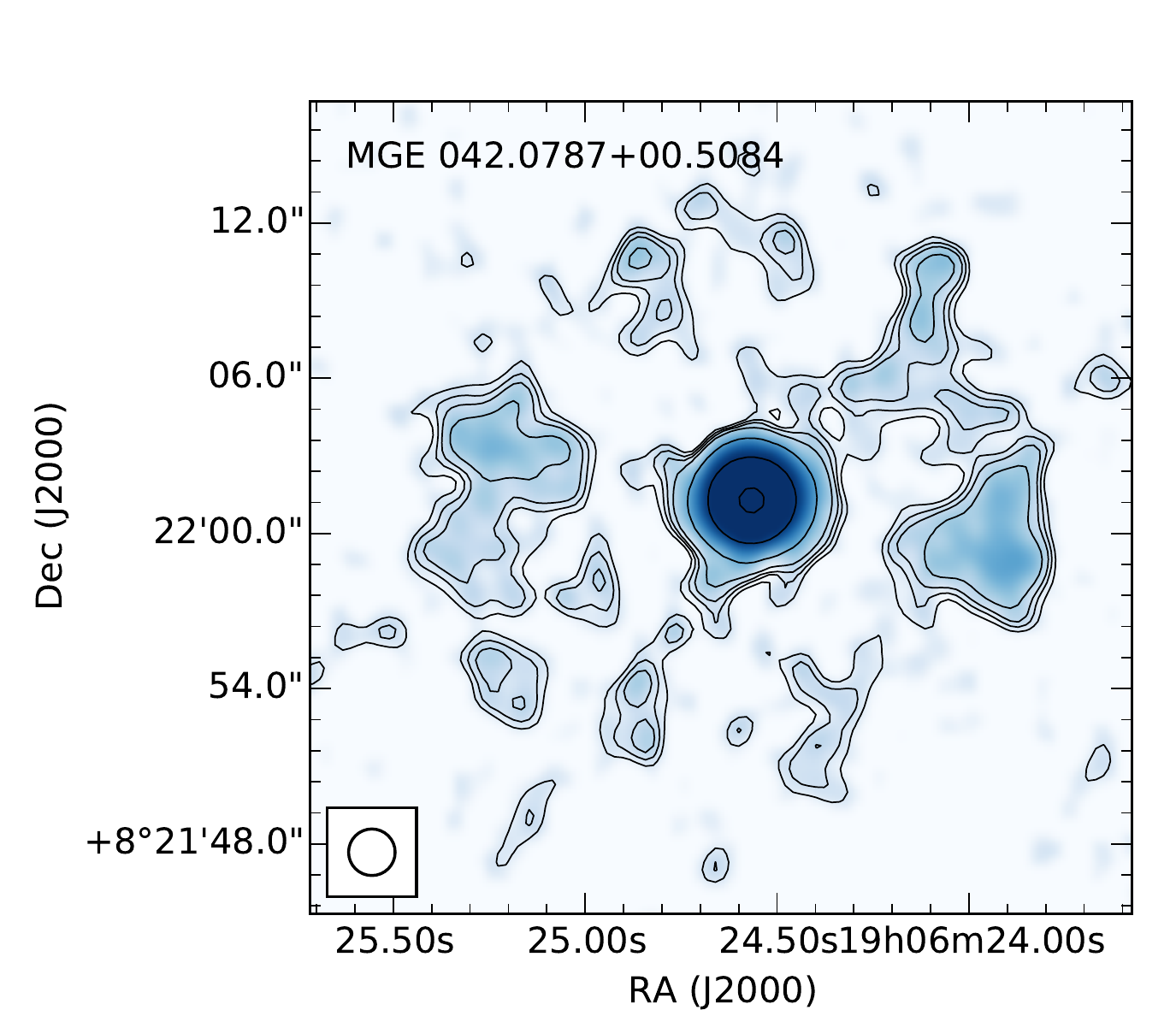}
\includegraphics[height=5cm]{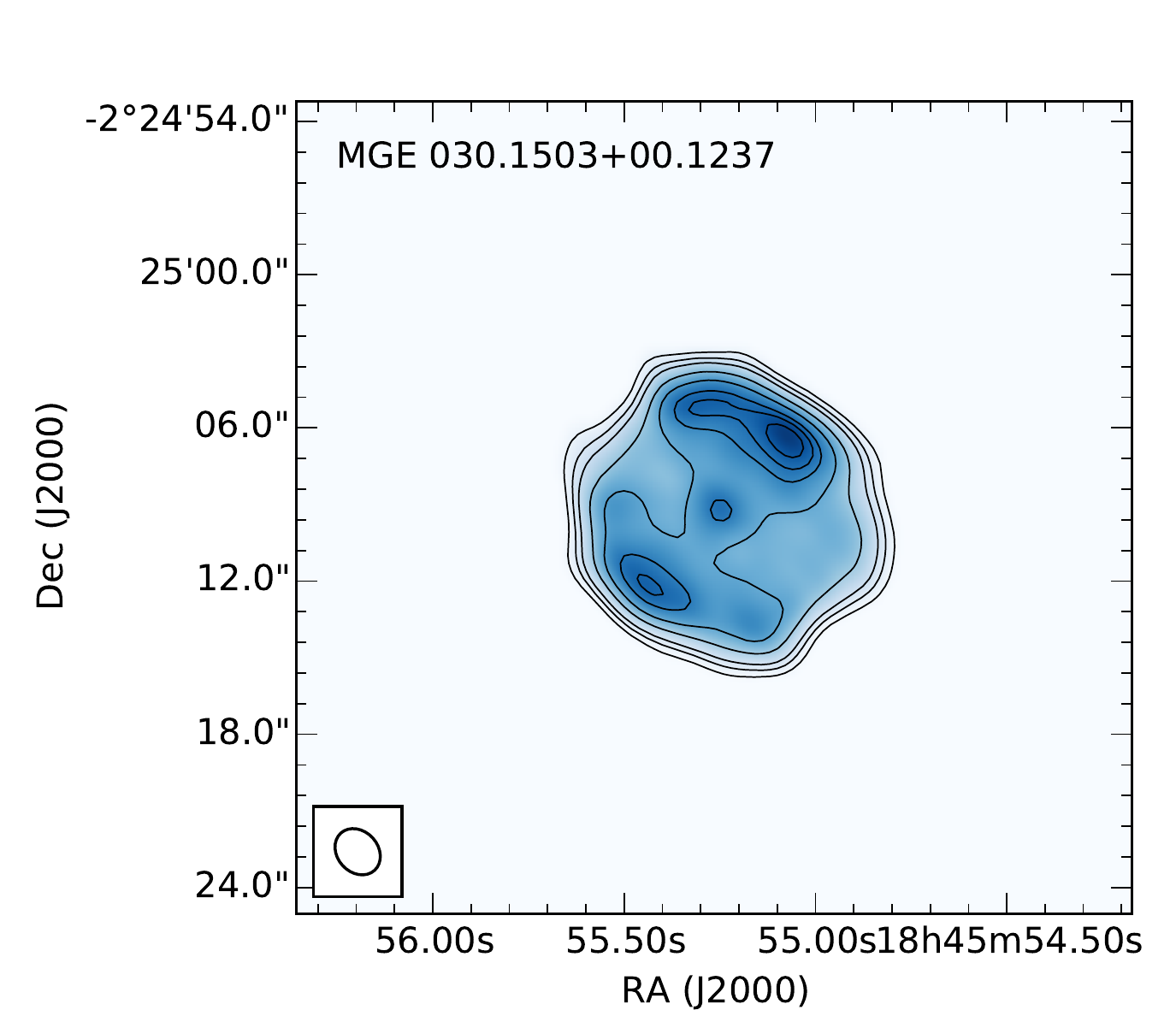}
\includegraphics[height=5cm]{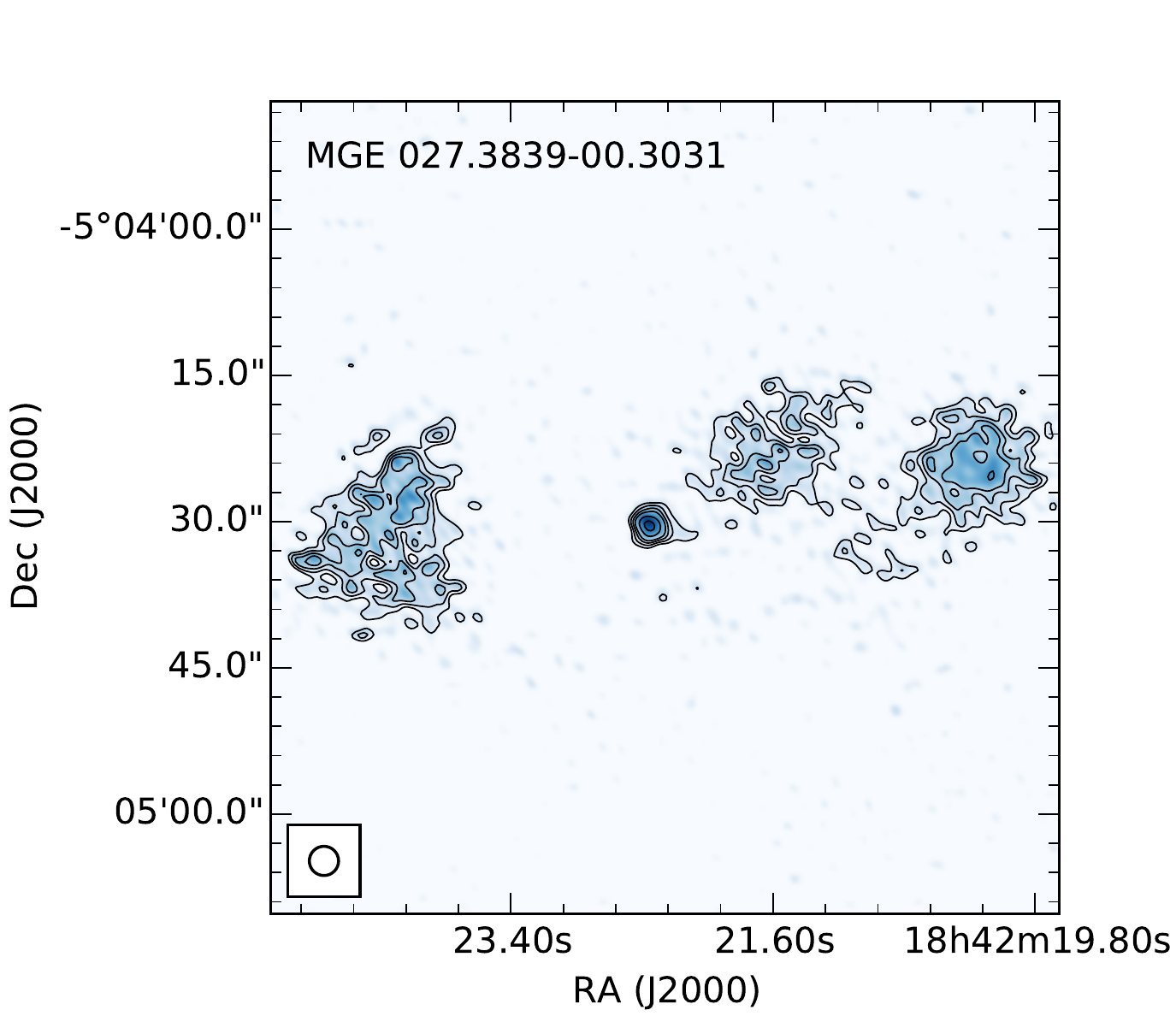}
\caption{6-cm images of the bubbles with central object (morphology `C' of Table \ref{tab:data}). From left to right: MGE 042.0787+00.5084, MGE 030.1503+00.1237 and MGE 027.3839-00.3031.}
\label{fig:bubbles_C}
\end{figure*}

\begin{figure*}
\includegraphics[height=5cm]{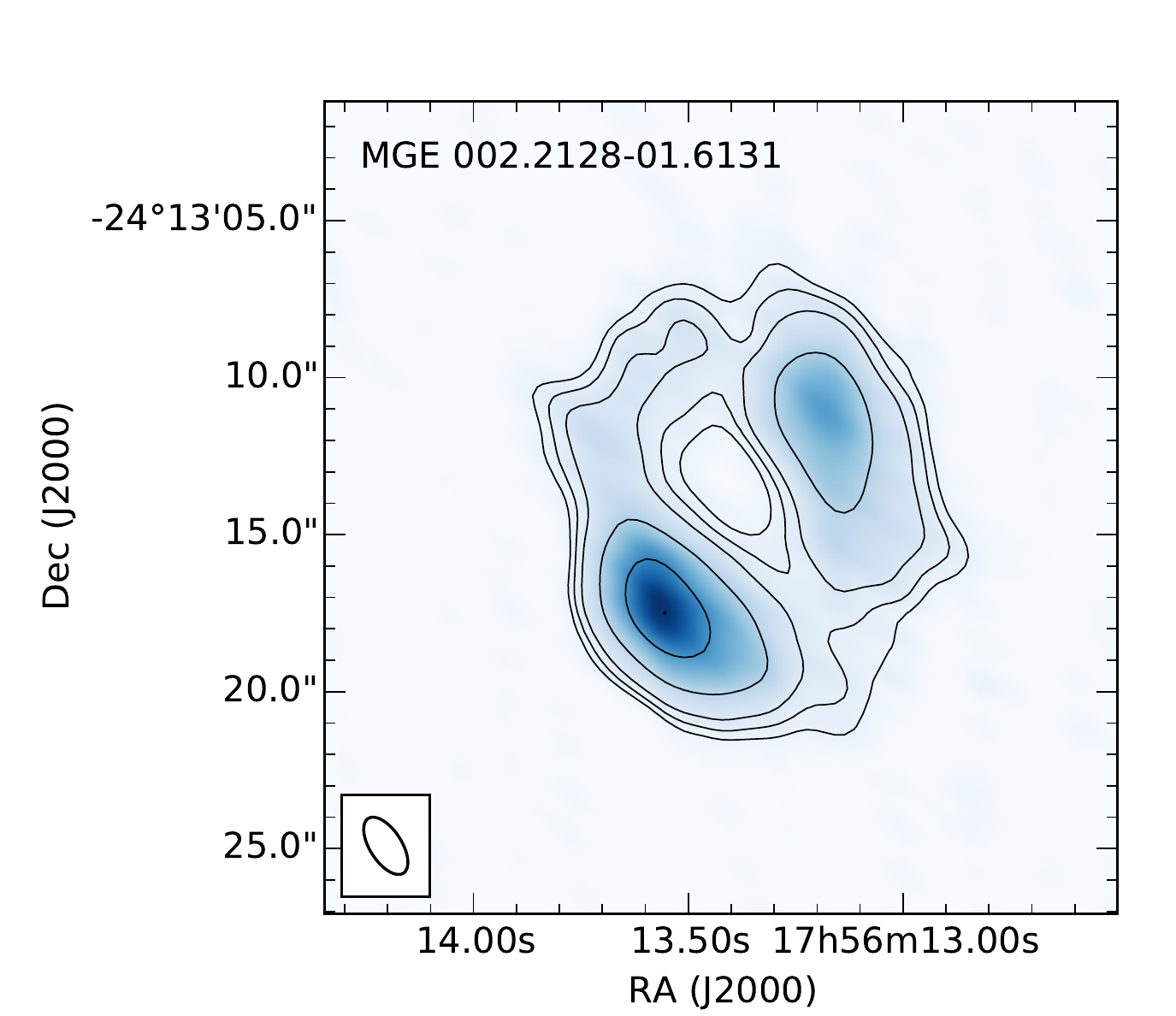}
\includegraphics[height=5cm]{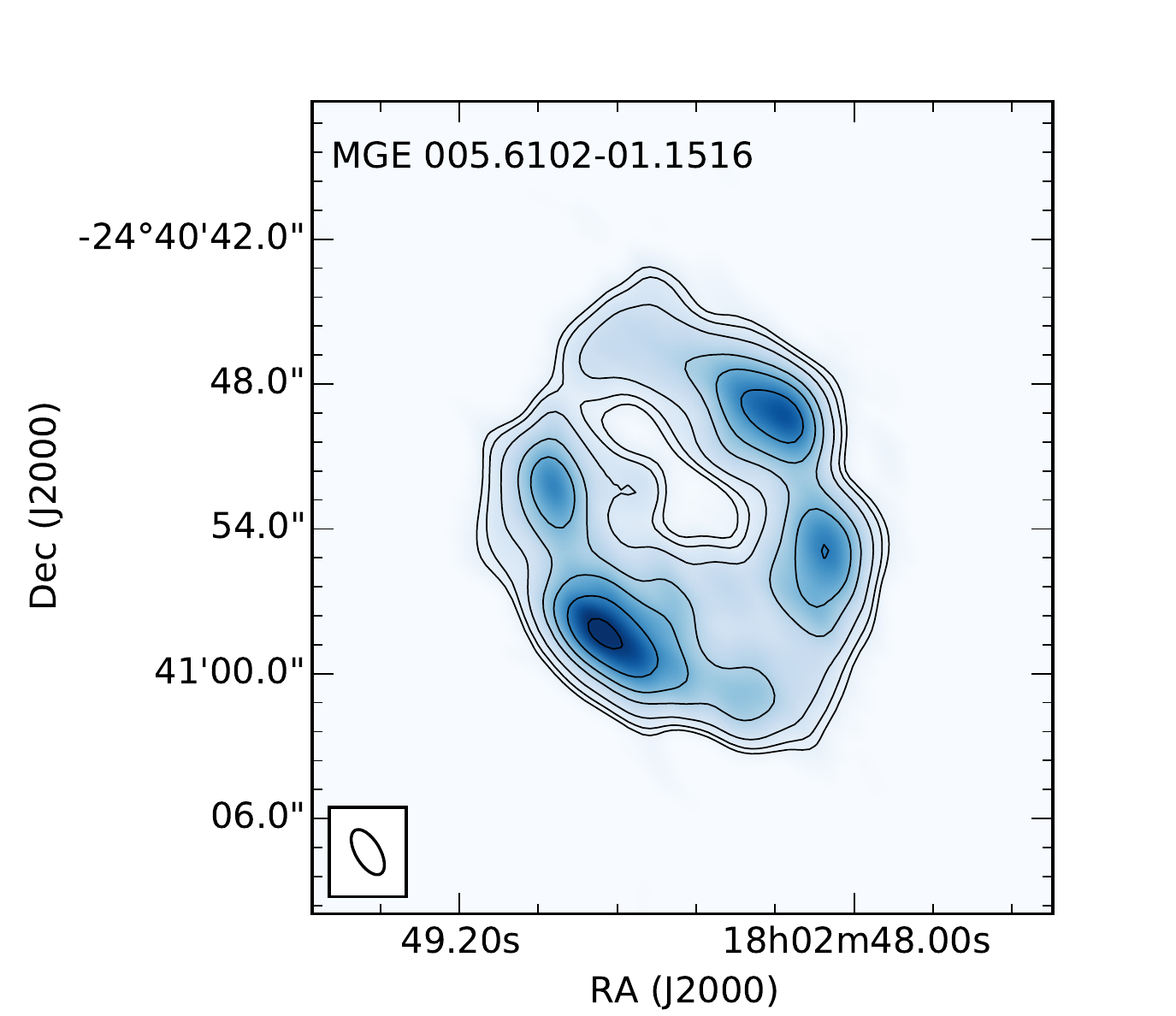}
\includegraphics[height=5cm]{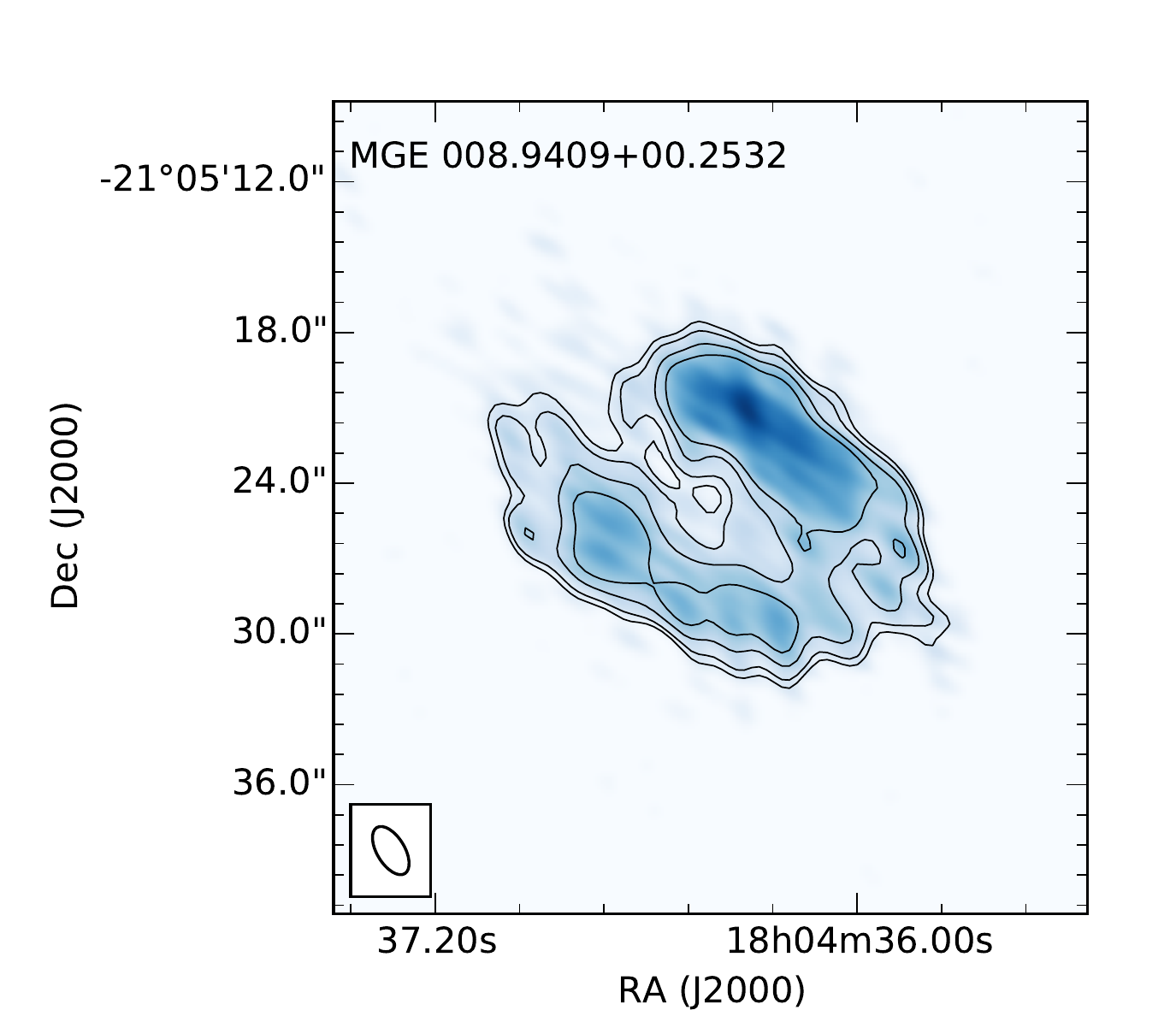}
\includegraphics[height=5cm]{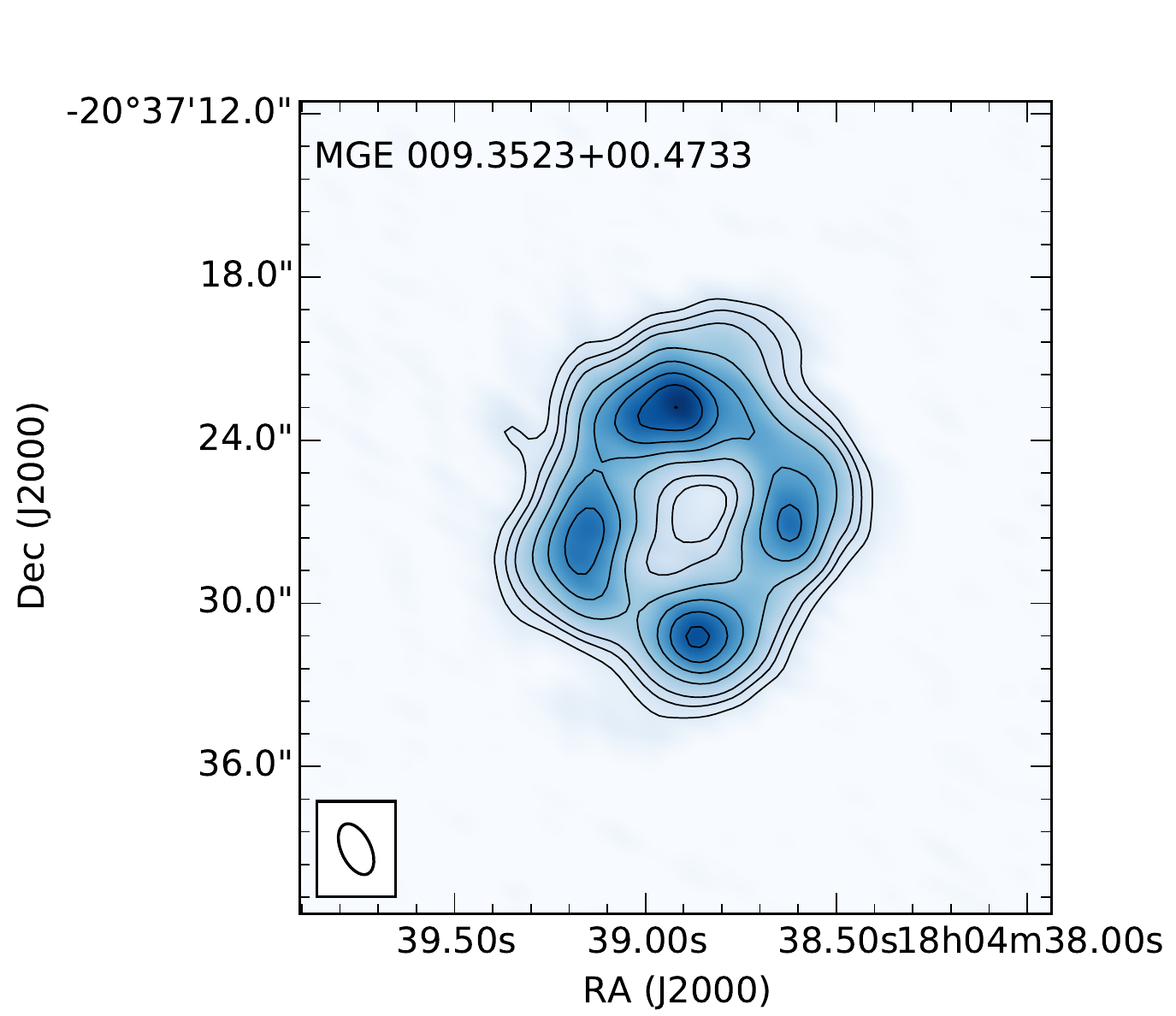}
\includegraphics[height=5cm]{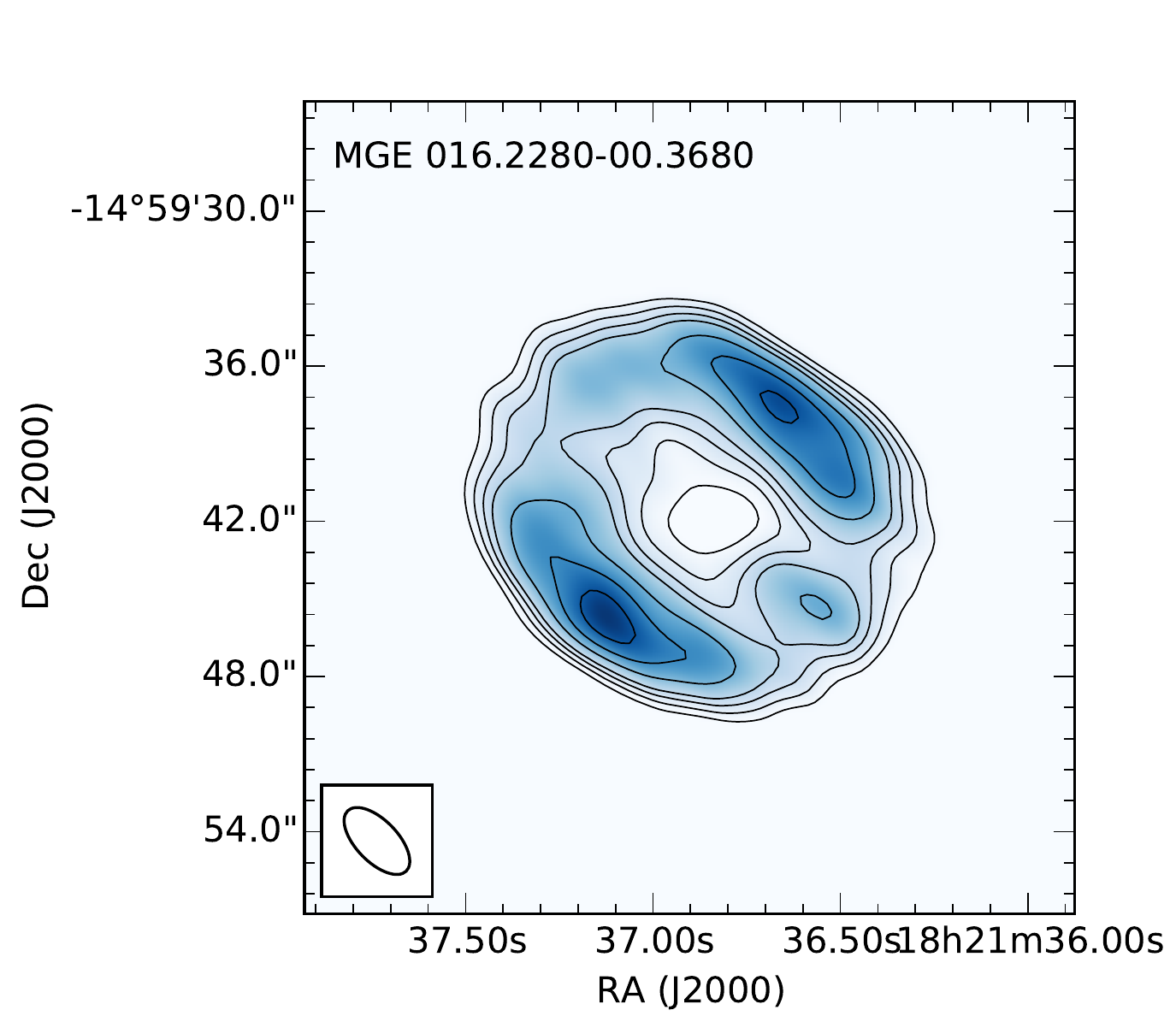}
\includegraphics[height=5cm]{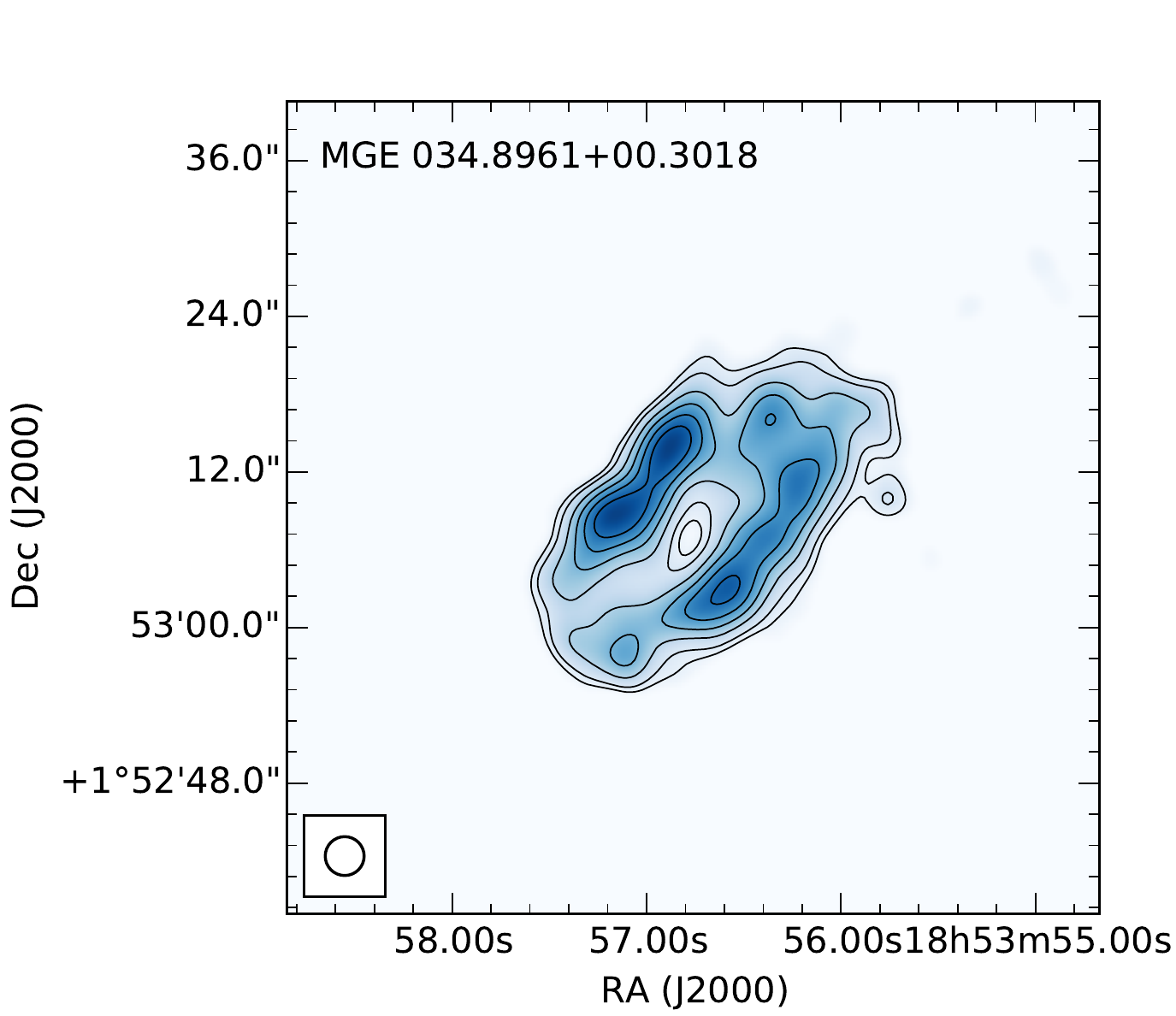}
\includegraphics[height=5cm]{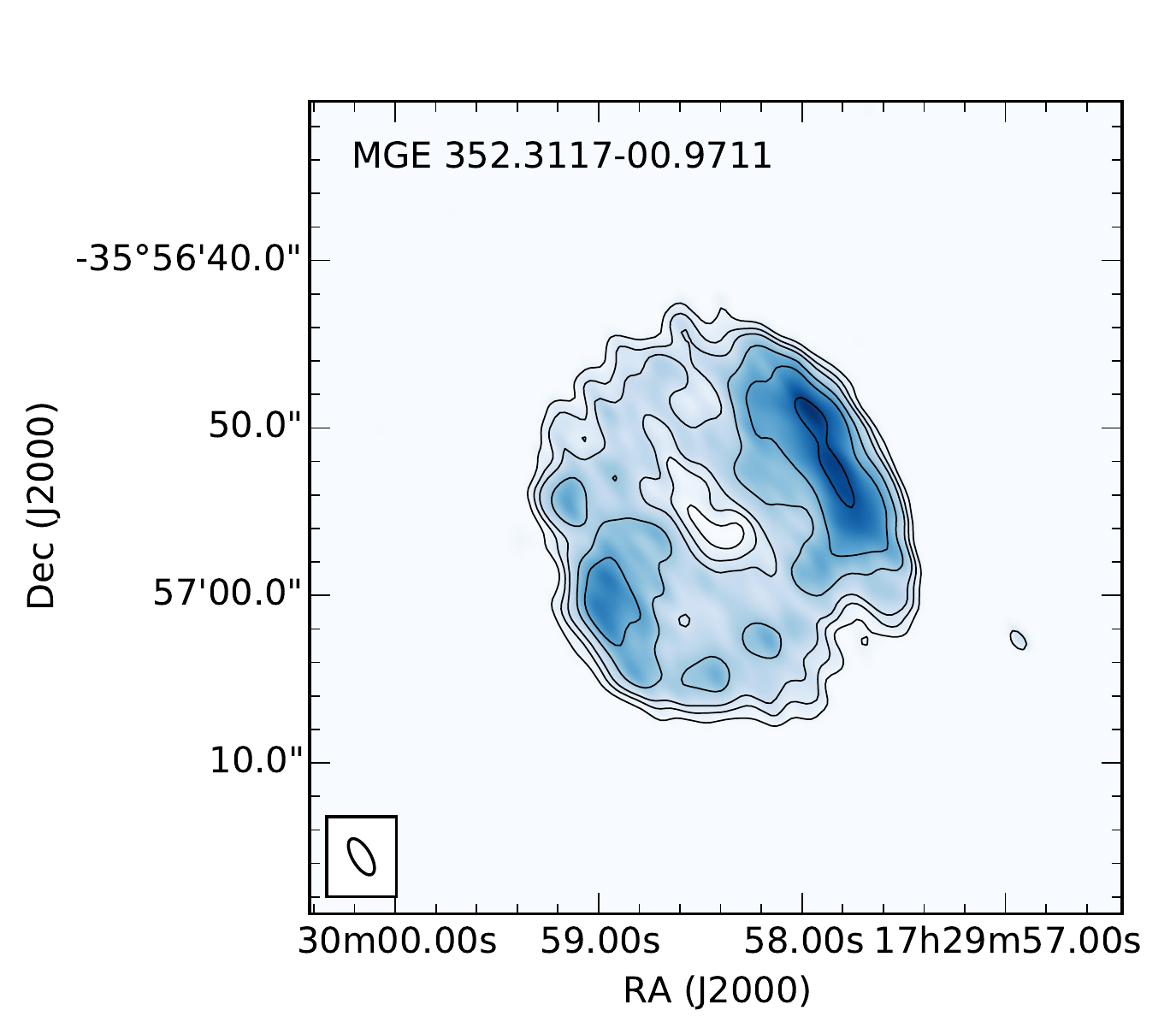}
\caption{6-cm images of the `elliptical' bubbles (morphology `E' of Table \ref{tab:data}). From left to right: (row 1) MGE 002.2128-01.6131, MGE 005.6102-01.1516, MGE 008.9409+00.2532; (row 2) MGE 009.3523+00.4733, MGE 016.2280-00.3680, MGE 034.8961+00.3018; (row 3) MGE 352.3117-00.9711.}
\label{fig:bubbles_E}
\end{figure*}

\begin{figure*}
\includegraphics[height=5.5cm]{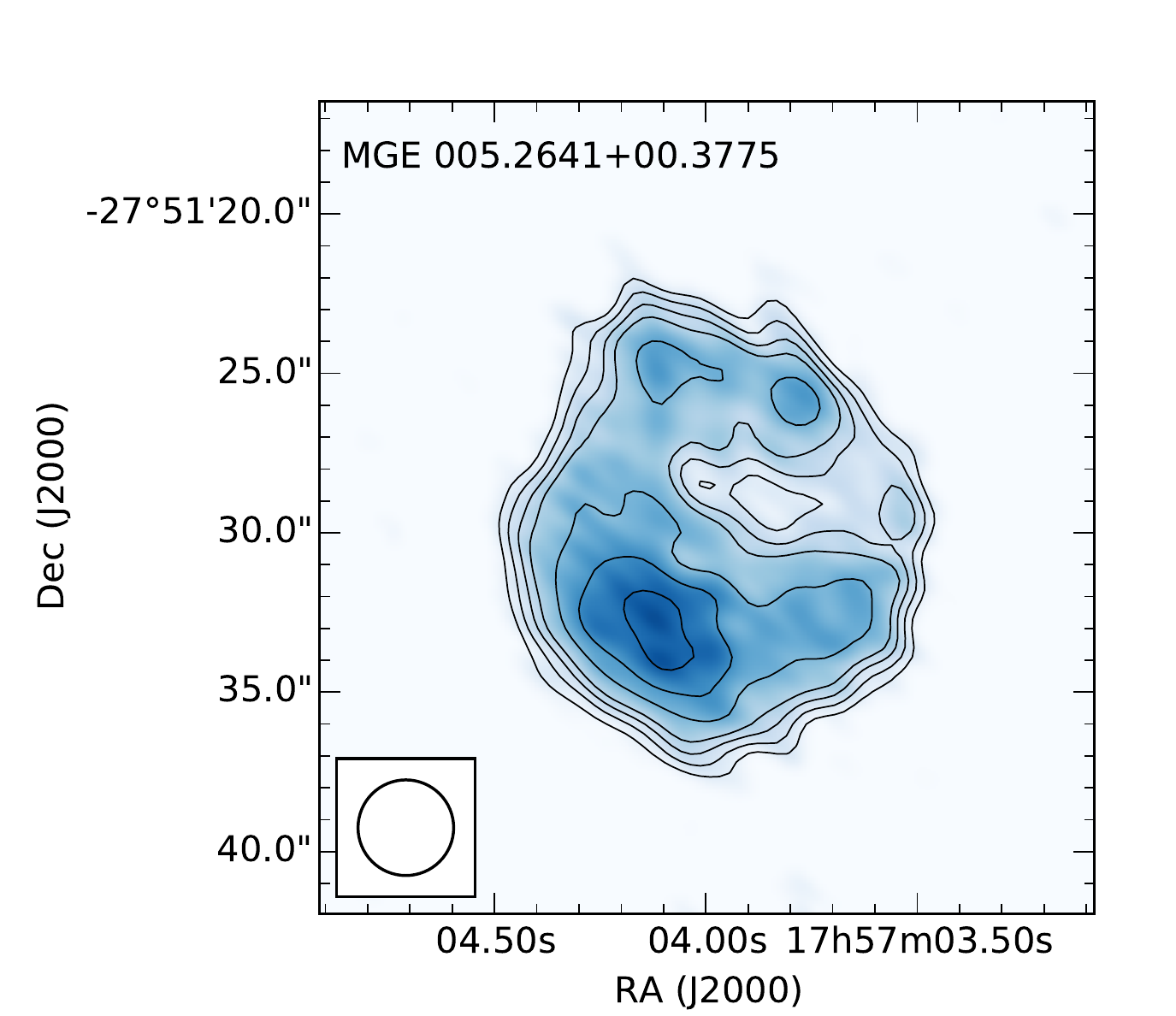}
\includegraphics[height=5.5cm]{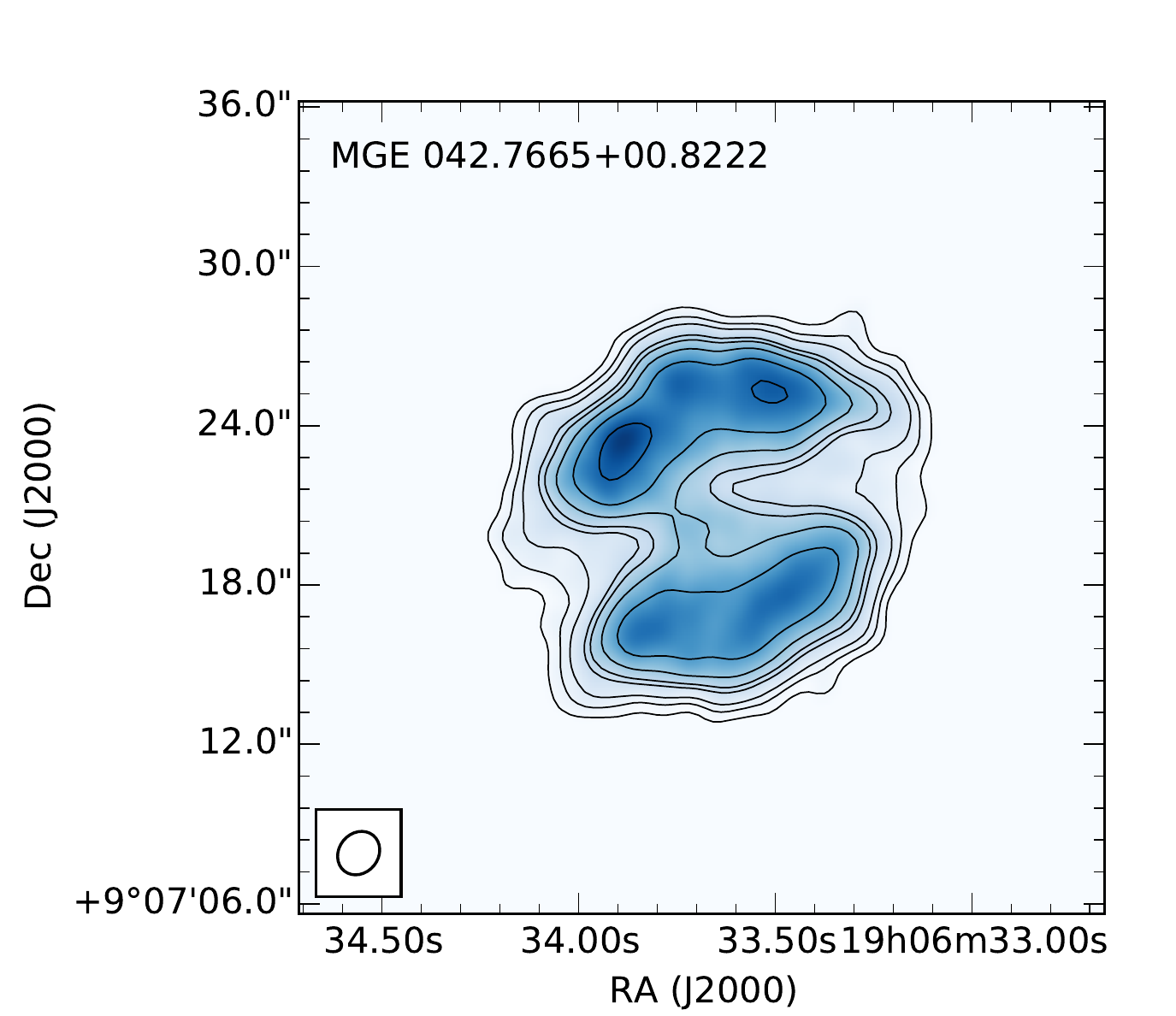}
\includegraphics[height=5.5cm]{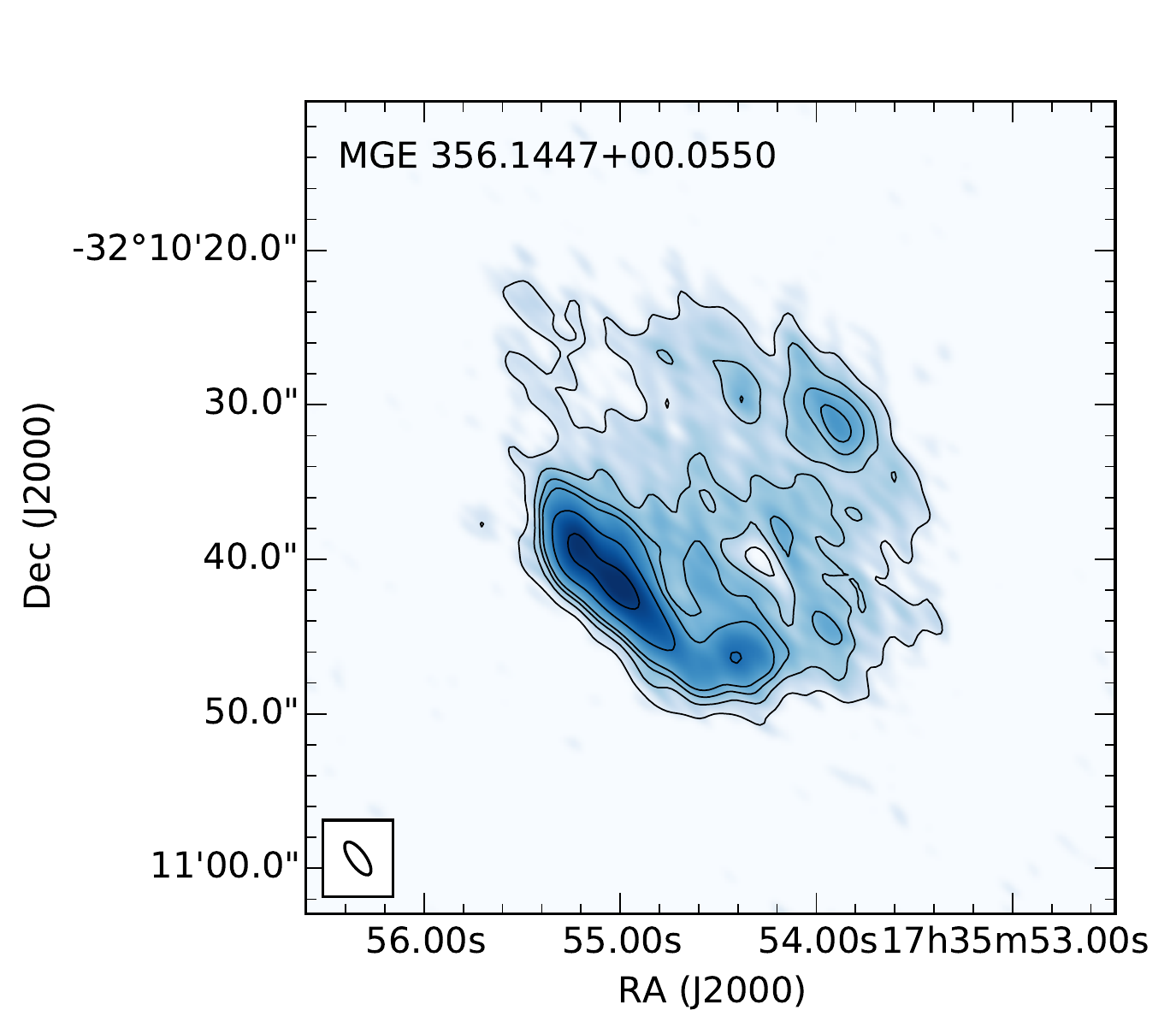}
\includegraphics[height=5.5cm]{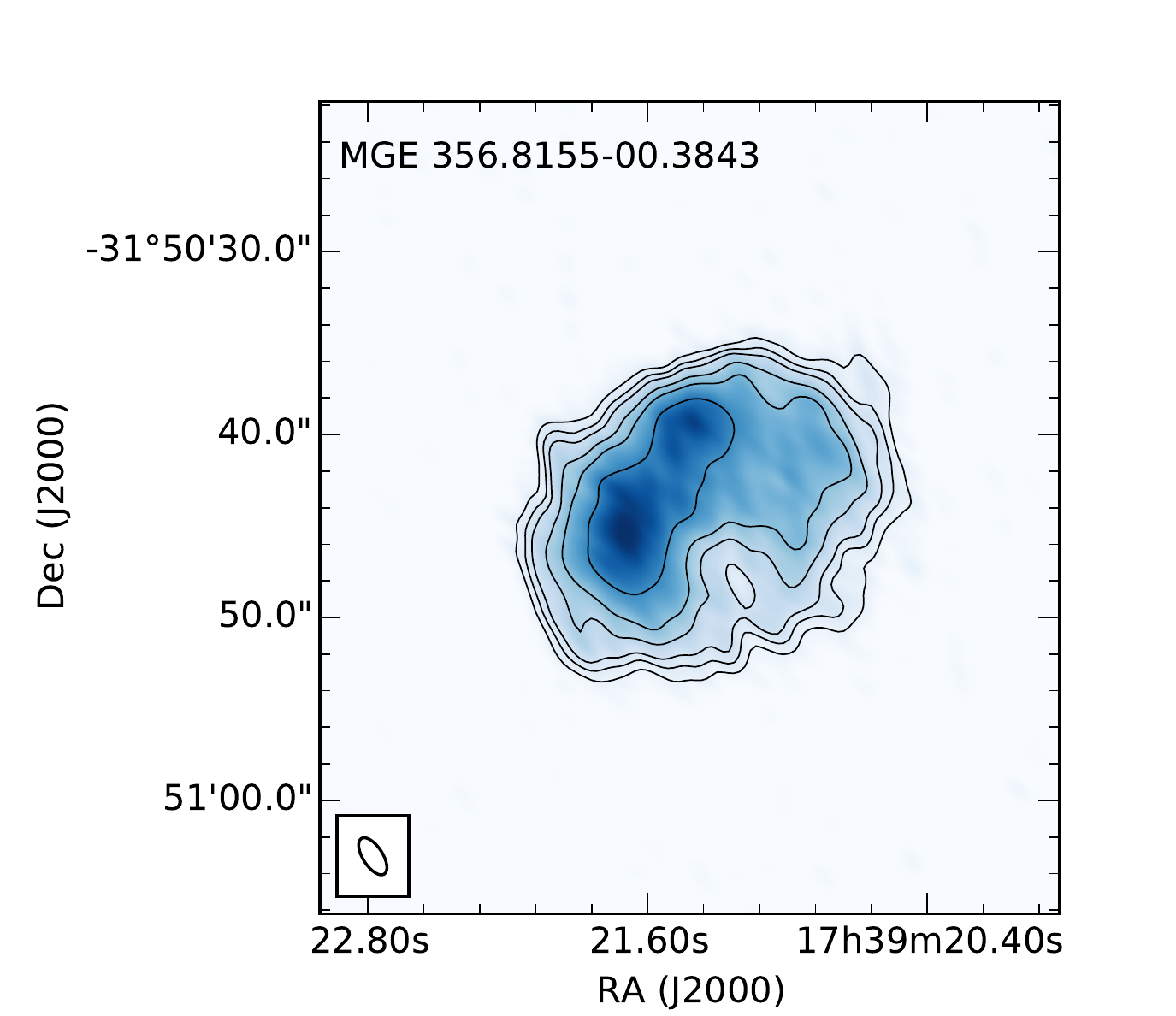}
\caption{6-cm images of the `filled elliptical' bubbles (morphology `F' of Table \ref{tab:data}). From left to right: (row 1) MGE 005.2641+00.3775, MGE 042.7665+00.8222; (row 2) MGE 356.1447+00.0550, MGE 356.8155-00.3843.}
\label{fig:bubbles_F}
\end{figure*}

\begin{figure*}
\includegraphics[height=5cm]{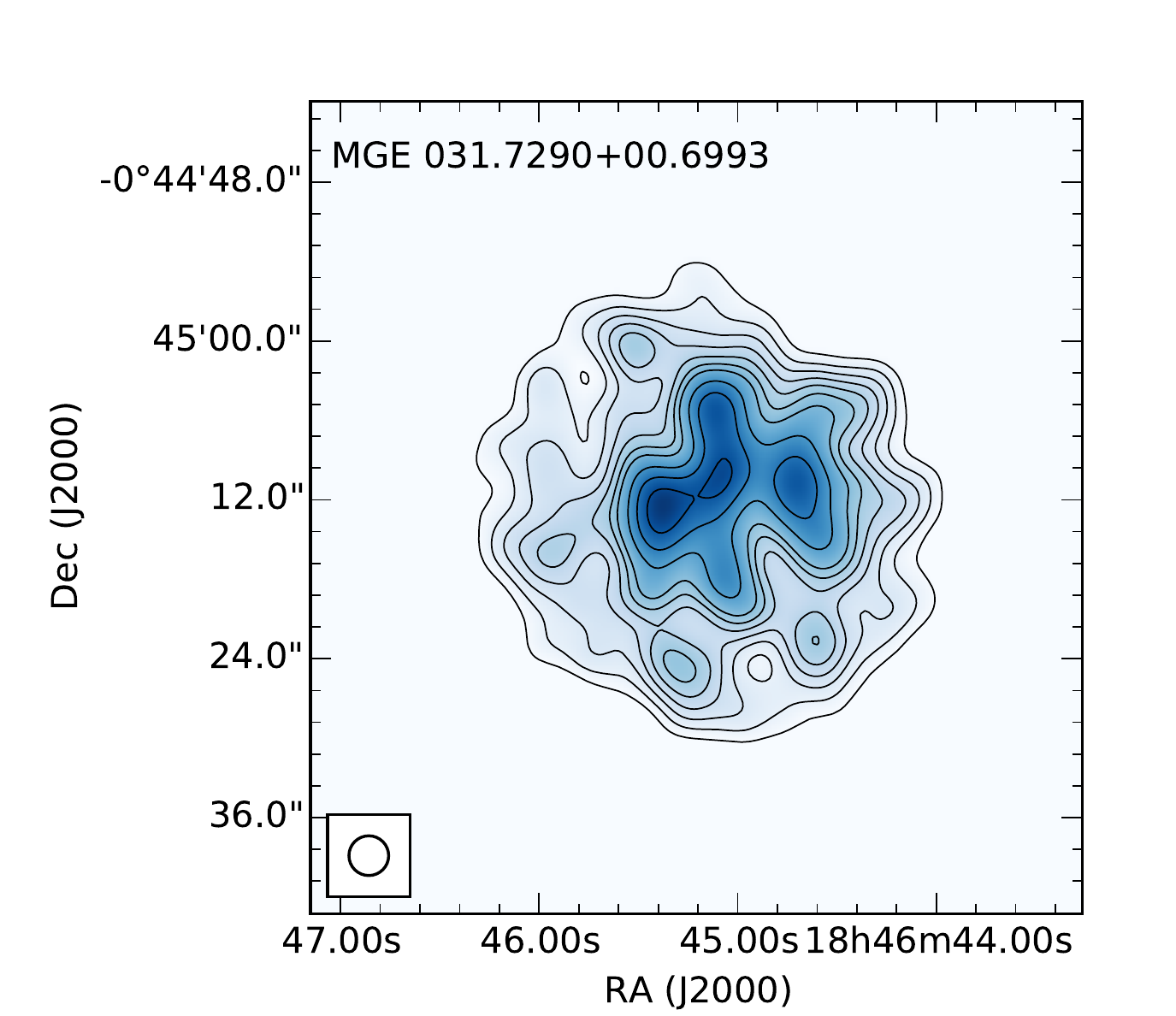}
\includegraphics[height=5cm]{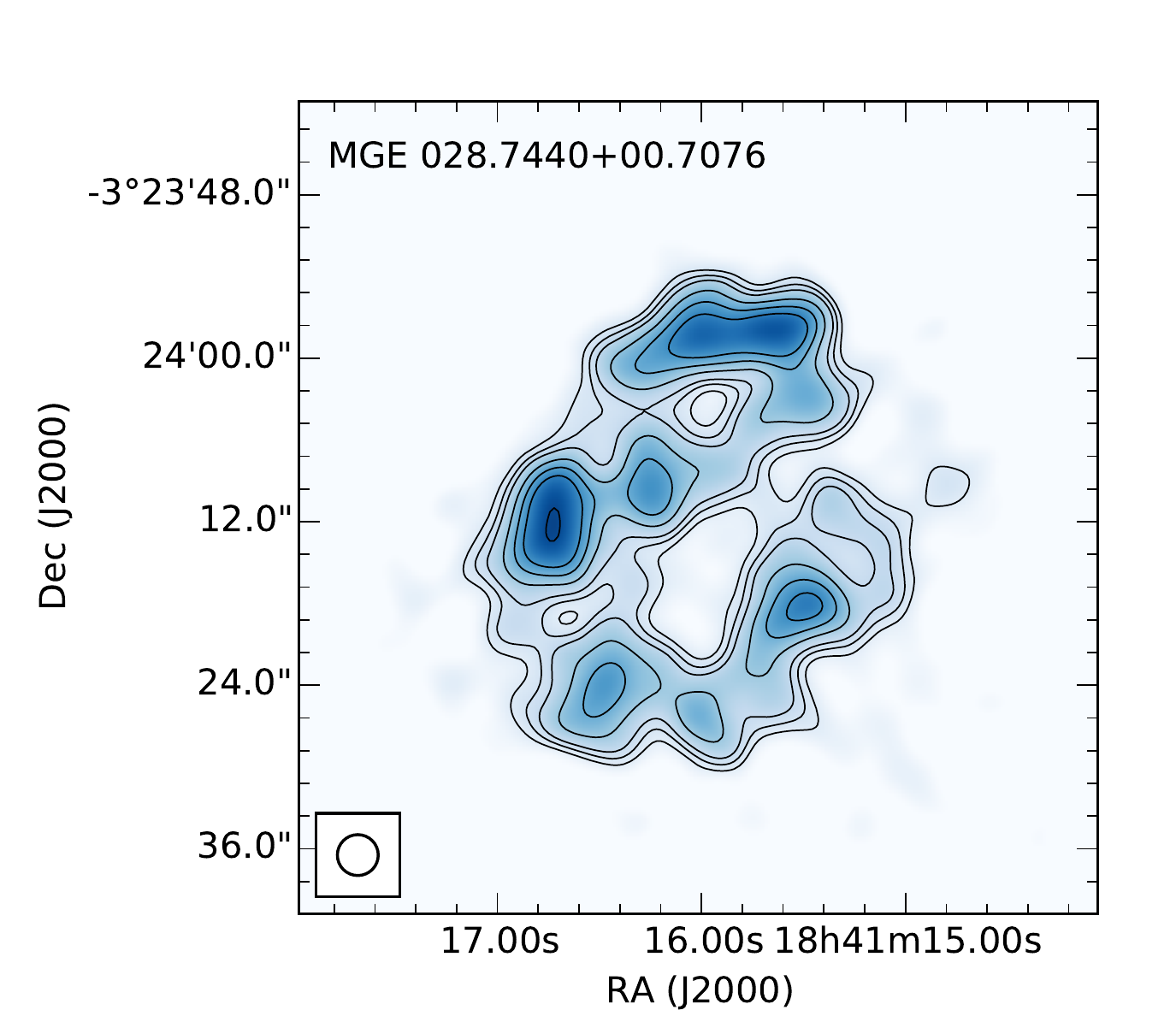}
\includegraphics[height=5cm]{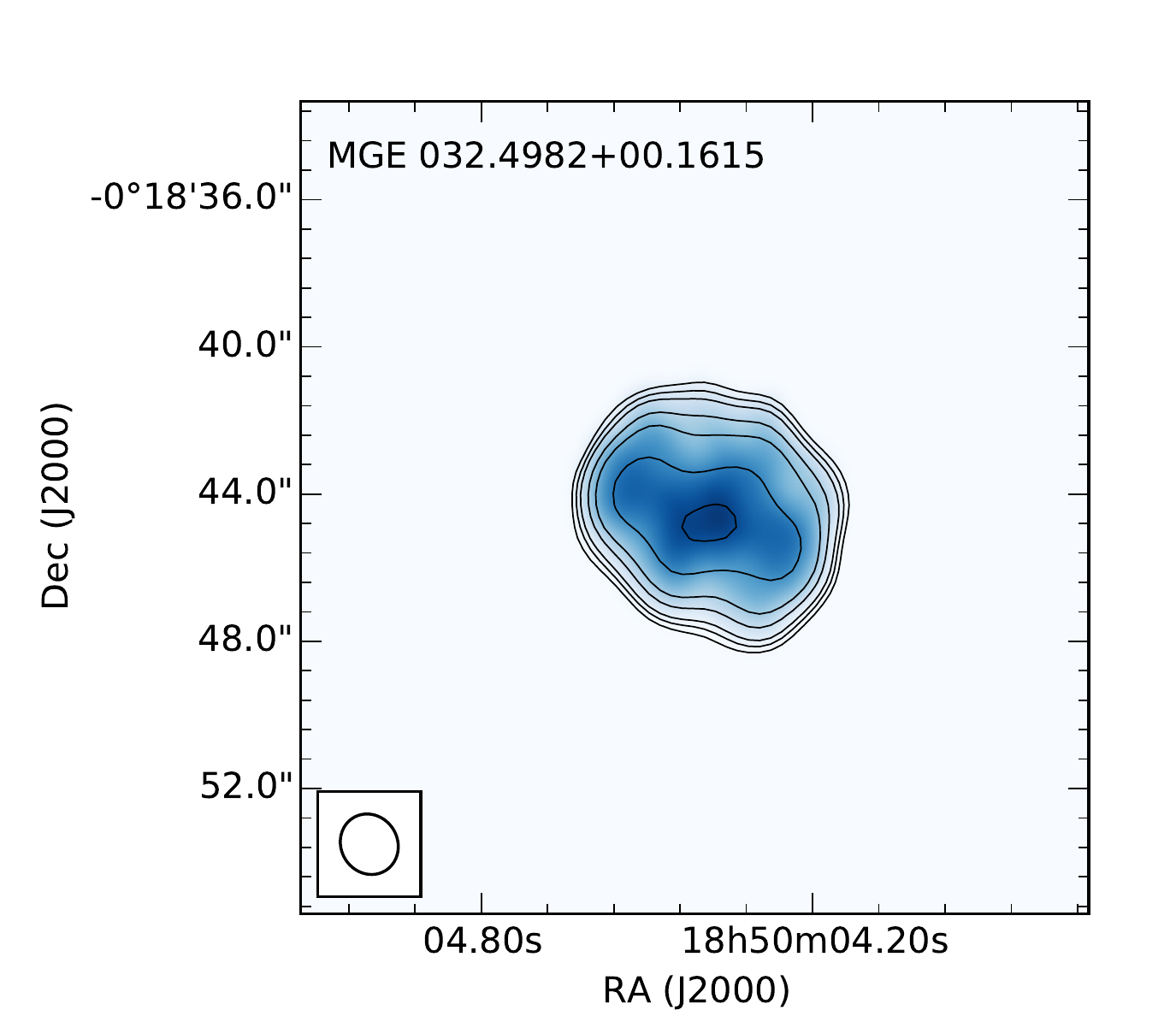}
\includegraphics[height=5cm]{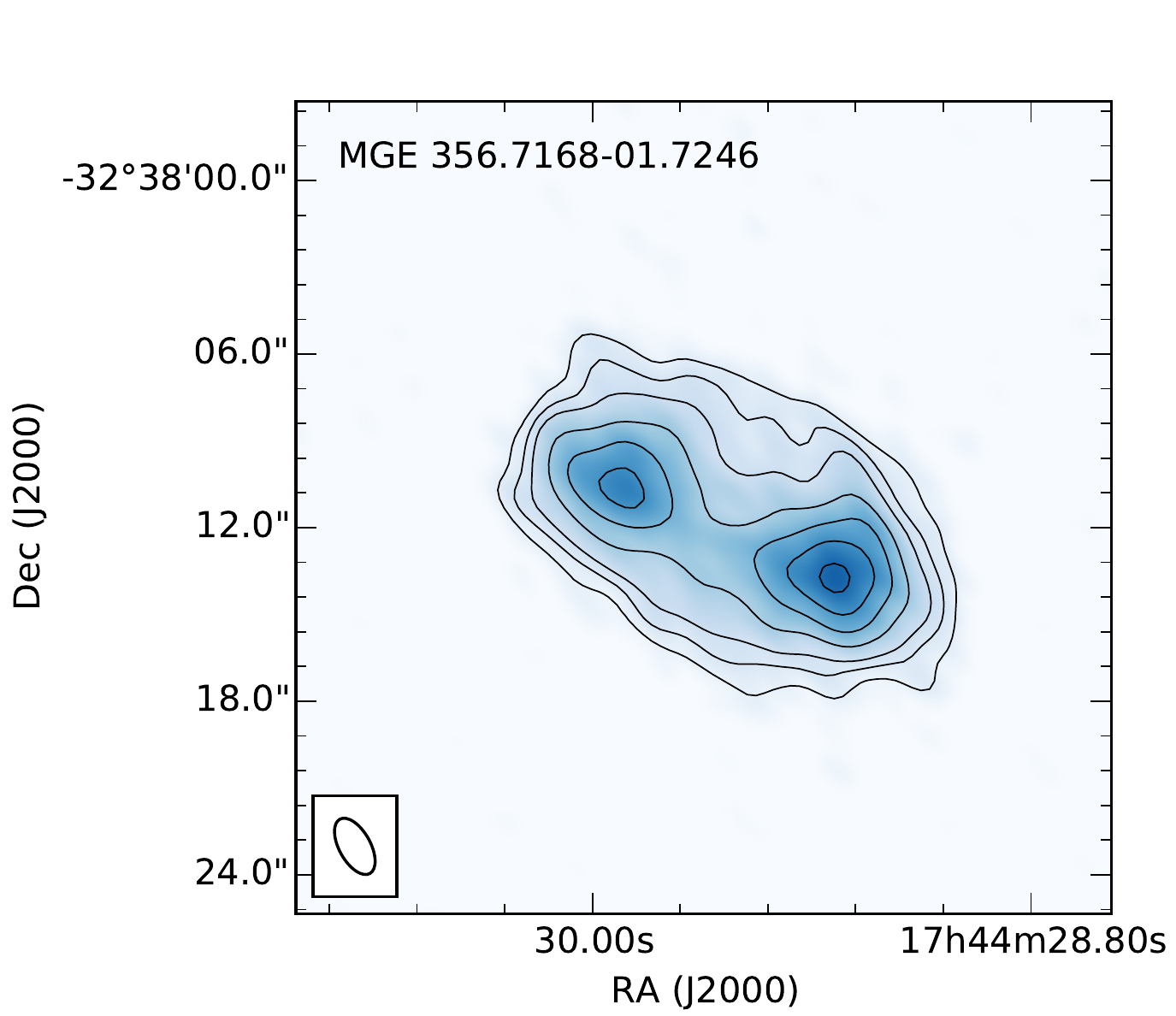}
\caption{6-cm images of the bubbles MGE 031.7290+00.6993, MGE 028.7440+00.7076 and MGE 032.4982+00.1615 (row 1, from left to right; morphology `O' of Table \ref{tab:data}); (row 2) MGE 356.7168-01.7246 (morphology `B' of Table \ref{tab:data}).}
\label{fig:bubbles_O}
\end{figure*}

The high-resolution radio maps reveal that the 18 bubbles are characterized by different shapes. From a visual morphological analysis we can group them into five categories. The first category consists of those bubbles with a clear evidence of a central point-like object (category `C' in Table \ref{tab:data}, see Figure \ref{fig:bubbles_C}). The second gathers all those bubbles with an elliptical ring shape and the total lack of a compact central object or diffuse emission toward their centre (category `E' in Table \ref{tab:data}, see Figure \ref{fig:bubbles_E}). A third category, dubbed `filled elliptical', groups those bubbles whose radio morphology is intermediate between the two previous categories, with diffuse emission toward the centre but without a clear central object (category `F' in Table \ref{tab:data}, see Figure \ref{fig:bubbles_F}). The remaining four bubbles show a morphology clearly different from the three categories introduced above. One of them, MGE 356.7168-01.7246 (second row in Figure \ref{fig:bubbles_O}), has a bipolar appearance (category `B' in Table \ref{tab:data}), while the others (first row in Figure \ref{fig:bubbles_O}) are somehow peculiar (category `O' in Table \ref{tab:data}) and will be discussed individually in Section \ref{sec:others}.

\subsection{Optical and infrared data}
The classification and characterization of the bubbles presented in this work take advantage of infrared and optical archive data. \textit{Spitzer} images were retrieved from the InfraRed Science Archive\footnote{http://irsa.ipac.caltech.edu} using the `cutout service' for the surveys GLIMPSE\footnote{Galactic Legacy Infrared Mid-Plane Survey Extraordinaire} and MIPSGAL. This service provides processed FITS images for which no further data reduction is needed. Optical FITS images (H$\alpha$ and $r$ filters) were retrieved from the SuperCOSMOS H-alpha Survey (SHS) \citep{Parker2005,Frew2014} website\footnote{http://www-wfau.roe.ac.uk/sss/halpha/}. Optical and infrared magnitudes from 2MASS\footnote{The 2-Micron All Sky Survey \citep{Skrutskie2006}.}, UKIDSS\footnote{UKIDSS-DR6 Galactic Plane Survey \citep{Lucas2008}.} and IPHAS\footnote{The INT Photometric H$\alpha$ Survey of the Northern Galactic Plane \citep{Drew2005}.} catalogues were taken from the VizieR catalogue access tool\footnote{http://vizier.u-strasbg.fr/viz-bin/VizieR-4}.

\section{Bubbles with a central object}
\label{sec:bub_cs}
The bubbles with a prominent central object at $24\mic{m}$ ($\sim\!15$ per cent of the total) are by far the most studied. Near- and mid-IR spectroscopy of the central object is able to impose strong constraints on the nature of the source. These bubbles have been found to be mostly evolved massive stars, mainly LBV candidates, O stars and WR stars, but also late-type giants \citep{Nowak2014,Wachter2010,Gvaramadze2010,Flagey2014}. At shorter wavelengths the CSE becomes fainter and only seldom is observed at $8\mic{m}$ \citep{Mizuno2010}. The central object is instead usually detected at wavelengths as short as $1\mic{m}$ or less, which makes near-IR spectroscopy possible.

\begin{figure*}
\includegraphics[height=5cm]{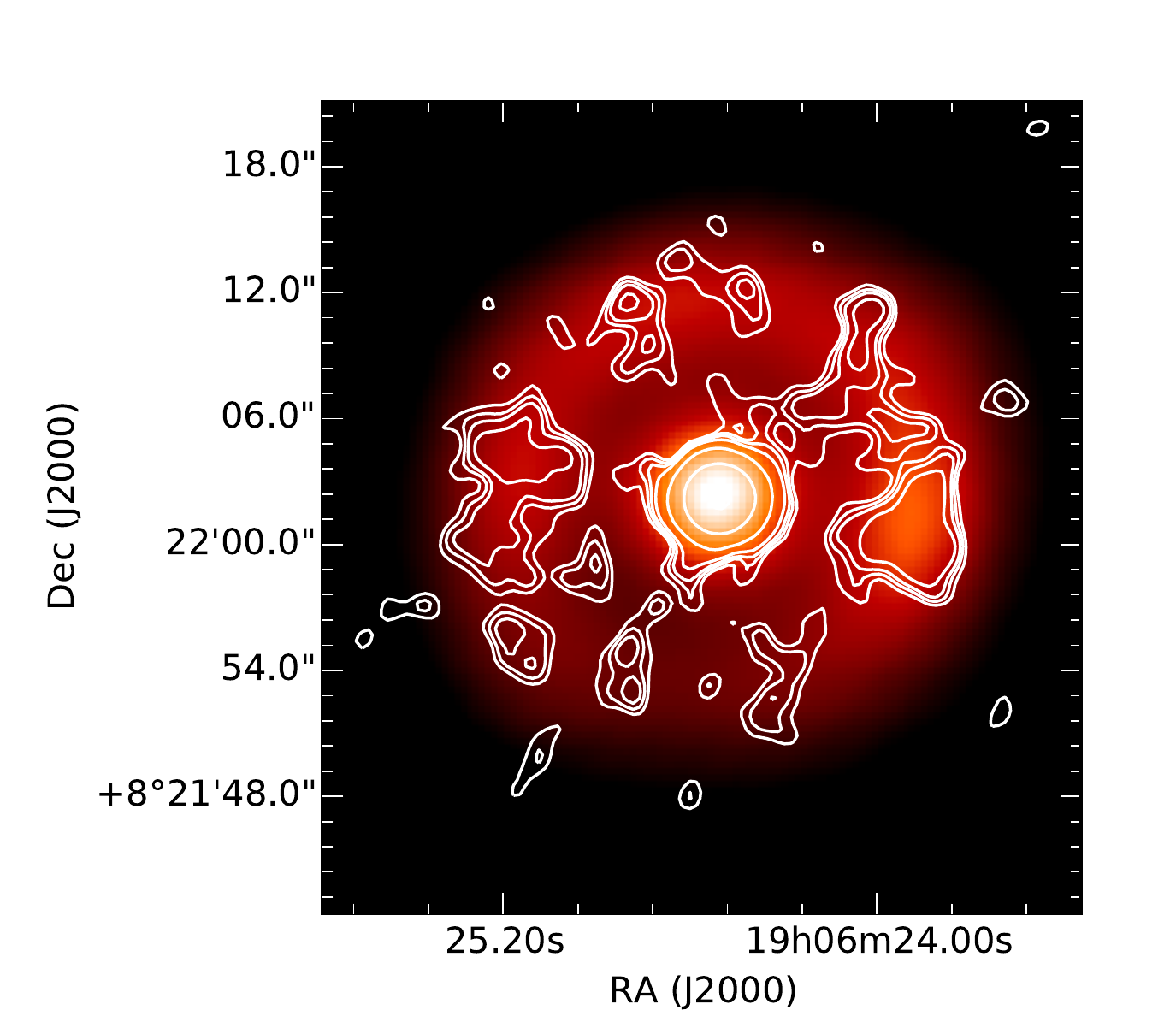}
\includegraphics[height=4.8cm]{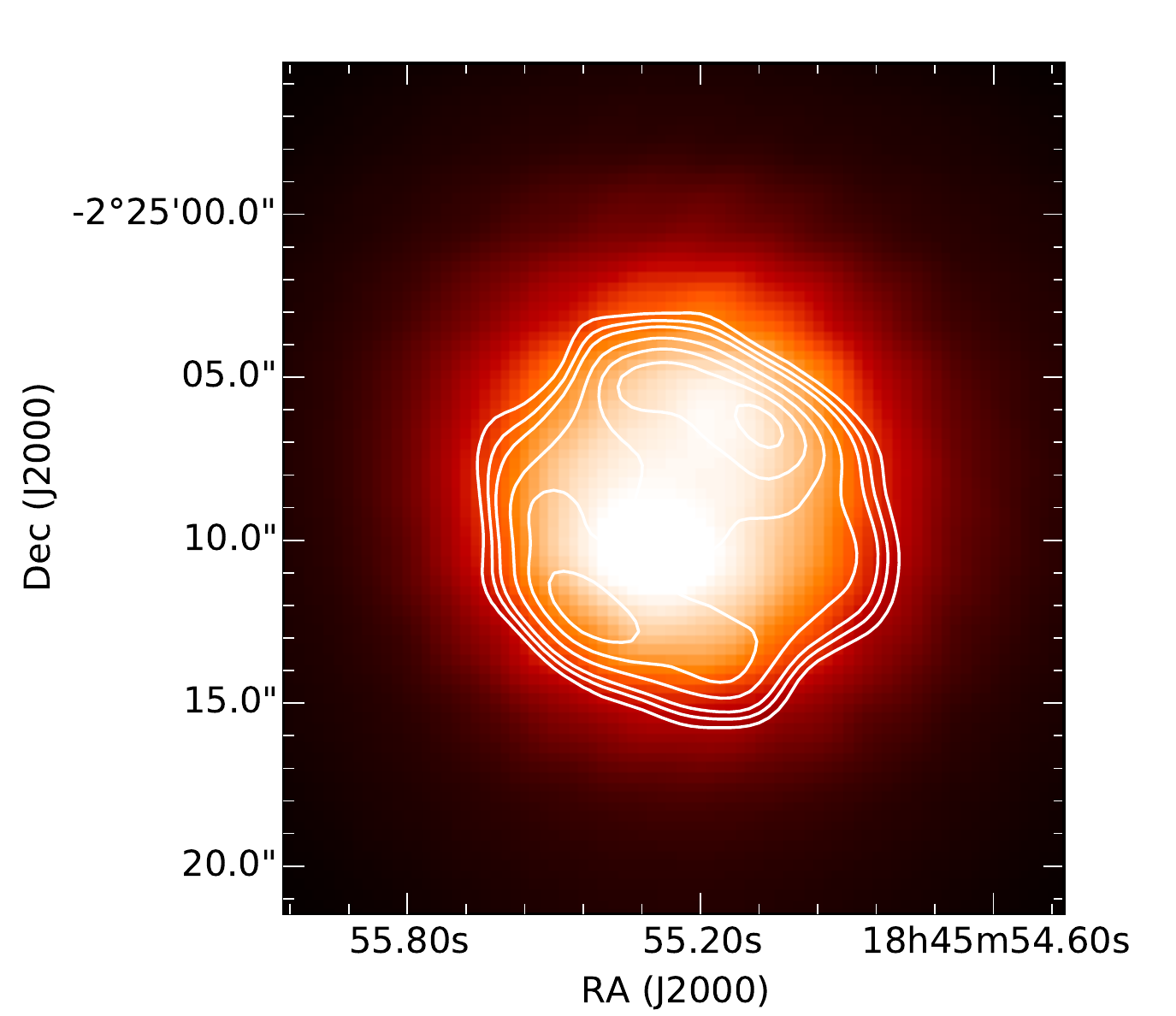}
\includegraphics[height=4.8cm]{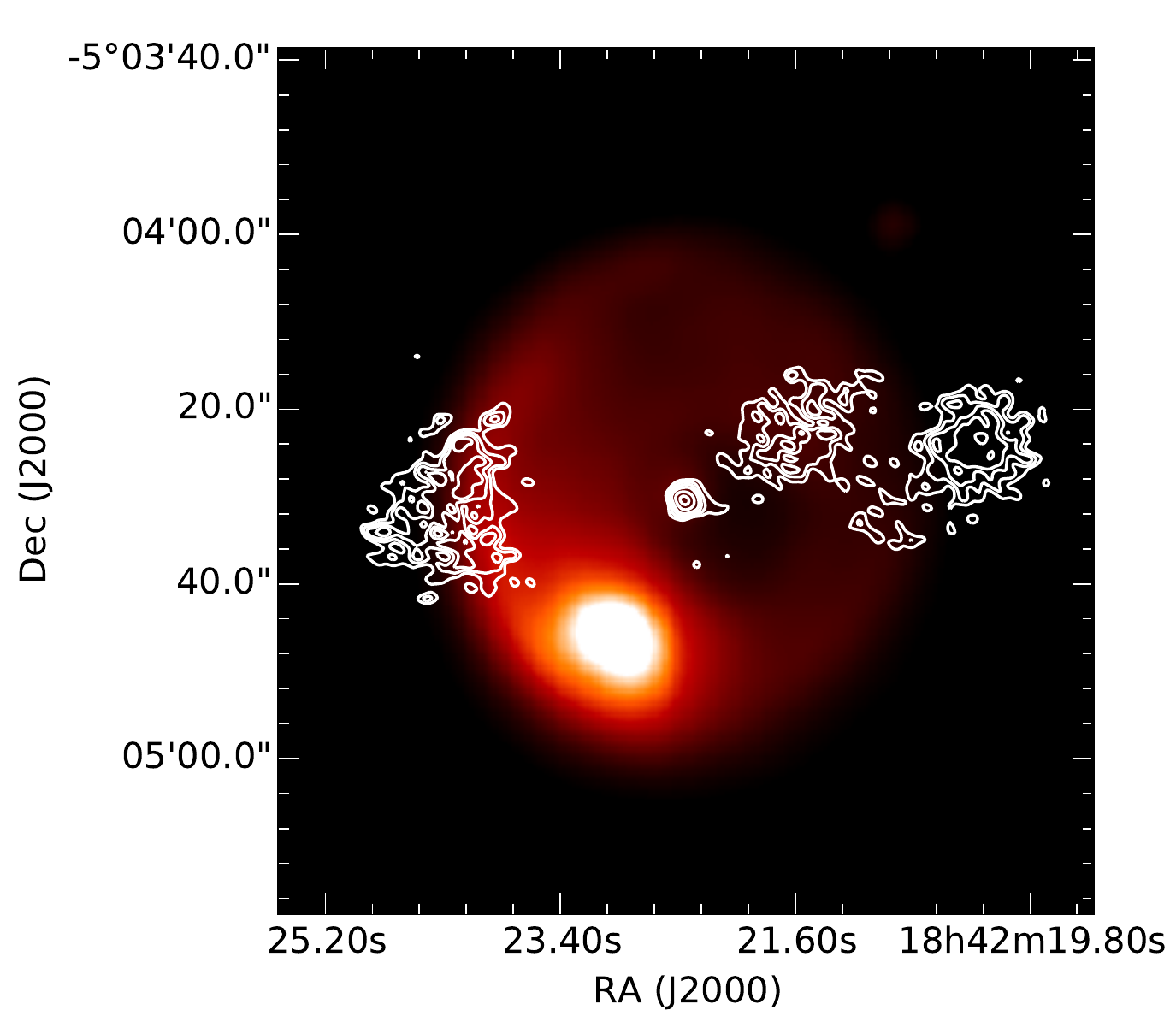}
\includegraphics[height=5cm]{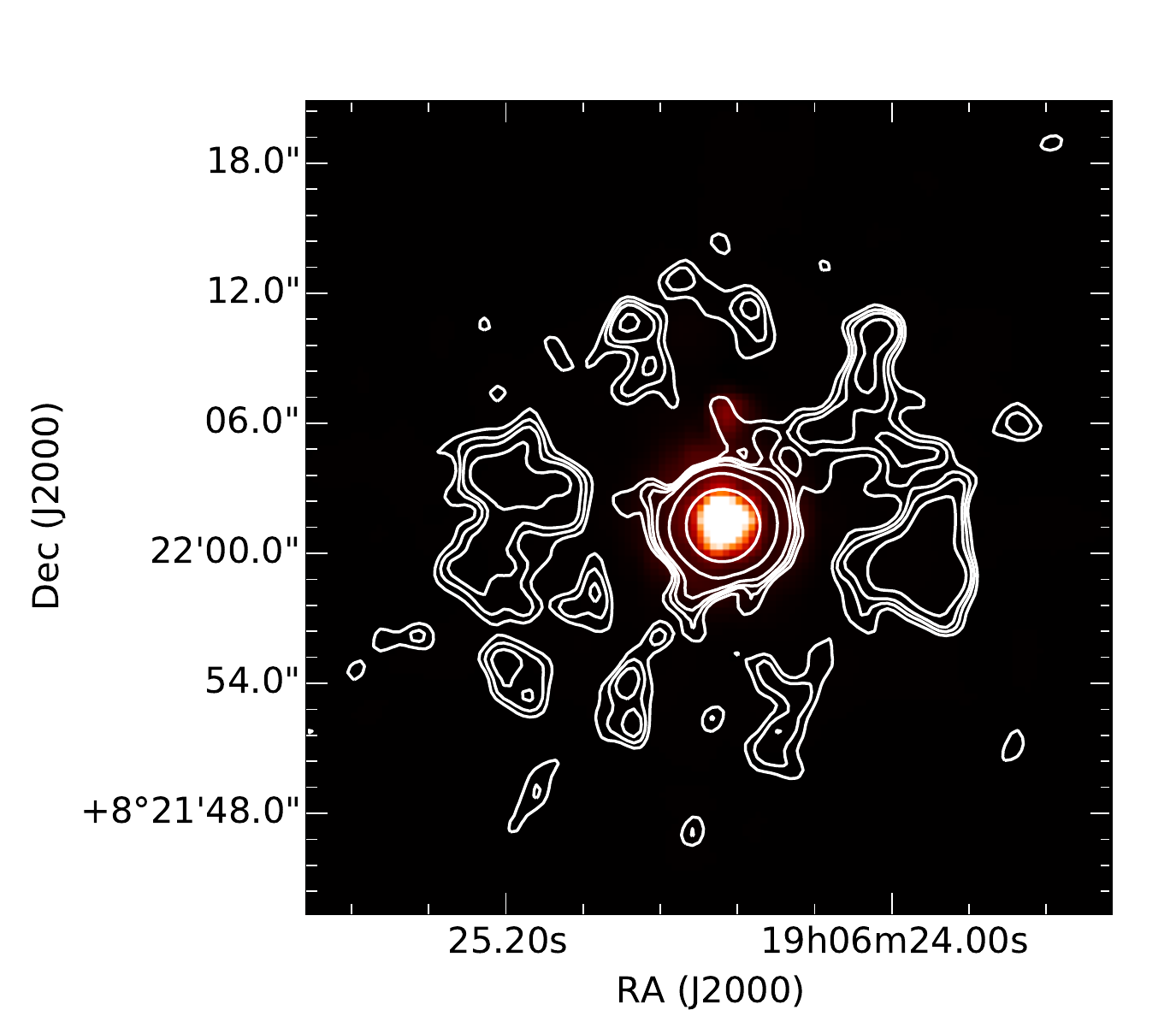}
\includegraphics[height=5cm]{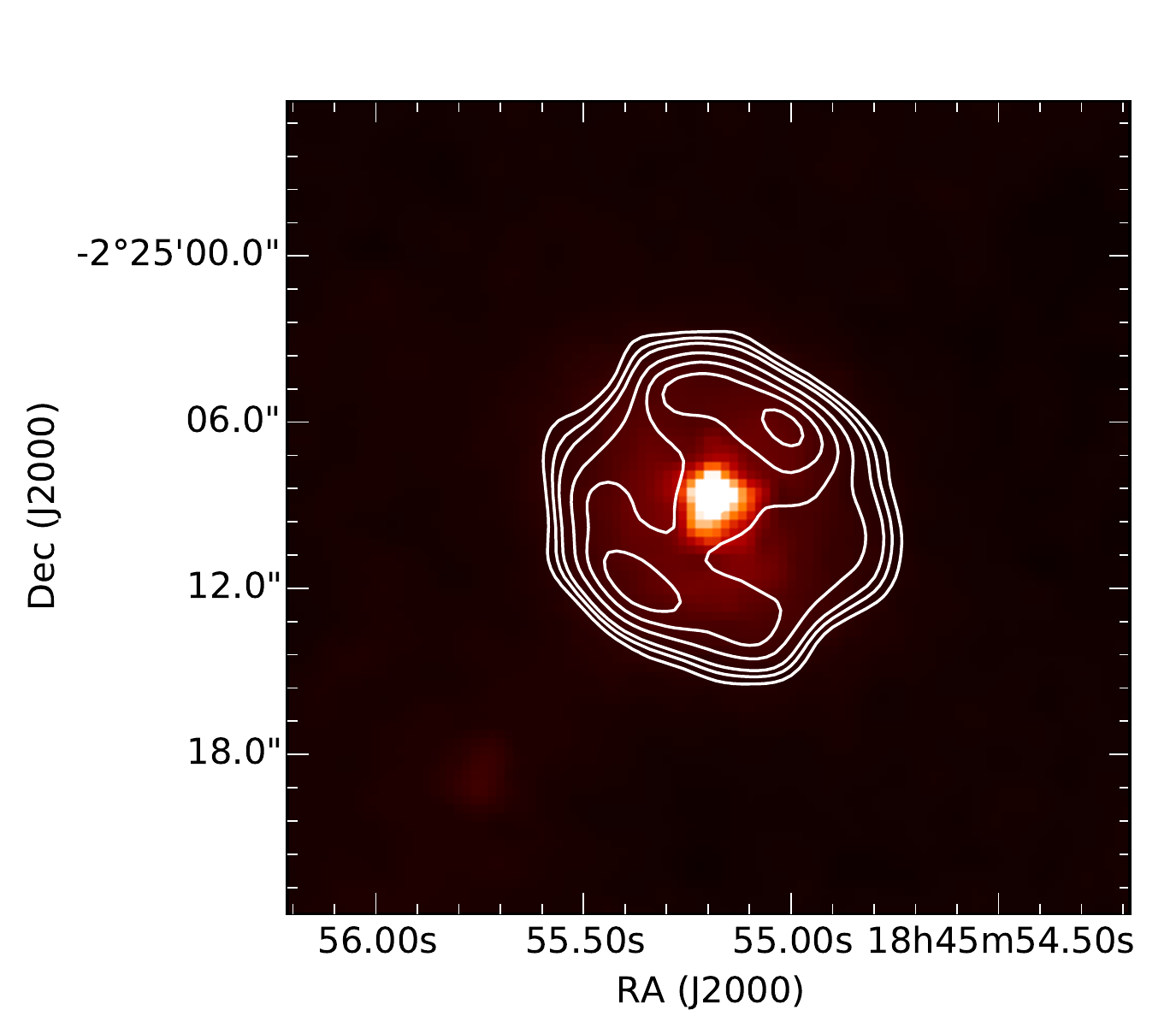}
\includegraphics[height=5cm]{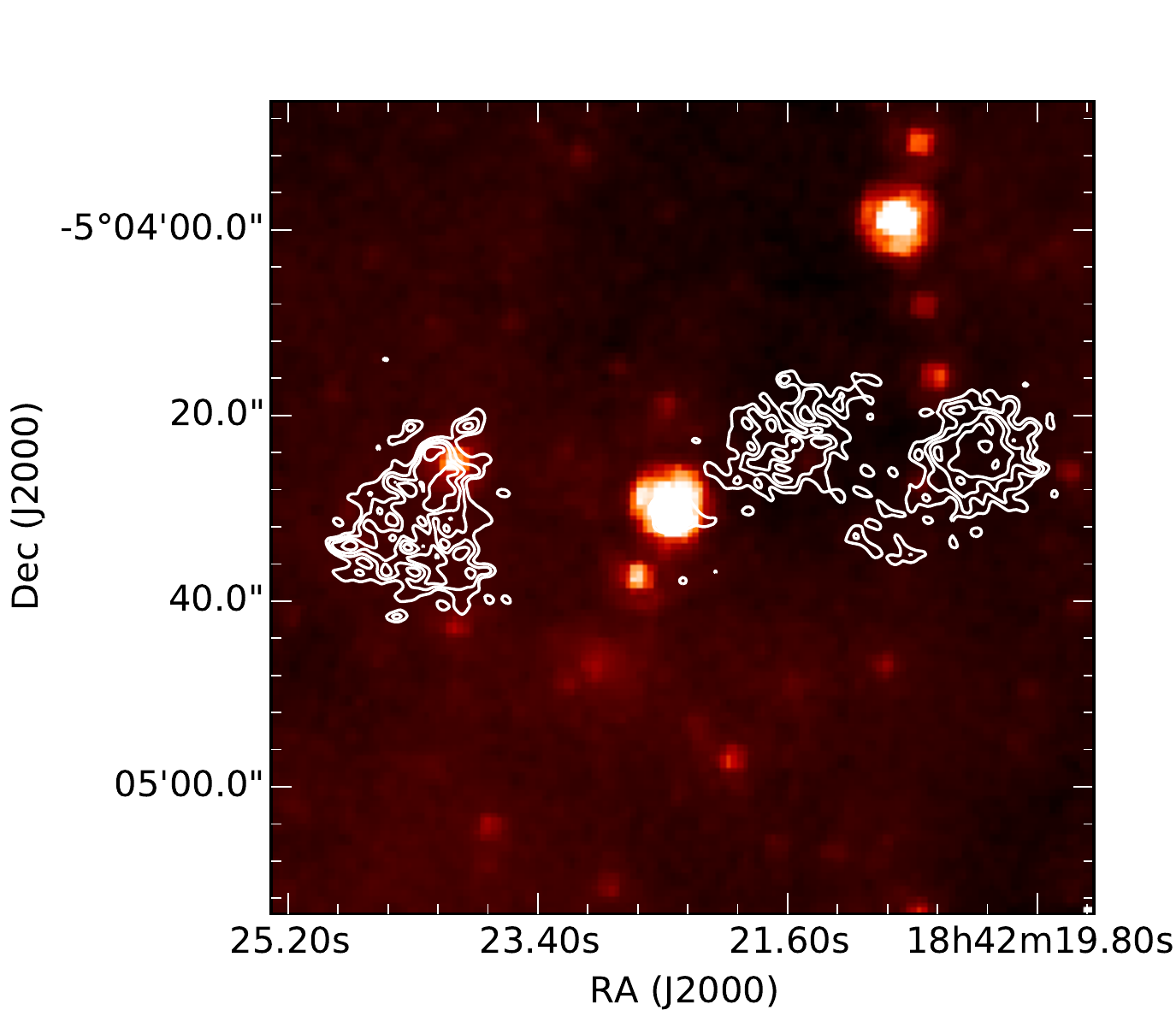}
\caption{Superposition of 6-cm contours on 24-$\umu$m (top) and 8-$\umu$m (bottom) images for the bubbles (from left to right) MGE 042.0787+00.5084, MGE 030.1503+00.1237, MGE 027.3839-00.3031.}
\label{fig:bubC_sup}
\end{figure*}

Among the bubbles observed in this work, three show a well-detected point-like central object in the radio (Figure \ref{fig:bubbles_C}). The central object is also detected at $24\mic{m}$ and in IRAC bands. In Figure \ref{fig:bubC_sup} we show the superposition between the radio contours and the 24-$\umu$m (top row) and 8-$\umu$m (bottom row) images.
The CSE morphology needs further attention. At $8\mic{m}$ we are able to detect only the CSE of MGE 030.1503+00.1237 (bottom centre in Fig. \ref{fig:bubC_sup}), which shows a ring structure slightly less extended than in the radio. At $24\mic{m}$ the central object dominates the emission of MGE 030.1503+00.1237 (top centre in Fig. \ref{fig:bubC_sup}), while for the other two bubbles it has approximately the same brightness of the ring-shaped CSEs. It is interesting to notice that while for MGE 042.0787+00.5084 and MGE 030.1503+00.1237 the radio is co-spatial with the 24-$\umu$m emission, this is not the case for MGE 027.3839-00.3031, where the radio nebula seems unrelated to IR.

From near-IR spectroscopy, \citet{Flagey2014} proposed MGE 042.0787+00.5084 and MGE 030.1503+00.1237 as LBV candidates. Ongoing spectroscopic studies on MGE 027.3839-00.3031 seem to support the same hypothesis (Silva et al. \textit{submitted}). Radio continuum emission is observed in different classes of massive evolved stars, not only in hot stars like LBVs and WR stars, but also in cool red supergiants and yellow hypergiants (when they are in binary systems with a hot companion; \citealt{Dougherty2010}). Our high-resolution radio images show that at $6\um{cm}$ these bubbles appear as a point source surrounded by a circular or a bipolar nebula. These images are in agreement with the hot massive evolved star scenario. In this case we expect that the radio emission originates from both the central object and from its CSE. The emission from the central object is mainly due to the stellar wind, a peculiar characteristic of the last stages of massive star evolution. The emission of the CSE derives from the ionization of the gas ejected by the star in previous moments. The shape of the radio nebula is greatly variable, with example of LBVs showing ring nebulae \citep{Umana2011,Agliozzo2014} or bipolar nebulae \citep{Umana2012}.

For MGE 042.0787+00.5084 and MGE 030.1503+00.1237 the radio spectral energy distribution (SED) reported in Paper I is approximately flat, with spectral index values of 0.02 and 0.05 respectively. For MGE 027.3839-00.3031 we have only a lower limit of 0.9. At centimetre and millimetre wavelengths, the presence of a stellar wind translates into a positive spectral index, typically around 0.6 for steady winds with a spherical electron density distribution scaling as the inverse of squared distance from the star \citep{Panagia1975}. The total radio emission of these sources is therefore reasonably the sum of different components, each of them characterized by a different value of the spectral index. The rarity of these stars and the limitation of instruments, both in resolution and in sensitivity, have prevented the possibility of studying the spectral index variation within the source, except for very few cases. The result is that, in general, these sources have been characterized only by an average spectral index, that could have hidden underlying poorly-studied phenomena. In recent years it has been pointed out that many WR stars do show a component of non-thermal emission. For example, \citet{Chapman1999} found that six out of nine WR stars exhibit a global spectral index around $-0.6$, typical of synchrotron emission. Further studies (see \citealt{Walder2012} for a review) demonstrated that, in many cases, the non-thermal component arises from the collision of the winds of the WR star and of a companion massive star. In single WR stars the possibility of synchrotron emission arising from the interaction between the stellar wind and the CSE has been discussed as possible, but not observed yet \citep{Cohen2006}. It is disputed whether this synchrotron emission can be really observed in single WR stars \citep{vanLoo2010,Blomme2010}. An analogous process was suggested by \citet{Cohen2006} to explain the non-thermal radio emission from the very young PN V1018 Sco. In this case the authors did not detect the central source but only spots of synchrotron radiation toward the edge of the optical nebula. In another work (Buemi et al. \textit{in prep.}) we are showing that the LBV star HR Car presents a differentiated spectral index too, with synchrotron-like values toward the edge. Very similar results are also found for the LBV candidates G26.47+0.02 \citep{Paron2012}, where a spectral index $\sim\!-0.9$ is found for the extended emission. In the next subsections we exploit the high-resolution and sensitivity of our radio data to study in detail the radio emission of these three bubbles, highlighting possible spectral index variations and their implications on the physics of these objects.

\subsection{MGE 042.0787+00.5084} 
\begin{figure}
\begin{center}
\includegraphics[width=\columnwidth]{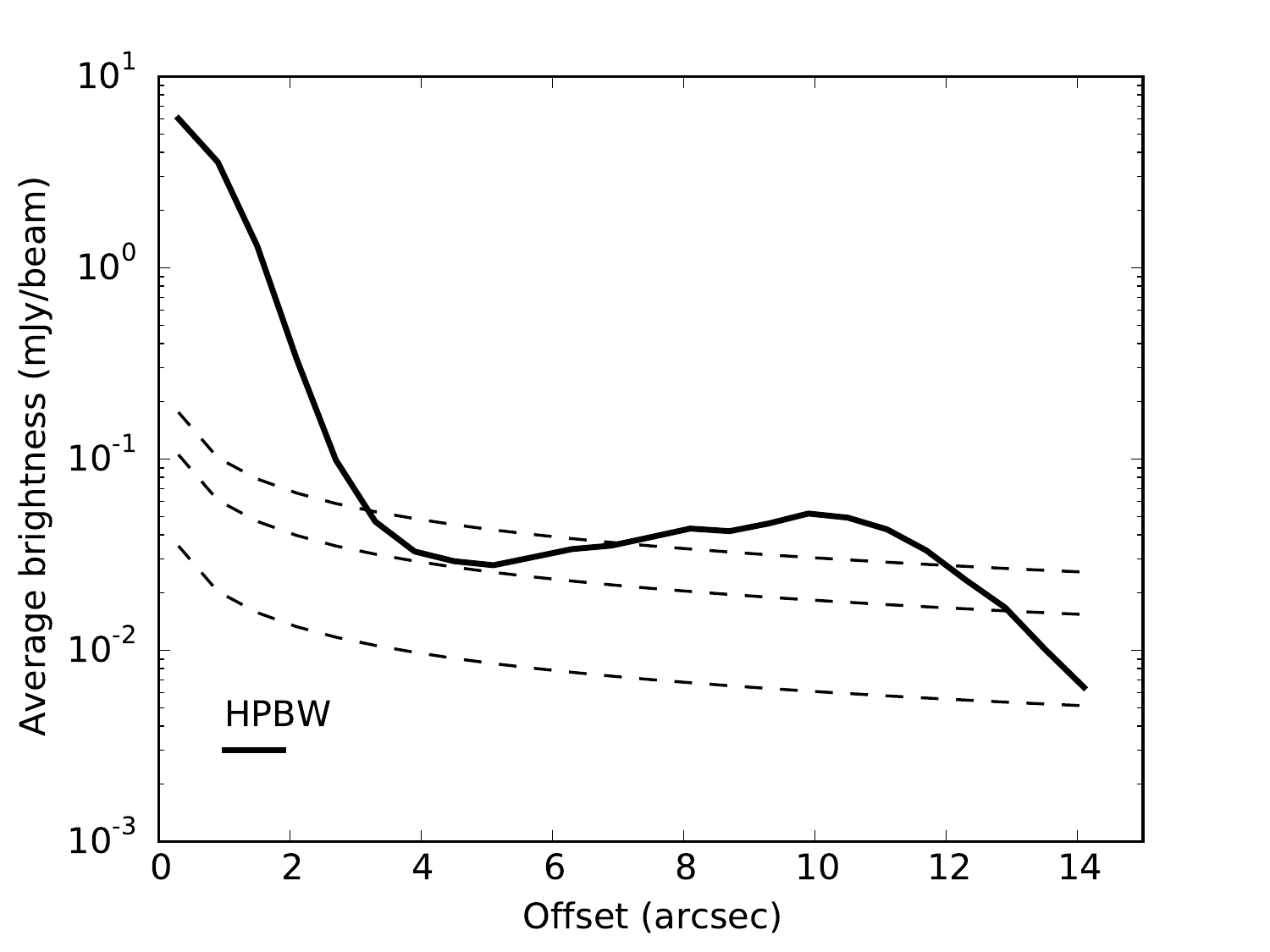}
\caption{Radial profile at $6\um{cm}$ of MGE 042.0787+00.5084. The solid line represents the average brightness, while the dashed lines are its error at, respectively, 1, 3 and $5\sigma$ level. The black rule on the bottom-left corner is the half power beam width (HPBW), reported to highlight that the profile of the central object is compatible with a point source.}
\label{fig:prof_3438}
\end{center}
\end{figure}

\begin{figure}
\begin{center}
\includegraphics[width=\columnwidth]{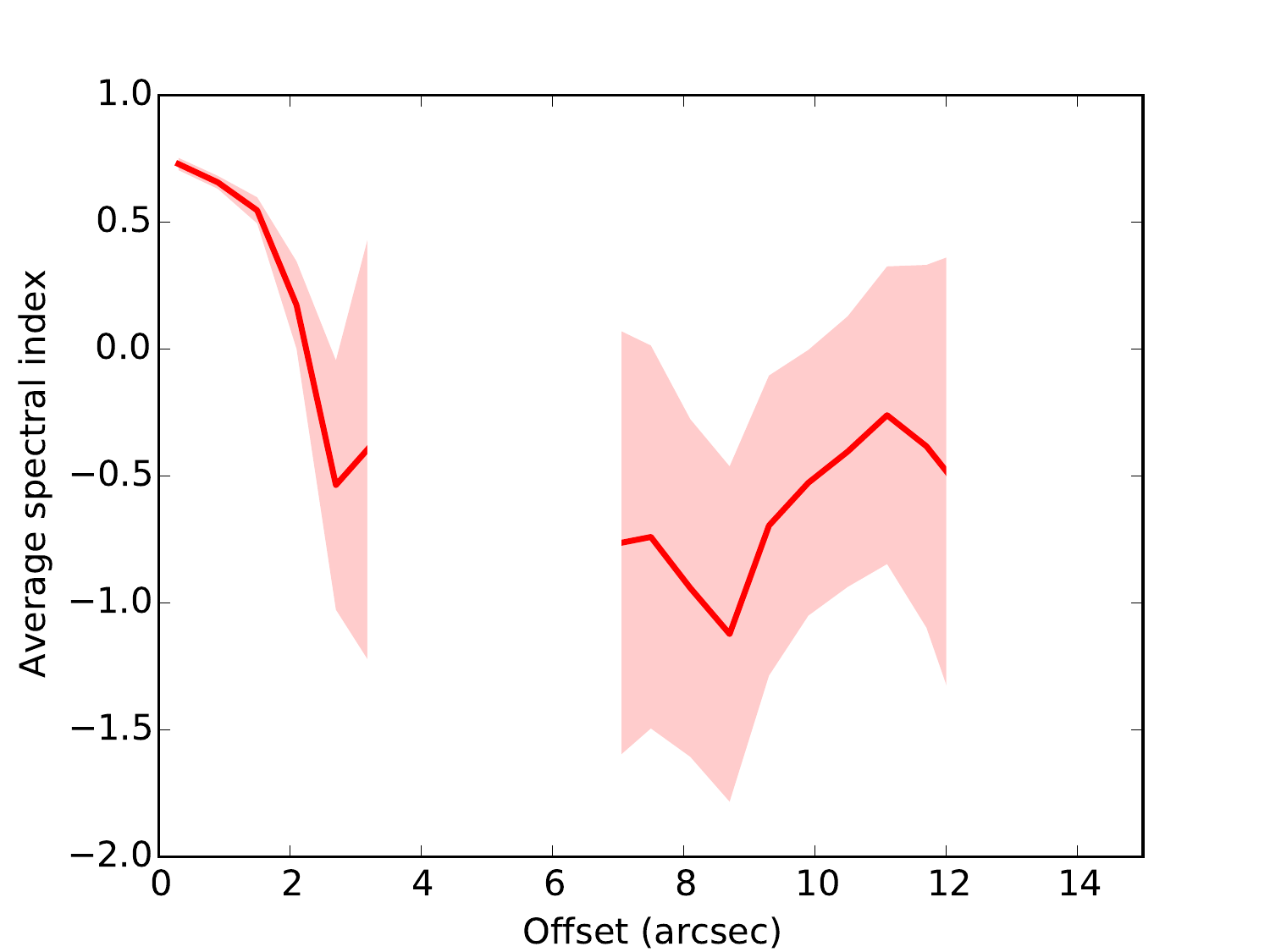}
\caption{Radial profile of the spectral index of MGE 042.0787+00.5084. The solid line represents the average spectral index, while the shaded area its error. The plot is masked where $I(r)/\Delta I(r)<5$.}
\label{fig:spix_3438}
\end{center}
\end{figure}

Spectral analysis of the central object in the near- and mid-IR showed that it has to be classified as a Be/B[e]/LBV star \citep{Flagey2014,Wachter2010}, so the star was proposed as an LBV candidate. This hypothesis is corroborated by the H$\alpha$ excess ($r-\mathrm{H}\alpha=1.77\pm0.03$), calculated from the IPHAS DR2 Source Catalogue \citep{Barentsen2014}. At radio wavelengths, this bubble is characterized by a central source with a flux density at $6\um{cm}$ of $7.0\pm0.2\um{mJy}$, calculated with a Gaussian fit. The flux density of the whole source reported in Paper I was $10.5\pm0.1\um{mJy}$, so we can estimate a total flux density of $3.5\pm0.1\um{mJy}$ for the CSE. In the radio the CSE shows limb brightening, likely due to a spherical shell nature, and its brightness, especially in the inner part, is only slightly above the background noise. In order to highlight the faintest details of the bubble we built its radial profile starting from a map obtained by concatenating the B- and D-configuration datasets, with a rms of $19.5\mic{Jy\ beam}^{-1}$ and an imposed circular restoring beam of $1.8\um{arcsec}$ FWHM. In particular we calculated the mean brightness of a series of circular annuli, centred on the central object, with a constant thick of $0.6\um{arcsec}$. The choice of a circular beam prevented the introduction of systematic errors deriving from the beam elliptical shape. To take into account fractional pixels we stretched the map by a factor 10 in both directions. The error associated with each computed mean brightness was estimated dividing the map rms by the square root of the number of independent beams per each annulus. The radial profile obtained is reported in Figure \ref{fig:prof_3438}.
Apart from the central source, it is possible to clearly distinguish the limb brightening of the CSE, culminating at a distance $r\sim\!11\um{arcsec}$. Though very faint, the inner part of the nebula is still above the $3\sigma$ level in this integrated view. As expected, the spectral index map, returned by \textsc{casa}, was too noisy toward the CSE. To estimate the spectral index variation across the bubble we used the same method of the radial profile described above, applied to the spectral index map, masked below $3\sigma$. The error on spectral index was calculated as
\begin{equation}
\Delta \alpha(r)=\left(\left(\frac{\sqrt{2}}{\displaystyle \frac{I(r)}{\Delta I(r)}\ln\frac{\nu_{\max}}{\nu_{\min}}}\right)^2+(\sigma_\alpha(r))^2\right)^{1/2},
\label{eq:delta_alpha}
\end{equation}
where $I(r)$ and $\Delta I(r)$ are the brightness radial profile and its error as derived above, $\nu_{\min}=4.2\um{GHz}$ and $\nu_{\max}=6\um{GHz}$ are the boundary frequencies of the net bandwidth (i.e. after flagging operations) and $\sigma_\alpha(r)$ is the standard deviation of the spectral index values inside the annulus at the distance $r$. The first term of the equation (\ref{eq:delta_alpha}) is obtained by error propagation, assuming that the signal-to-noise ratio is constant across the entire bandwidth. In Figure \ref{fig:spix_3438} we show the spectral index profile.
The central object has a spectral index $\alpha_\mathrm{cob}=0.70\pm0.03$, perfectly compatible with a stellar wind, likely with a mean electron density decreasing more rapidly than $r^{-2}$ \citep{Panagia1975}. The spectral index of the most statistically significant part of the CSE ($I(r)/\Delta I(r)>5$), from $r\sim\!7$ to $r\sim\!12\um{arcsec}$, varies between $-0.25$ and $-1.10$ with a mean value of $-0.65$. Despite a quite large error ($\Delta\alpha\sim\!0.6$), this is a significant clue of non-thermal emission, which implies that an electron acceleration mechanism is taking place in the outer part of the CSE. The error on this mean value across the entire CSE can be calculated dividing the average $\Delta\alpha$ by the square root of the number of points of the CSE for which $I(r)/\Delta I(r)>5$ and it results to be $0.23$. 

\begin{figure}
\begin{center}
\includegraphics[width=\columnwidth]{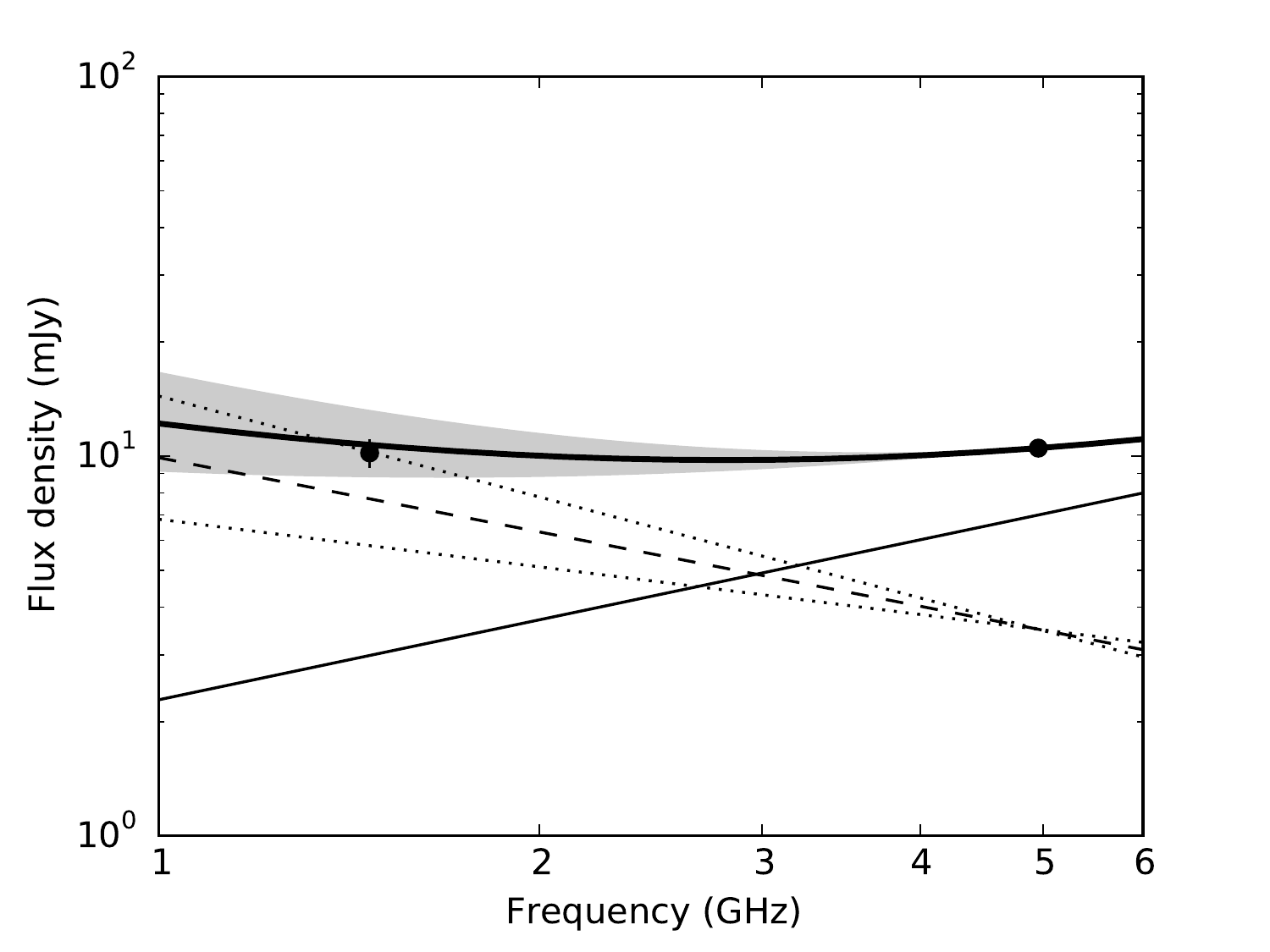}
\caption{SED of MGE 042.0787+00.5084, highlighting the contributions of the central object and of the CSE. The solid thin line is the SED of the central object, extrapolated from the measured flux at $4.959\um{GHz}$ with a spectral index of $\alpha_\mathrm{cob}=0.7$. The dashed line is the SED of the CSE, considering a spectral index $\alpha_\mathrm{neb}=-0.65$, with the dotted line representing the same SED but with spectral indices $-0.42$ and $-0.88$ (these limits are obtained from $-0.65\pm0.23$, where $0.23$ is the error on $\alpha_\mathrm{neb}$ as calculated in the text). The thick line is the sum of the two SEDs, representing therefore the total flux density of the bubble, with the shaded area representing the uncertainties of $\alpha_\mathrm{neb}$. The two solid circles are the flux density values reported in Paper I. The thick line passes through the $4.959\um{GHz}$ point by construction, while it is possible to appreciate how the 1.468-GHz point lies inside the shaded area (best fitted with $\alpha_\mathrm{neb}\sim\!-0.6$).}
\label{fig:sed_3438}
\end{center}
\end{figure}

Once the flux densities of the central object and of the CSE were determined, we extrapolated these values toward low frequencies, to check if the independent 20-cm flux density value reported in Paper I is consistent with our model. In Figure \ref{fig:sed_3438} we report the inferred SED of the bubble, plotting the two components (central object and CSE) and their sum (i.e. the bubble total flux density). We can see that the flux density measure at $20\um{cm}$ is consistent, within the errors, with our model.

All our calculations on this bubble implicitly assumed that the flux density did not vary significantly between 2010 ($6\um{cm}$, D configuration), 2012 ($20\um{cm}$) and 2015 ($6\um{cm}$, B configuration). Indeed, we are not able to exclude that such a variation could have occurred, because we are supposing that this bubble could be an LBV. In fact, at $6\um{cm}$, 2010 data have a poor resolution so we cannot distinguish between the point-like central object and the CSE, while in 2015 data the CSE is likely resolved out, so the total flux density would be underestimated. However, since dramatic variations would invalidate our model, we are confident that the flux density of this bubble has not varied significantly in the last years.

\subsection{MGE 030.1503+00.1237} 
Despite the huge number of observations in near- and mid-IR, the nature of this object is controversial. \citet{Lumsden2013} presented the catalogue of the Red MSX Sources survey searching for massive young stellar objects (MYSO) in our Galaxy. The authors excluded that MGE 030.1503+00.1237 could be a MYSO from near-IR selection criteria and from the high radio flux density at $6\um{cm}$ \citep{Urquhart2009}. They suggested that Galactic sources with 6-cm flux densities above $\sim\!10\um{mJy}$ are likely to be PN or H \textsc{ii} regions, since massive stars do not show such flux densities. However, it has been shown that LBVs do show high radio flux densities (e.g. \citealt{Umana2012}). More recently near-IR spectroscopy of the central object revealed that it is a Be/B[e]/LBV star \citep{Flagey2014}, but with some uncertainties deriving from its faintness, and the star was proposed as an LBV candidate.

\begin{figure}
\begin{center}
\includegraphics[width=\columnwidth]{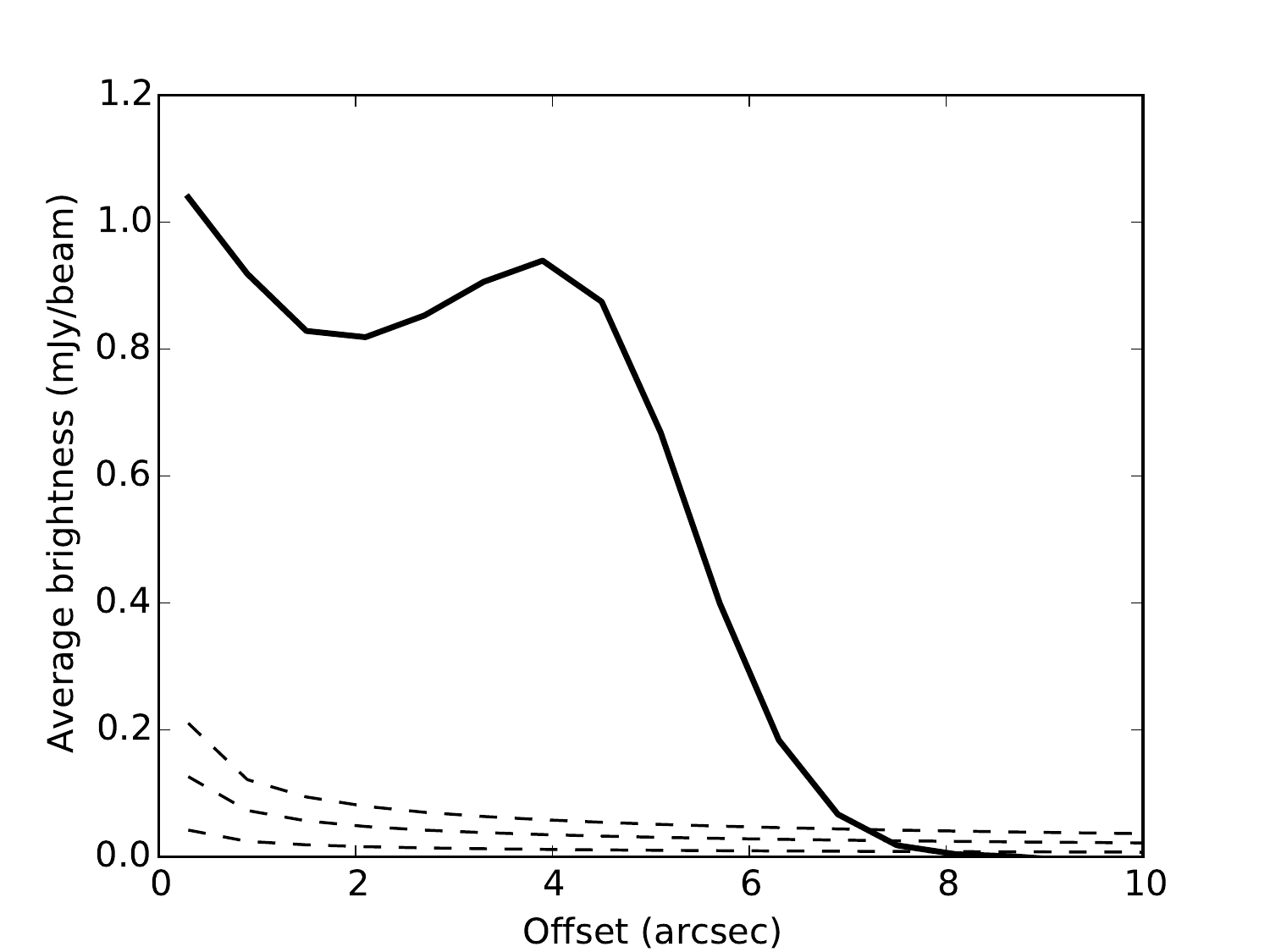}
\caption{Radial brightness profile of MGE 030.1503+00.1237. The solid line represents the average brightness, while the dashed lines are its error at, respectively, 1, 3 and $5\sigma$ level.}
\label{fig:prof_3222}
\end{center}
\end{figure}

At radio wavelengths the central object and the edge of the CSE are characterized by approximately the same brightness, in contrast with the previously discussed MGE 042.0787+00.5084 where the central object is dominant. In Figure \ref{fig:prof_3222} we show the radial profile of this bubble, calculated as for MGE 042.0787+00.5084.
We can clearly see how the average brightness toward the central object is roughly the same of the CSE edge brightness. We can also see how the inner part of the nebula is much brighter than that of MGE 042.0787+00.5084. 

\begin{figure}
\begin{center}
\includegraphics[width=\columnwidth]{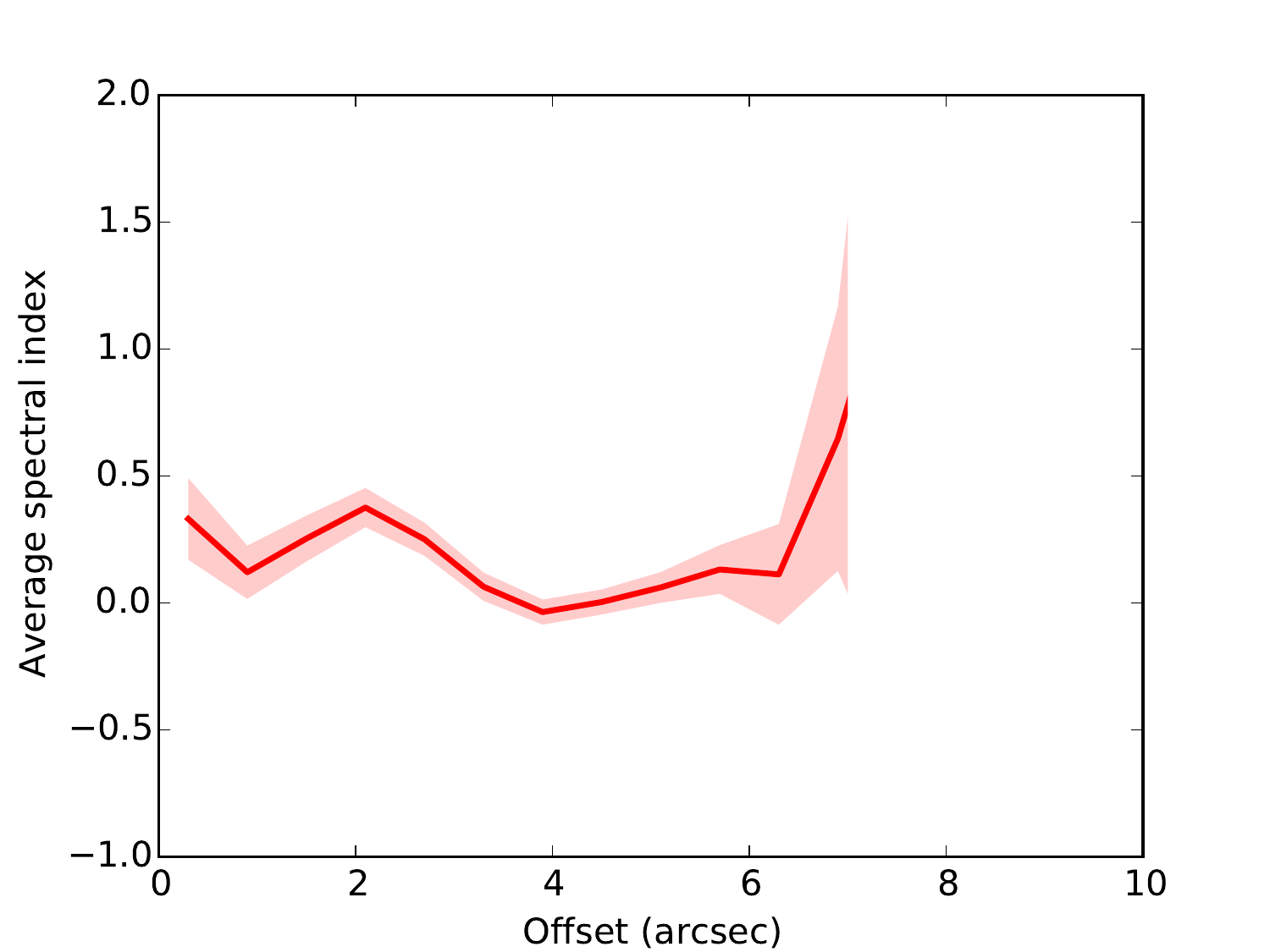}
\caption{Radial profile of the spectral index of MGE 030.1503+00.1237. The solid line represents the average spectral index, while the shaded area its error. The plot is masked where $I(r)/\Delta I(r)<5$.}
\label{fig:spix_3222}
\end{center}
\end{figure}

The spectral index does not show significant local departures from the mean value $0.05\pm0.13$ reported in Paper I. In Figure \ref{fig:spix_3222} we report the radial profile calculated as previously described.
It is possible to notice that the spectral index varies between 0 and 0.4 for $r\lesssim6\um{arcsec}$. The spectral index behaviour does not show a clear evidence of a stellar wind. However \citet{Flagey2014} estimate that there is a stellar wind with a terminal velocity of $200\um{km\ s}^{-1}$. From the plot in Figure \ref{fig:prof_3222}, extrapolating the CSE brightness toward the centre of the bubble, we can estimate the flux density of the central object as $\sim\!0.4\um{mJy}$, with the nebula brightness at the centre of the bubble $\sim\!0.65\um{mJy\ beam}^{-1}$. This means that a canonical spectral index of a stellar wind (0.6) would only barely affect what we see in Figure \ref{fig:spix_3222}, where indeed we can see a minimum hint of a rising spectral index for $r\rightarrow0$. It would have no effect at all to the global spectral index reported in Paper I, given a total 6-cm flux density of $22.9\pm1.5\um{mJy}$.

\subsection{MGE 027.3839-00.3031}
This bubble has received very little attention despite its very peculiar morphology at $24\mic{m}$. \citet{Gvaramadze2010} listed this bubble in their catalogue of massive star candidates unveiled by \textit{Spitzer}. They report its ring shape and show that the source is also detected at $20\um{cm}$ with a similar morphology. Near-IR spectroscopy by Silva et al. (\textit{submitted}) shows that the central star is very likely an LBV. The star is also detected by 2MASS in $H$- and $K$-band (with magnitudes $11.101\pm0.036$ and $8.684\pm0.025$ respectively). From UKIDSS-DR6 Galactic Plane Survey we found a $J$ magnitude of $14.238\pm0.003$. Both 2MASS and UKIDSS photometries suggest that the star is highly extincted.

\begin{figure}
\begin{center}
\includegraphics[width=\columnwidth]{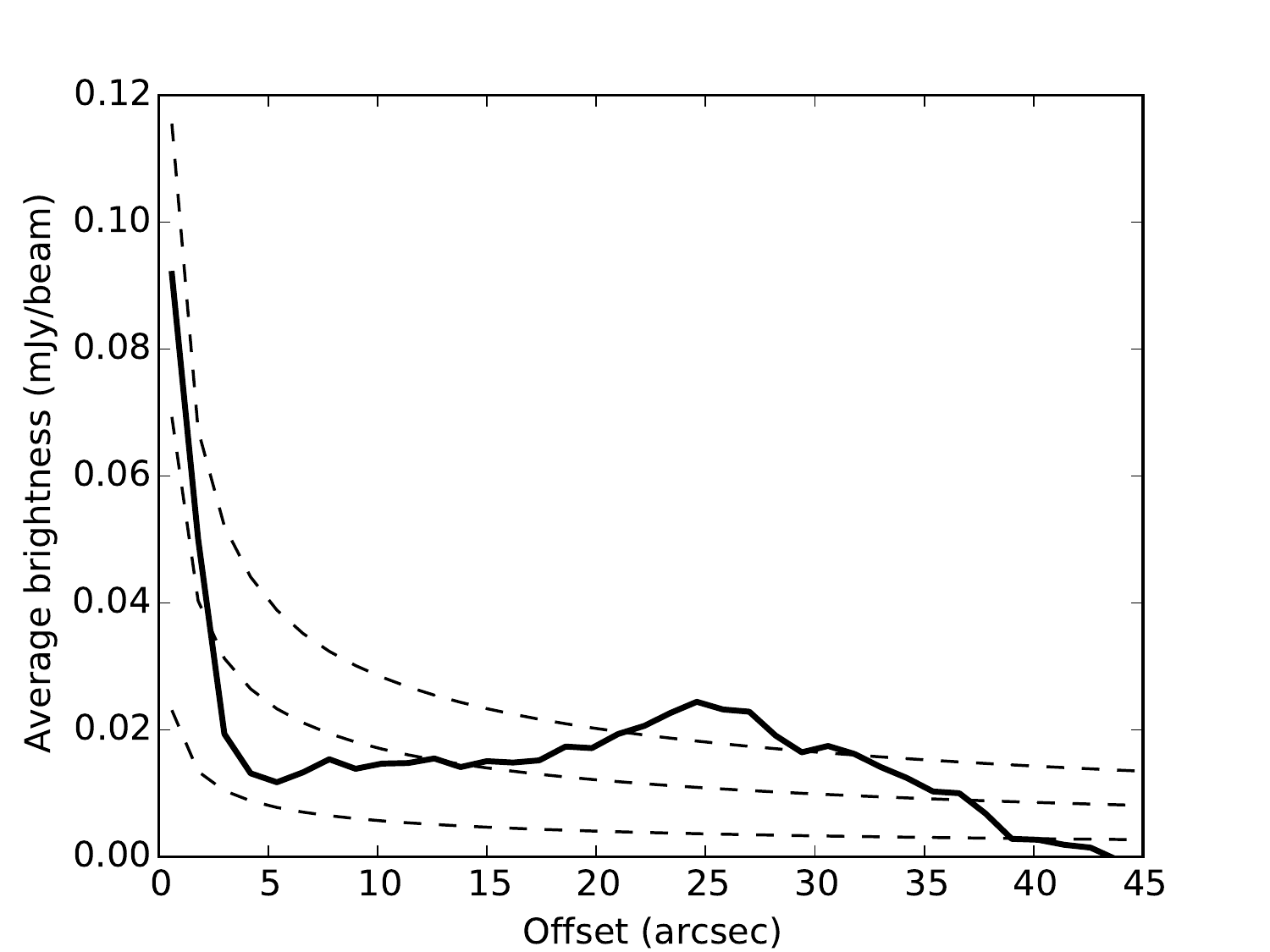}
\caption{Radial brightness profile of MGE 027.3839-00.3031. The solid line represents the average brightness, while the dashed lines are its error at, respectively, 1, 3 and $5\sigma$ level.}
\label{fig:prof_3736}
\end{center}
\end{figure}

The integrated radio flux density of this bubble, as reported in Paper I, is $18.1\pm0.5\um{mJy}$ at $6\um{cm}$. However the source has a remarkable angular extension ($\sim\!1\um{arcmin}$ in diameter) and then its brightness temperature is very low. The overall radio morphology of this bubble is similar to that of MGE 042.0787+00.5084, with a point-like central object and a faint CSE. However, the central symmetry of its CSE is less evident, and a bipolar shape seems to emerge. Beside a real enhancement of the emission toward these two lobes, it is possible that imaging artefacts greatly affect its appearance. Taking into account this weak symmetry, the radial profile was calculated limiting the area to $\pm45\um{deg}$ from the RA axis, where most of the CSE emission is located. In Figure \ref{fig:prof_3736} we plot the obtained profile.


Though very faint, this bubble shows a profile with a brighter central object and a limb-brightened CSE. For this bubble it is not possible to derive a reliable spectral index profile. Even for the central object, with a flux density $\sim\!0.12\um{mJy}$ (calculated from the map using a Gaussian fit), this calculation is difficult. For it we obtain $\alpha_\mathrm{cob}=0.9\pm1.0$. This index is still compatible with a stellar wind, but given the extremely high error, this is no more than a hint.

In Paper I we were not able to detect the source at $20\um{cm}$. Indeed, the GPS image from MAGPIS shows the presence of the nebula at $20\um{cm}$, but the bubble is immersed in a very noisy background and a precise flux density measurement is not possible. The morphological similarity with the previous bubbles supports the hypothesis that MGE 027.3839-00.3031 could be an LBV.

\section{Elliptical bubbles}
\label{sec:ell}
As reported in Section \ref{sec:data_red}, seven bubbles show an `elliptical' shape. Taking a look at Figure \ref{fig:bubbles_E} we can sketch a common morphological appearance. The basic shape of these bubbles is an elliptical ring, presenting different elongations. There is the total lack of a central object, with the brightness of the bubble centre comparable to background. The nebula brightness is not uniform along the ring. Furthermore, the brightness shows a rough bilateral symmetry, the axis of symmetry being its major axis. The brightness is enhanced perpendicularly to the axis of symmetry. A partial exception is represented by MGE 009.3523+00.4733 (Figure \ref{fig:bubbles_E}, first picture of the second row) and, to a lesser extent, MGE 005.6102-01.1516, where, beside the (slightly hidden) bilateral symmetry, a clumpiness is evident.

The elliptical shape in Galactic objects is a common feature of PNe (e.g. \citealt{Zhang1998}). It is likely produced by the interaction of a fast wind with the previous slow wind of the AGB phase. \citet{Aaquist1996} and \citet{Zhang1998} showed that radio and optical morphologies could be well reproduced by the so-called `prolate ellipsoidal shell' model, where the winds interaction is modelled in an environment characterized by a radial and latitude-dependent density gradient. The PN appears, simply from projection, as a spherical shell, an ellipsoidal shell or butterfly-shaped. For spherical and ellipsoidal shells, the brightness toward one axis can be greater with respect to the orthogonal axis. The radio and optical emissions can be reproduced by the same model because they both arise from the circumstellar gas.

Among the seven elliptical bubbles, MGE 009.3523+00.4733 is classified as a PN \citep{Pottasch1988} and MGE 016.2280-00.3680 is a proposed PN candidate \citep{Anderson2011}. The remaining five bubbles are not classified yet. At $24\mic{m}$, following the morphological classification scheme proposed by \citet{Mizuno2010}, they all appear as a disk nebula, except MGE 005.6102-01.1516 classified as a ring nebula. \citet{Nowak2014} pointed out that, among the MIPSGAL bubbles, about 90 per cent of classified disk nebulae and 60 per cent of ring nebulae are PNe, and conversely about 95 per cent of bubbles classified as PN are disk (65 per cent) or ring (30 per cent) nebulae. As we discussed in another work \citep{Ingallinera2015}, the difference between disk and ring nebula may be an artefact of the limited resolution of MIPSGAL and many disk nebula would appear as rings in higher-resolution images. Taking into account the radio and MIPSGAL morphology, we propose the five unclassified elliptical bubbles as PN candidates.

\begin{figure*}
\includegraphics[height=4.9cm]{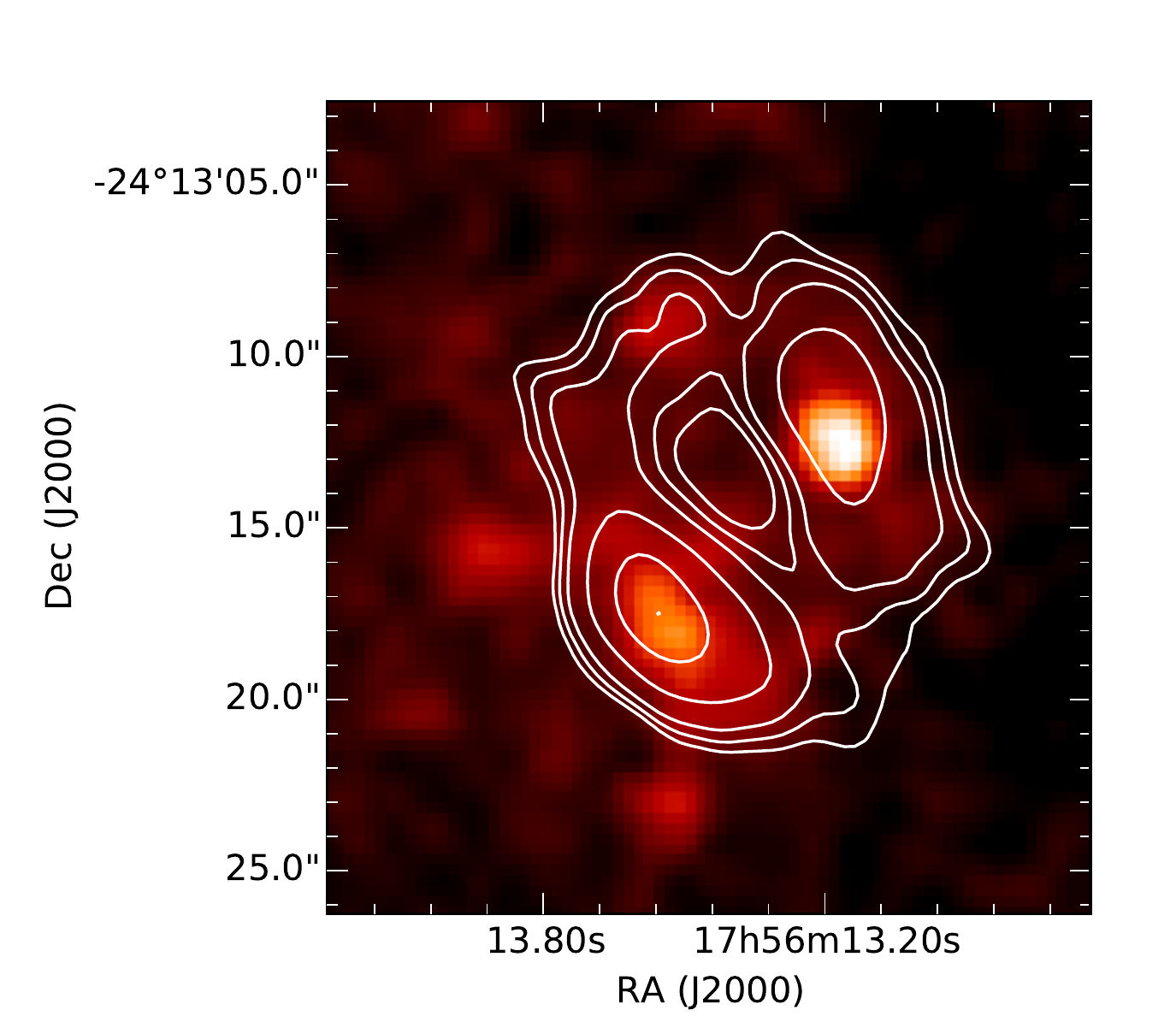}
\includegraphics[height=4.9cm]{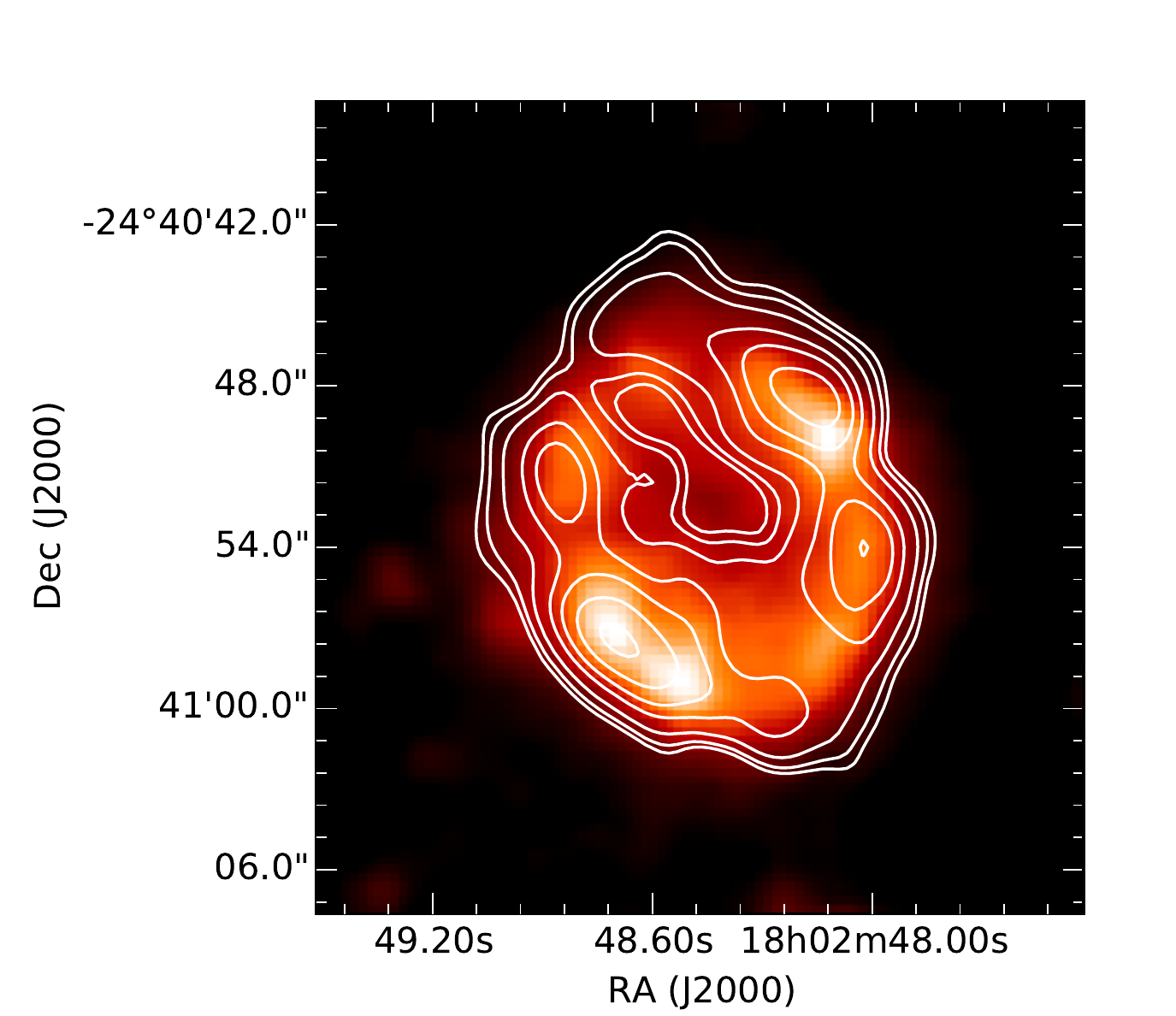}
\includegraphics[height=4.9cm]{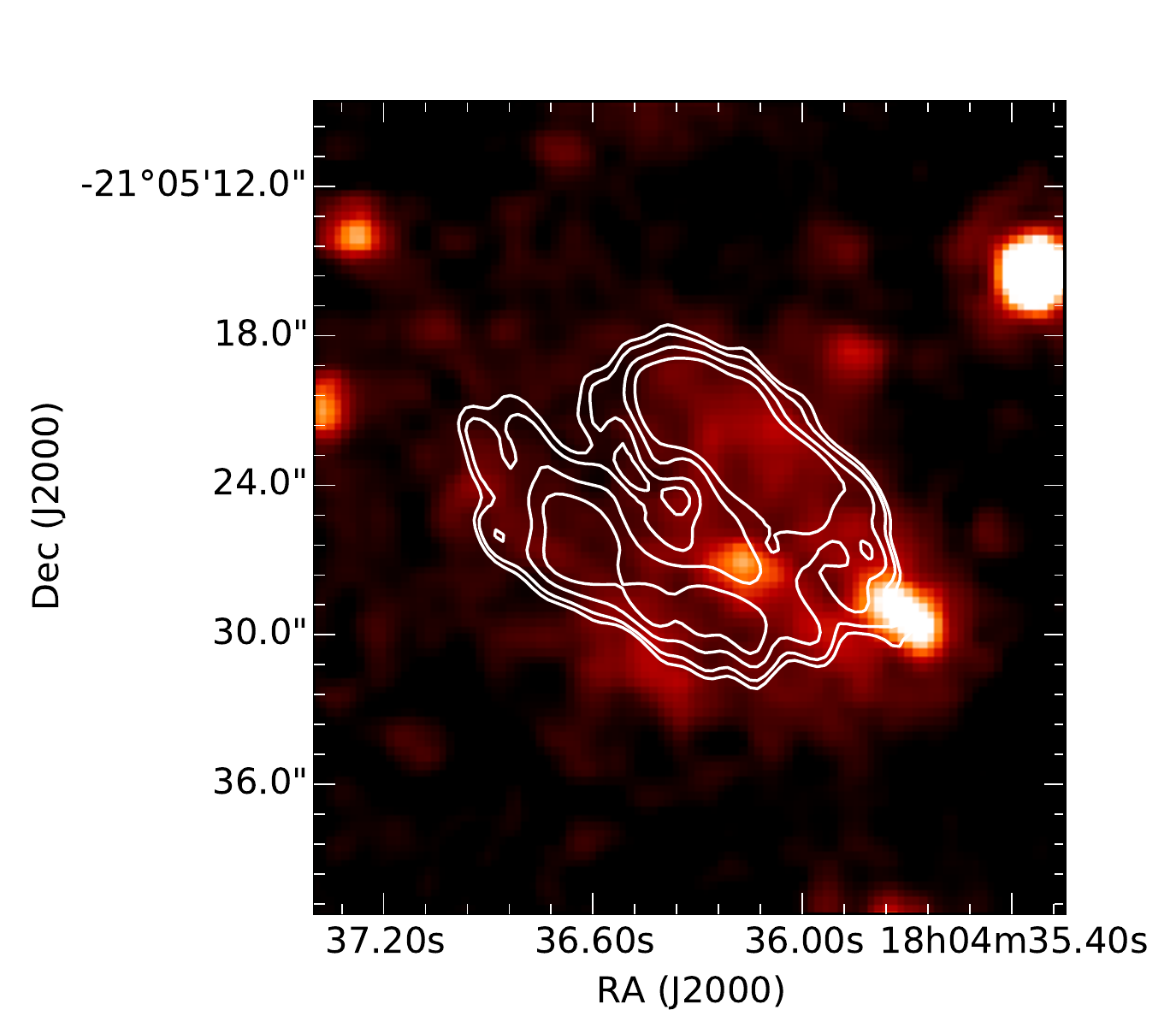}
\includegraphics[height=5.1cm]{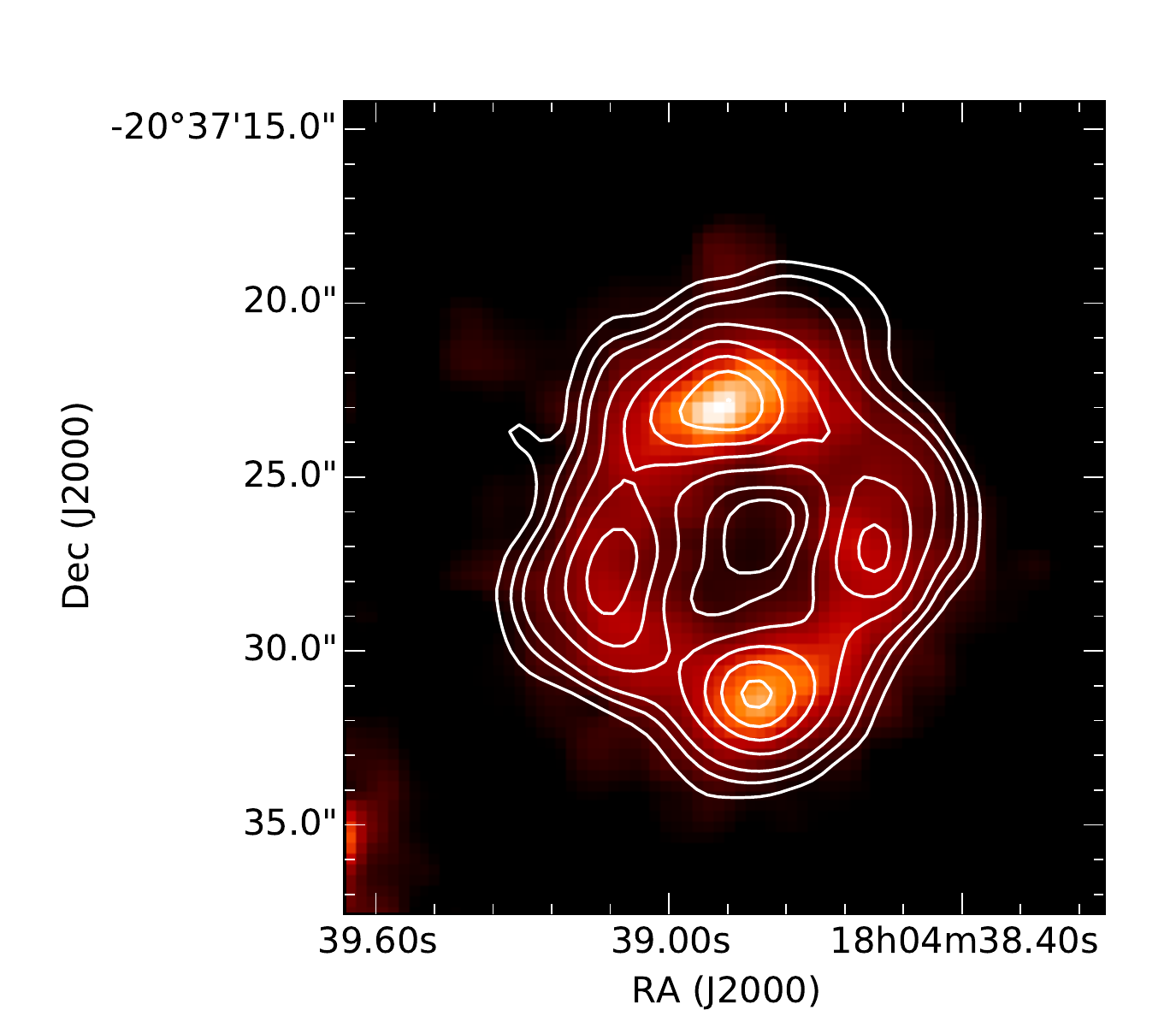}
\includegraphics[height=4.8cm]{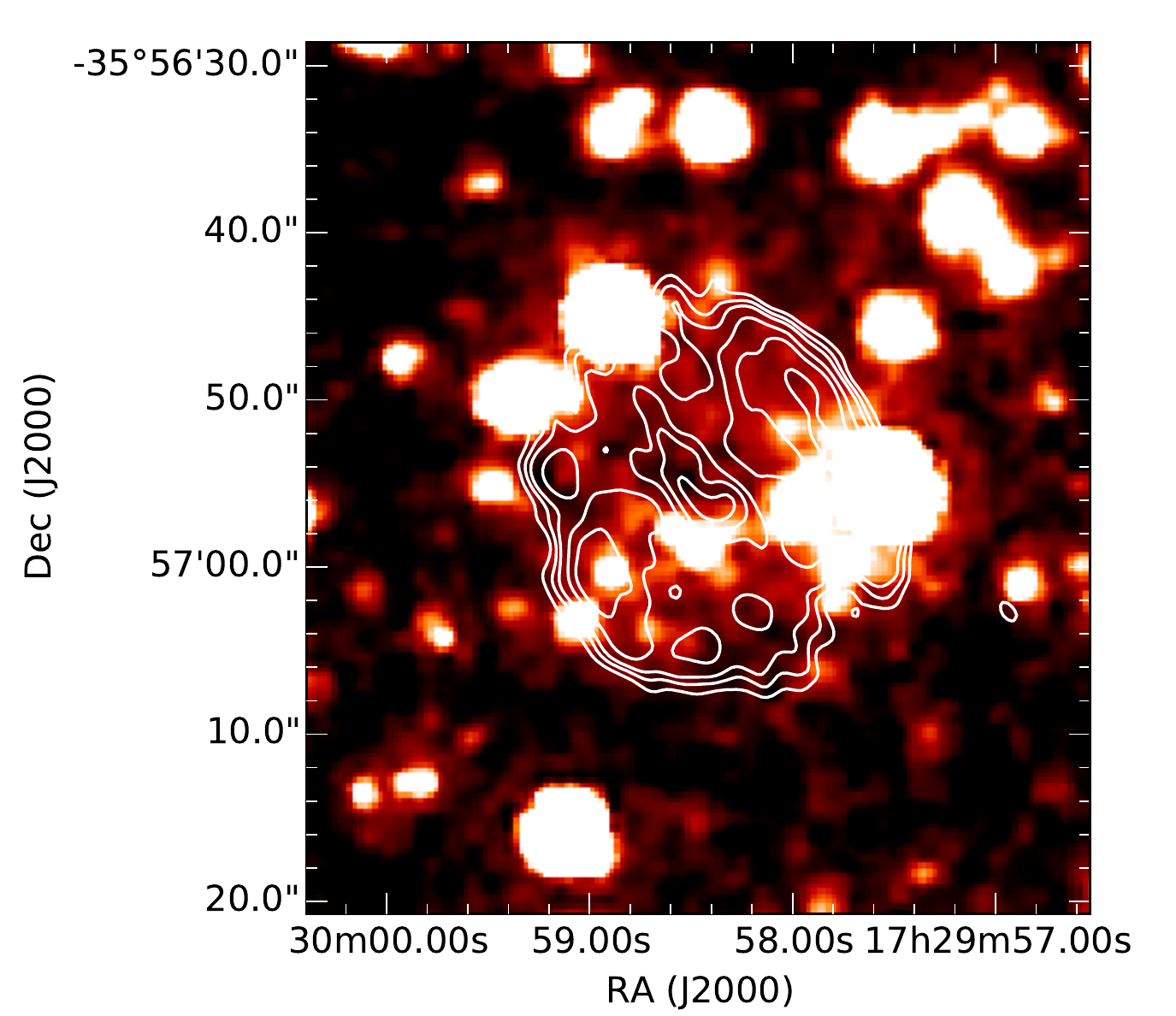}
\caption{Superposition of IRAC 8-$\umu$m images and 6-cm contours of the `elliptical' bubbles. From left to right: (row 1) MGE 005.2641+00.3775, MGE 005.6102-01.1516, MGE 008.9409+00.2532; (row 2) MGE 009.3523+00.4733, MGE 352.3117-00.9711 (for this last bubble 4.5$\umu$m image is used instead.}
\label{fig:sup_E}
\end{figure*}

\begin{figure*}
\includegraphics[height=7cm]{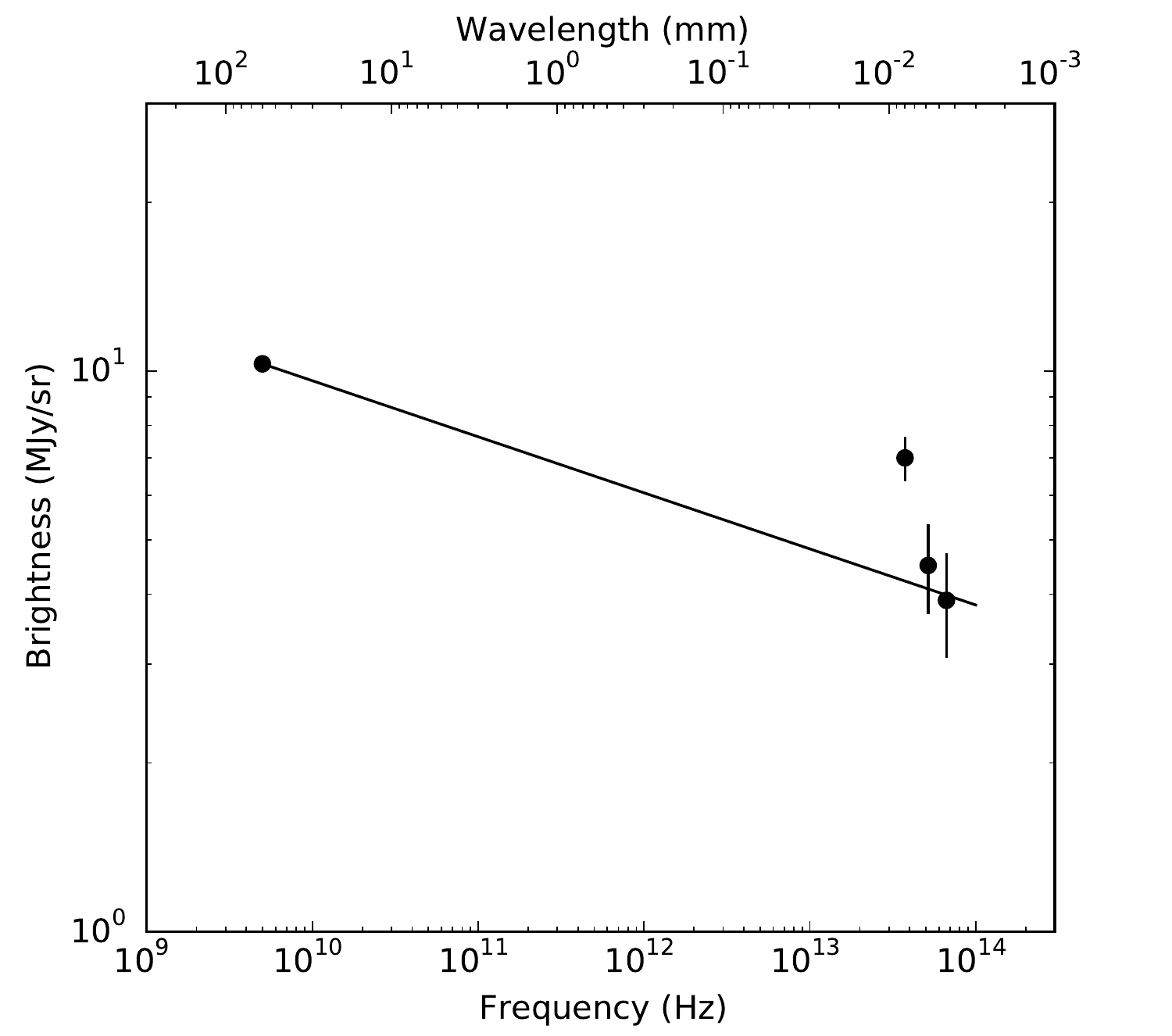}
\includegraphics[height=7cm]{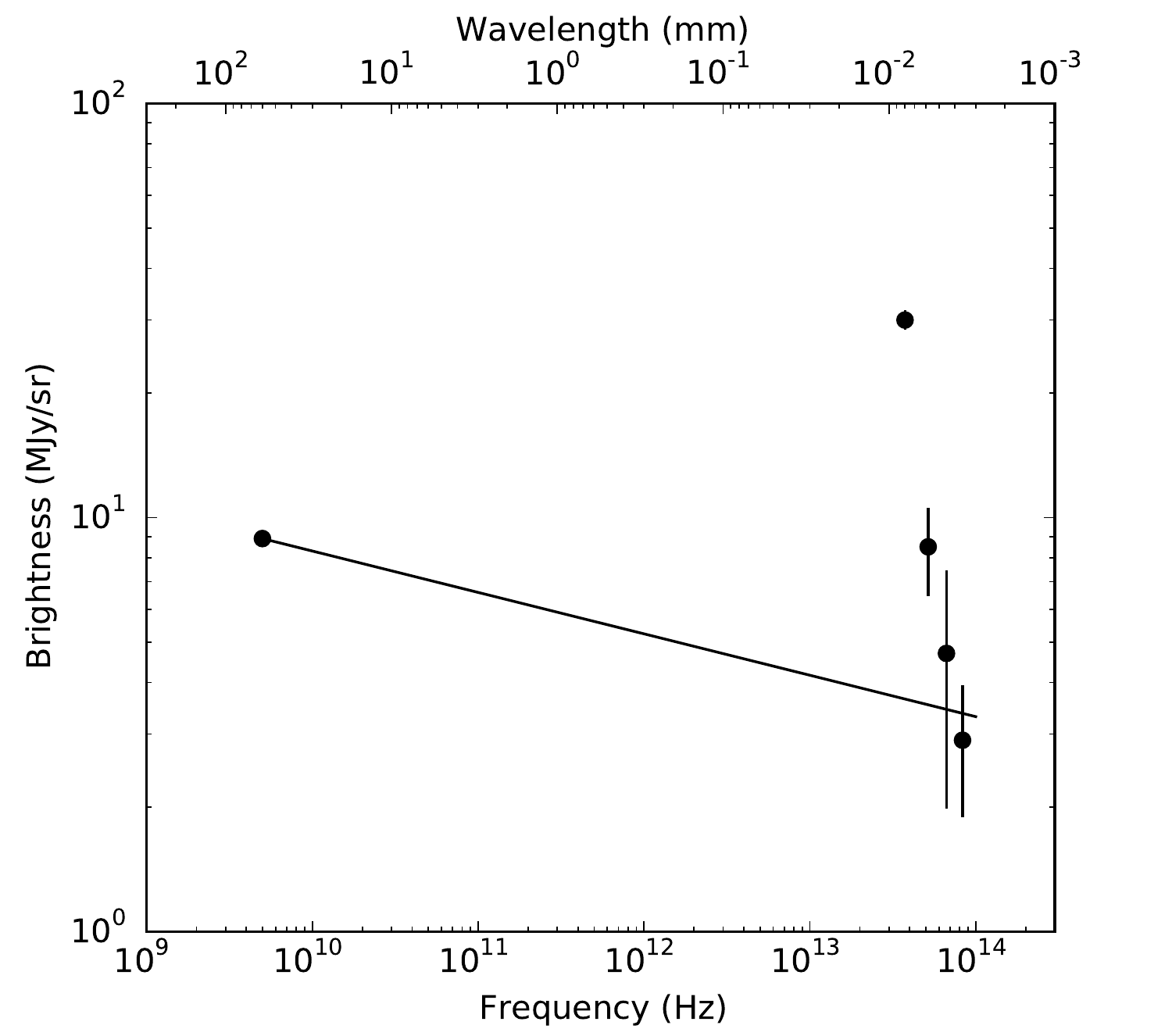}
\includegraphics[height=7cm]{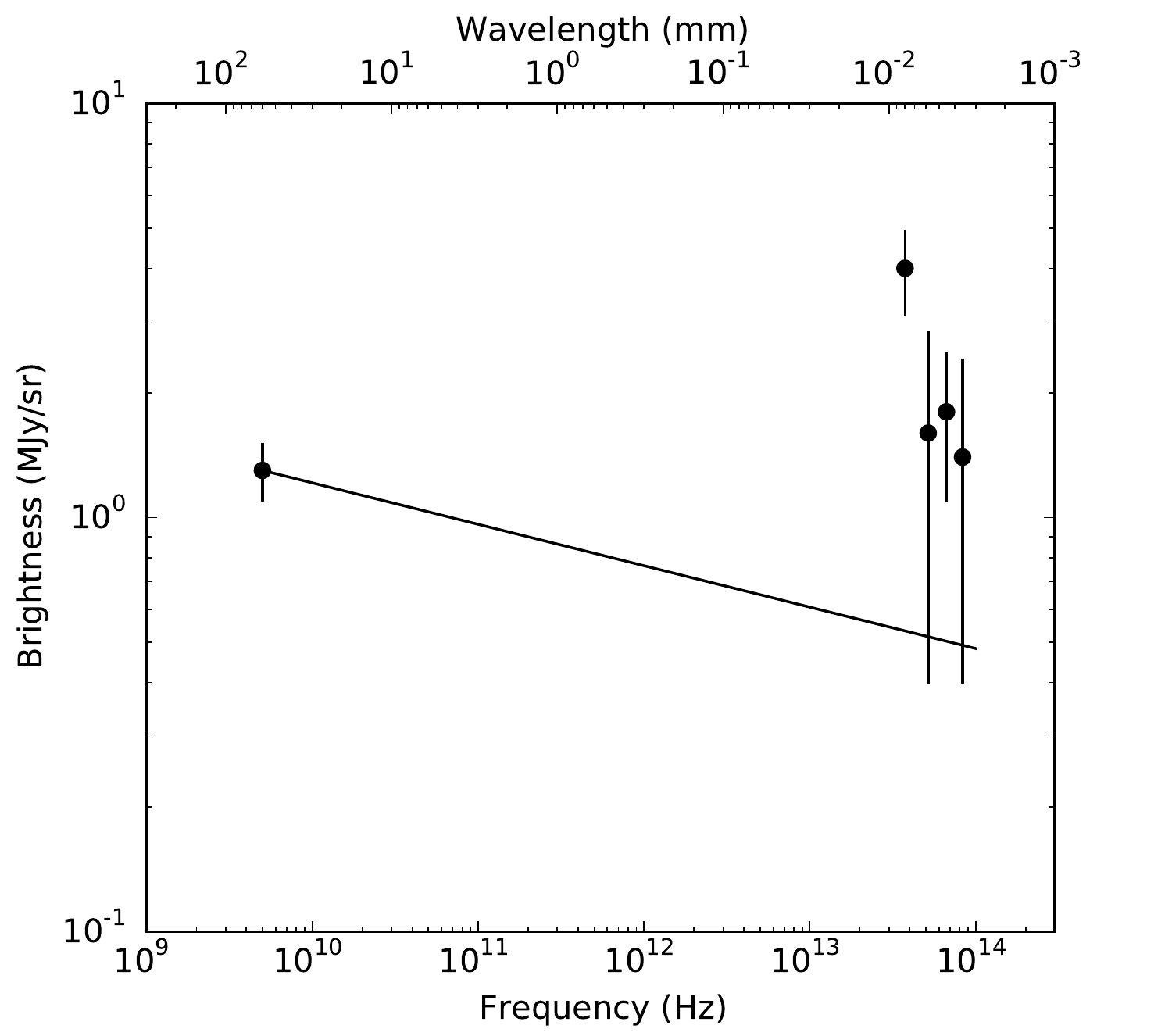}
\includegraphics[height=7cm]{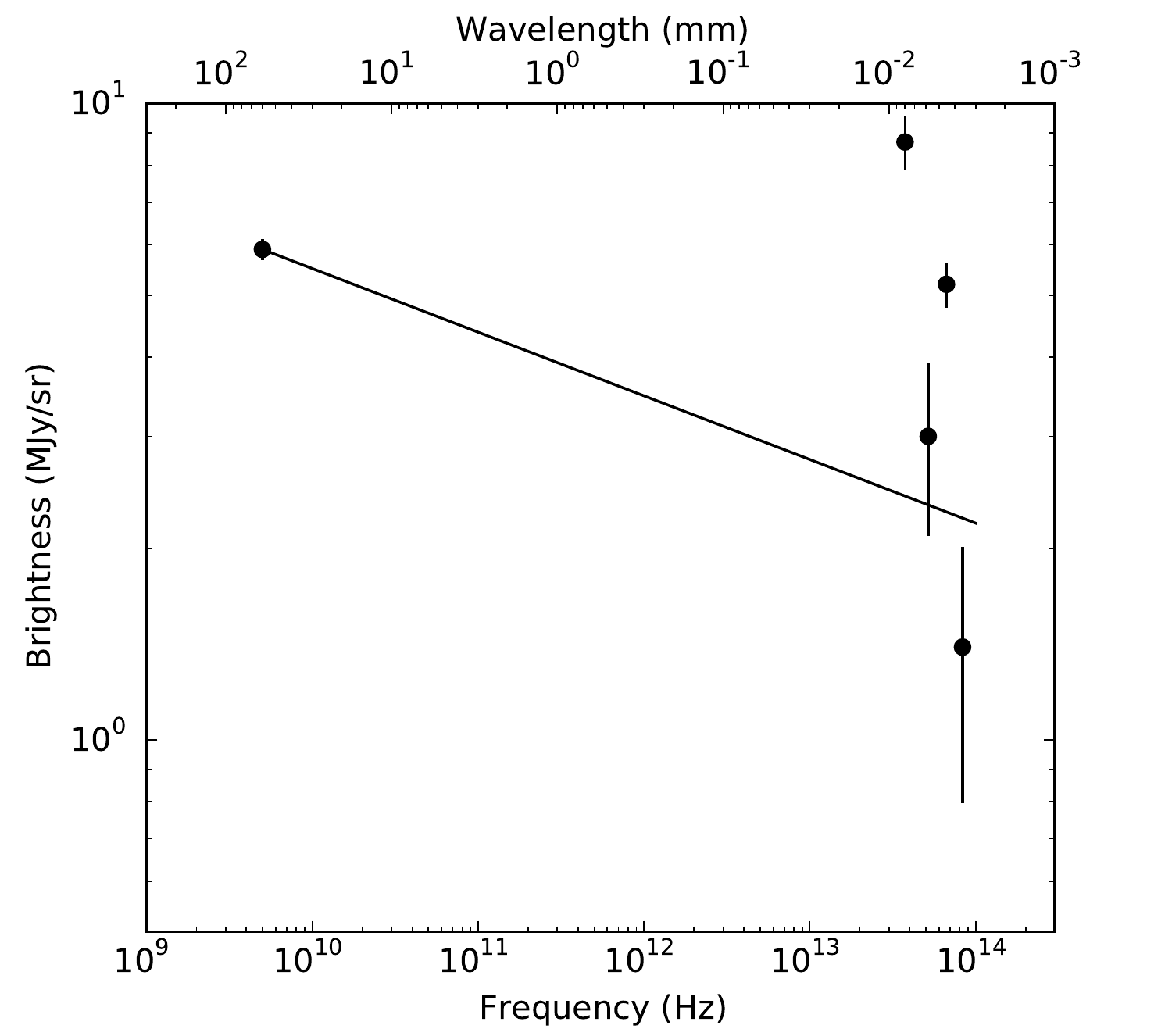}
\includegraphics[height=7cm]{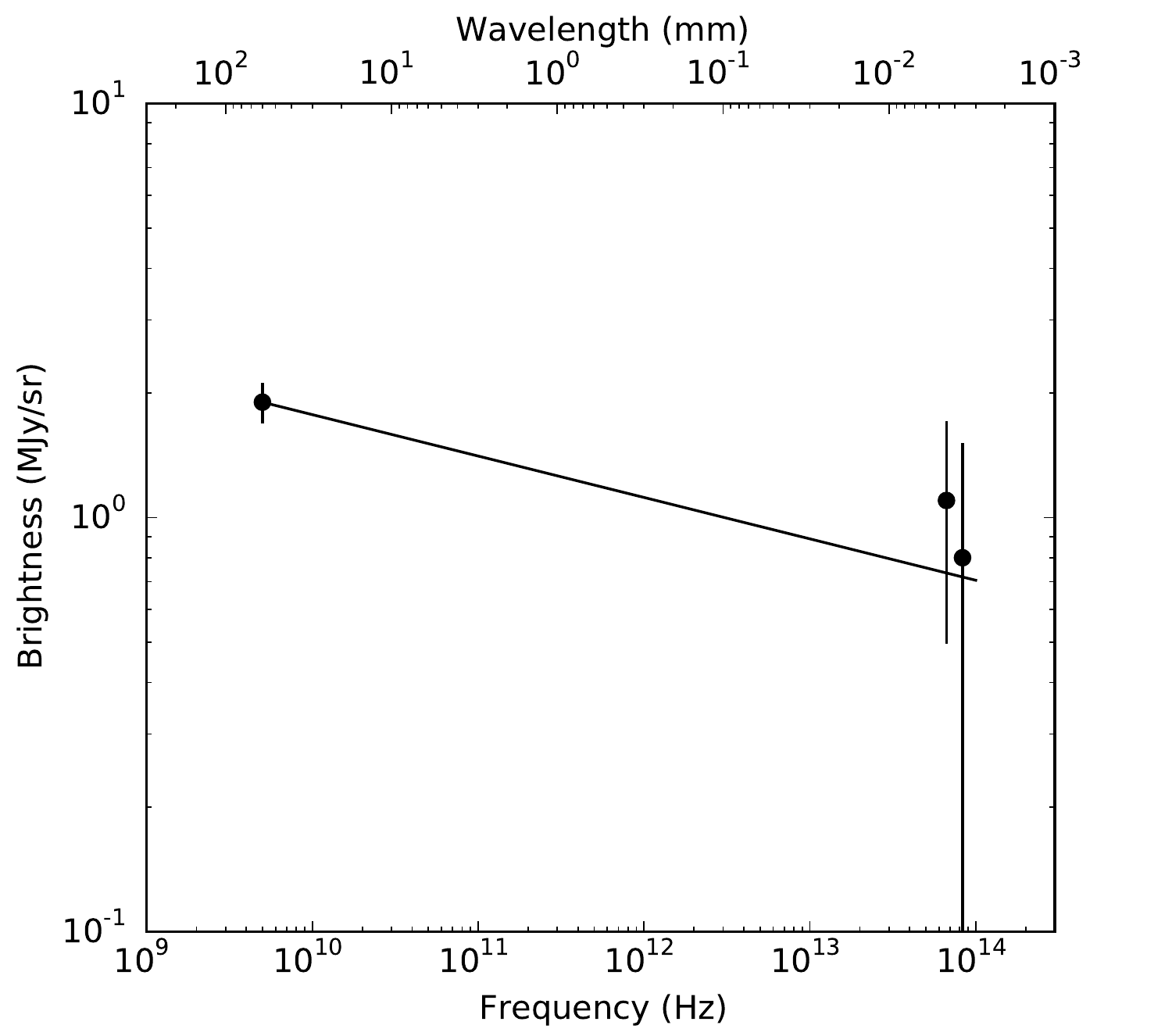}
\caption{SED of the `elliptical' bubbles. From left to right: (row 1) MGE 005.2641+00.3775, MGE 005.6102-01.1516; (row 2) MGE 008.9409+00.2532, MGE 009.3523+00.4733; (row 3) MGE 352.3117-00.9711. The solid line is an extrapolation of the free-free emission from the 6-cm point, considering a spectral index $\alpha=-0.1$. Please notice that for some bubbles it was not possible to derive their brightness in some bands.}
\label{fig:sed_E}
\end{figure*}

A more meaningful radio-IR comparison can be made using GLIMPSE images instead of MIPSGAL. At the IRAC wavelengths the image resolution is around $3\um{arcsec}$, only slightly worse than our VLA map resolution (around $2\um{arcsec}$), and twice as good as MIPSGAL. The heavy drawback is that GLIMPSE only seldom detects the MIPSGAL bubbles. This is mainly due to an intrinsic lower brightness at IRAC wavelengths and to a higher image noise both from background and foreground (Galactic diffuse emission, stars). As a result, only for five elliptical bubbles we were able to detect a counterpart in IRAC bands, while for the other two, namely MGE 016.2280-00.3680 and MGE 034.8961+00.3018, this was not possible. Superpositions of the 8-$\umu$m images and 6-cm contours are presented in Figure \ref{fig:sup_E}, except for MGE 352.3117-00.9711 where 4.5-$\umu$m image is reported instead, since for this bubble the 8-$\umu$m is characterized by an extremely high background. The comparison between radio and 8-$\umu$m images shows a substantial coincidence of the two emission, with the 8-$\umu$m emission slightly less extended than radio, except for MGE 008.9409+00.2532. As far as we know, there are not extensive studies on the morphological comparison between radio and the infrared wavelengths covered by IRAC. \citet{Zhang2009} compared GLIMPSE and H$\alpha$ which, tracing the excited gas, is usually co-spatial with radio emission. They found that the greatest differences between H$\alpha$ and $8\mic{m}$ mainly characterize the bipolar PNe, where the former is much more extended. This kind of PNe is mostly excluded \textit{a priori} from the MIPSGAL bubbles, since the sample was selected based on the roundish morphology (ellipsoidal disks are included, especially taking into account that many bubbles are only barely resolved). For the other types of PNe some difference still exist. \citet{Zhang2009} then show the IR SED of some nebulae discussing the behaviour at IRAC bands. The result is that the large bands collect both continuum emission and lines from silicates, molecules (like PAH or H$_2$) and atomic species (e.g. Br$\alpha$). This makes the SED of each PN somehow peculiar, with no standard reference. In IRAC images, our elliptical bubbles, being close to Galactic plane, are located in fields particularly crowded with stars. Aperture photometry to derive their IR flux density is extremely hard and prone to errors. Exploiting the fact that both in radio and in GLIMPSE our bubbles are resolved, we built their SEDs starting from the brightness. To obtain a significant value for the brightness we selected an area of the nebula far from contaminant stars in the image at the shortest wavelength where the bubble is still detected. We then calculated the median value of the brightness inside the same area, both for radio and IR. For GLIMPSE images we estimated the background picking the median value of an opportune region close to the nebula and free of stars. In Figure \ref{fig:sed_E} we show the SEDs obtained, overplotting a pure free-free emission extrapolated from the 6-cm brightness using a spectral index $\alpha=-0.1$. The 8-$\umu$m brightness is usually much higher than the extrapolated free-free value, very likely because at this wavelength the thermal emission from dust becomes dominant. The brightness at shorter wavelengths is closer to the free-free values, a sign that the emission in now dominated by gas continuum. However several deviation can be observed, likely to be ascribed to the line transitions reported above. For example the 4.5-$\umu$m is higher than 5.8-$\umu$m in MGE 008.9409+00.2532 and MGE 009.3523+00.4733 (SEDs in the second row of Figure \ref{fig:sed_E}), a clear sign that the emission at this wavelength for these bubbles has a relevant line component.

In this scenario a rough similarity between IR and radio image is then remarkable. Indeed in very young PNe the dust component, whose composition is related to the original metallicity of the star, is prominent and is tightly linked to the evolutionary phase of the transition of the nebula (see, for example, \citealt{Stanghellini2012}). But as the CSE expands and the UV stellar radiation increases, most of the dust is destroyed and the optical and IR emission becomes dominated by gas lines from highly excited atomic species \citep{Flagey2011}. In particular, as the PN evolves, the 24-$\umu$m emission begins to be dominated by [O \textsc{iv}] line at $25.9\mic{m}$ and to a lesser extent by [Ne \textsc{v}] line at $24.3\mic{m}$ \citep{Chu2009}. The [O \textsc{iv}] line at $25.9\mic{m}$ is a tracer of He \textsc{ii}. In fact the He \textsc{ii} $\lambda$4686 line derives from the recombination of He \textsc{iii}, whose excitation potential ($54.4\um{eV}$) is close to that of O \textsc{iv}. The radio emission, arising from the ionized gas, mostly traces the same ionized and highly-excited gaseous nebula and the PN retains the same overall morphology in all these parts of the electromagnetic spectrum. An exception is represented by optically thick evolved PNe where the diameter of the He \textsc{ii} zone (and therefore the 24-$\umu$m emission) is usually smaller than the ionised hydrogen zone (i.e. the radio nebula). It is probable that the component of the IR emission arising from thermal dust ($8\mic{m}$) is co-spatial with the radio emission because the dust is heated by the same UV photons responsible for gas ionization. This morphological similarity is not observed in other kinds of Galactic sources, nor in evolved massive star, as discussed in Section \ref{sec:bub_cs}, nor in H \textsc{ii} regions, as we are going to discuss in \ref{sec:bub_hii}.


\section{Filled elliptical bubbles}
\label{sec:fill}
\begin{figure}
\includegraphics[width=\columnwidth]{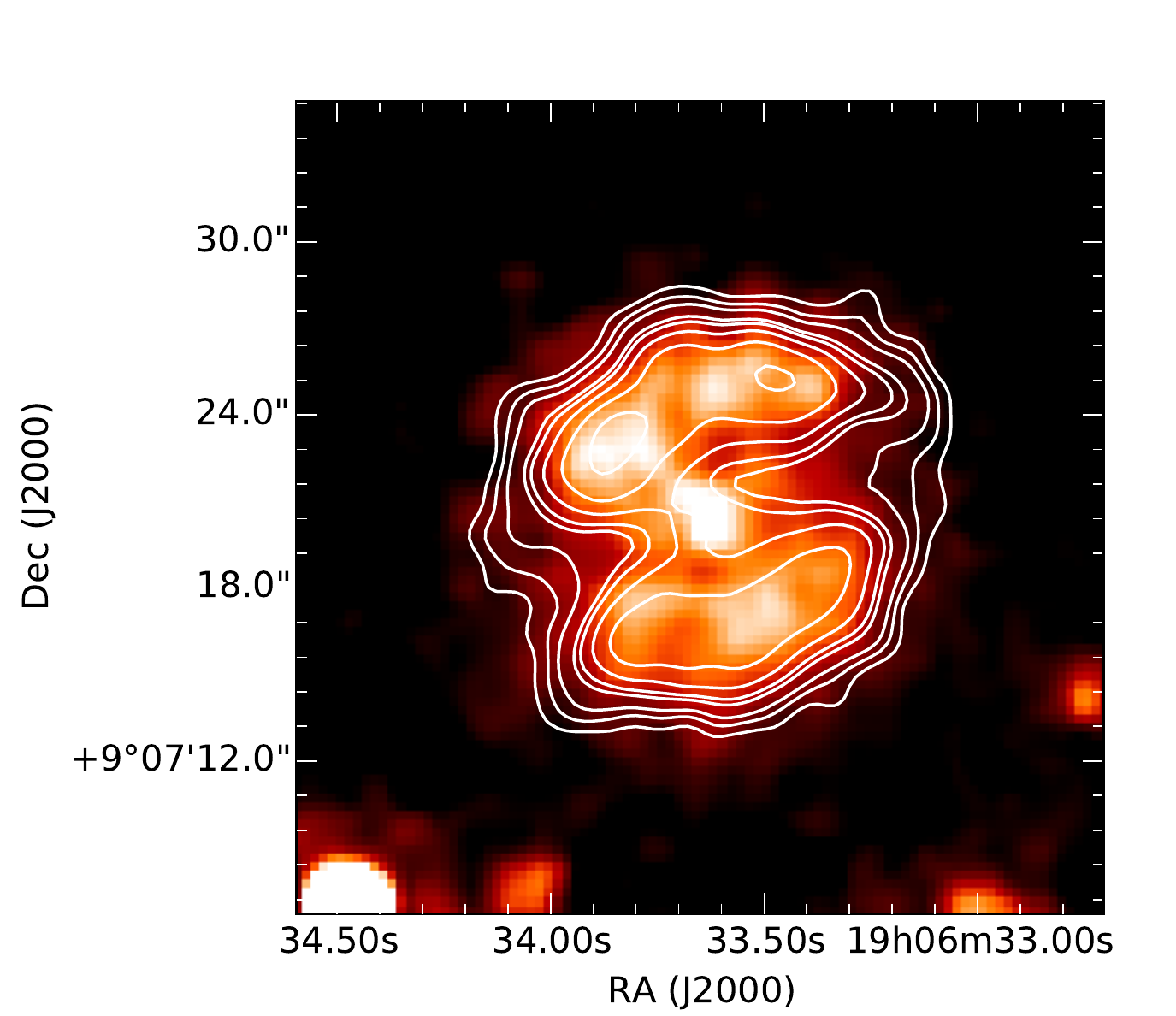}
\caption{Superposition of 8-$\umu$m image and 6-cm contours of MGE 042.7665+00.8222.}
\label{fig:3448}
\end{figure}

Four bubbles, labelled with `F' in Table \ref{tab:data} and dubbed `filled elliptical bubbles', show a morphology intermediate between the elliptical bubbles and those with central object. In the radio, these four bubbles are characterized by an overall roundish shape and at their centre there is not a clear point source but a low-brightness diffuse emission filling almost all the nebula and clearly above the background level. \citet{Mizuno2010} classified them as disks, except MGE 042.7665+00.8222 classified as two-lobed nebula. None of these bubbles is definitely classified. The most studied of them is MGE 042.7665+00.8222, proposed by \citet{Clark2005} as a PN candidate. This interpretation seemed to be confirmed by \citet{Gvaramadze2010} who spectroscopically found that the central star is a [WC], while \citet{Flagey2014} suggested that it is indeed a WC5-6 star. To shed light on the nature of these four bubbles we search for their counterpart at $8\mic{m}$. For MGE 042.7665+00.8222 the nebula and a central object are clearly detected (Figure \ref{fig:3448}), while for the others we found only hints of the nebula and no clear central objects. The central object is detect also in IPHAS (IPHAS2 J190633.67+090720.7) with a $r$ magnitude of $20.64\pm0.08$ \citep{Barentsen2014}.

\begin{figure}
\includegraphics[width=\columnwidth]{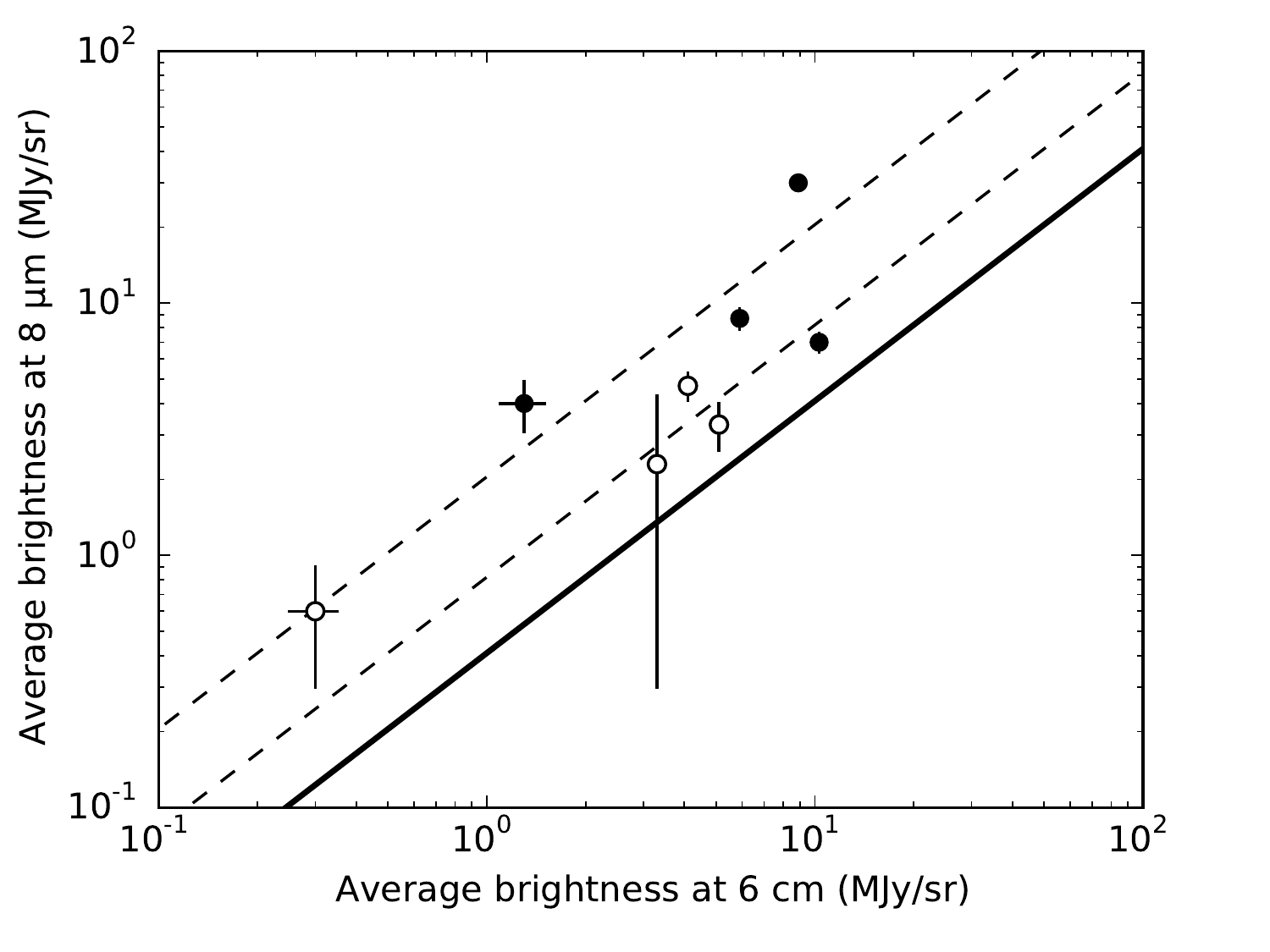}
\caption{Average brightness at $8\mic{m}$ against average brightness at $6\um{cm}$ for the `elliptical' bubbles (solid circles) and for the `filled elliptical' bubbles (open circles). The solid line represent a pure free-free emission, while the two dotted lines represent a brightness respectively twice and five times greater.}
\label{fig:8_6}
\end{figure}

The striking difference between MGE 042.7665+00.8222 and the bubbles with central object is that the 8-$\umu$m emission of the latter is largely dominated by the central object (Figure \ref{fig:bubC_sup}), while in MGE 042.7665+00.8222 it has approximately the same brightness as the nebula. Indeed the radio morphology is still compatible with a PN, closely resembling NGC 6439 shown by \citet{Zhang1998}. And this is the case also for the other three bubbles, showing a strongly asymmetric elliptical shape resembling, for example, the PN Jonckheere 900 (PN VV 28). Bearing in mind what we discussed in Section \ref{sec:bub_cs} and that massive evolved stars are characterized by strong stellar winds, we suspect that these four bubbles could all be PN candidates. For all of them we were able to estimate the average brightness at $8\mic{m}$ as described in Section \ref{sec:ell}, an operation not possible for the other IRAC bands because the low brightness and high background prevent their detection. In Figure \ref{fig:8_6} we plot the 8-$\umu$m brightness against the 6-cm brightness for the four filled elliptical bubbles and for the elliptical bubbles detected at $8\mic{m}$.
For filled elliptical bubbles the mean ratio between the 8-$\umu$m measured brightness and that expected extrapolating from radio a pure free-free emission is $2.7\pm1.3$, while for the elliptical bubbles it is $5.2\pm2.7$. This lower ratio explains the difficulty of their detection at $8\mic{m}$. It is not possible, at least with our data, to say whether the discrepancy between these two morphological groups is real or it is just due to the poor statistics. If it were real it could be ascribed to a dust deficiency and/or to less dominating line emission.

\section{Other bubbles}
\label{sec:others}

\subsection{The H \textsc{ii} region candidate MGE 031.7290+00.6993} 
\label{sec:bub_hii}
It is possible that a fraction of the MIPSGAL bubbles is not represented by evolved star CSEs but is related to young massive stars. These stars are usually embedded in clouds of pre-existing material (gas and dust) that becomes heated and excited. If the star is hot enough (spectral types O and early B) the circumstellar gas is ionized and can be observed in radio as an H \textsc{ii} region.

The mechanism of radio emission from H \textsc{ii} region is the same of PNe (i.e. thermal free-free), characterized by a flat SED ($\alpha=-0.1$ if optically thin). As previously said, massive evolved stars show a flat SED, at least globally. There is then a degeneracy between H \textsc{ii} region and evolved stars in radio, and therefore it is not possible to discriminate between these kinds of sources from their SED. IR colours are sometimes used, since H \textsc{ii} regions and PNe have different dust temperature and different IR emission mechanism. In Paper I we showed that one of the most important IR band for this goal is the $8\mic{m}$. The comparison between radio and $8\mic{m}$, for example, effectively discriminates between H \textsc{ii} regions and PNe. The peculiar element in H \textsc{ii} regions is that the emission at $8\mic{m}$ traces the PAH outside the ionized nebula \citep{Deharveng2010}. This explains the 8-$\umu$m excess in the comparison with radio with respect to PNe. However, since radio is a tracer of the ionized region, provided enough resolution, the reciprocal distribution of radio and 8-$\umu$m emission is an immediate tool to spot H \textsc{ii} regions.

\begin{figure}
\includegraphics[width=\columnwidth]{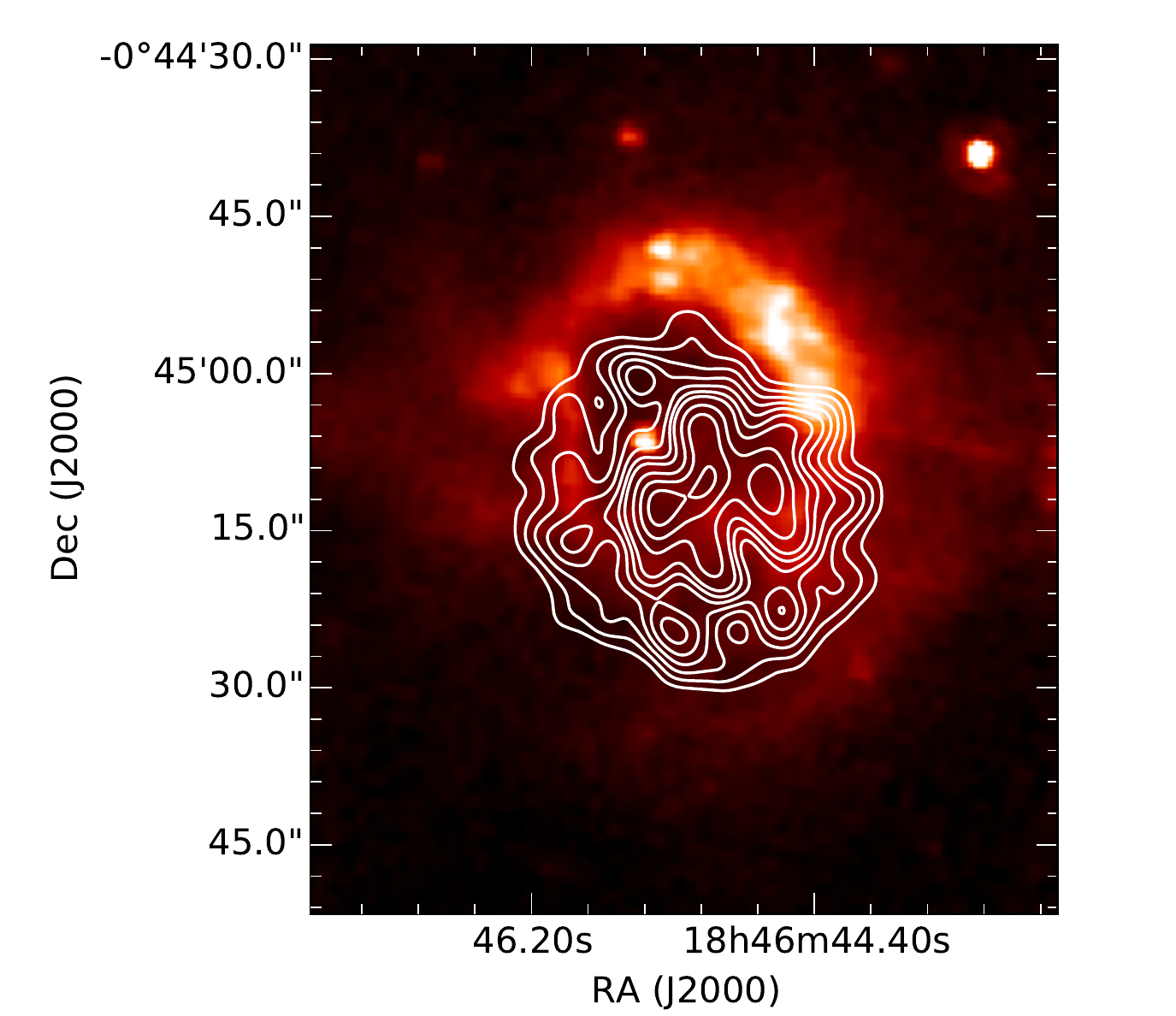}
\caption{Superposition of 8-$\umu$m image and 6-cm contours of MGE 031.7290+00.6993.}
\label{fig:3354}
\end{figure}

In our sample, the bubble MGE 031.7290+00.6993 was proposed as an H \textsc{ii} region by \citet{Anderson2011} on the basis of radio recombination lines studies. In Figure \ref{fig:3354} we report a superposition of the 8-$\umu$m and radio emission.
It is possible to notice how the 8-$\umu$m emission is clearly placed outside the ionized gas nebula traced by radio. \citet{Anderson2011} notice that 208 (over 603) of their H \textsc{ii} region candidates were associated to the `GLIMPSE bubbles' found by (\citealt{Churchwell2006}; not to be confused with the MIPSGAL bubbles discussed in this work), objects that present a bubble structure at $8\mic{m}$ very similar to MGE 031.7290+00.6993. Our conclusion is therefore that MGE 031.7290+00.6993 is very likely an H \textsc{ii} region. The most promising candidate for its ionising star is probabily the point source GLIMPSE G031.7269+00.6988. It is located at the position of the 24-$\umu$m maximum and at the geometrical centre of the radio emission. This star is likely detected in UKIDSS and IPHAS.

\subsection{MGE 028.7440+00.7076} 
\begin{figure}
\includegraphics[width=\columnwidth]{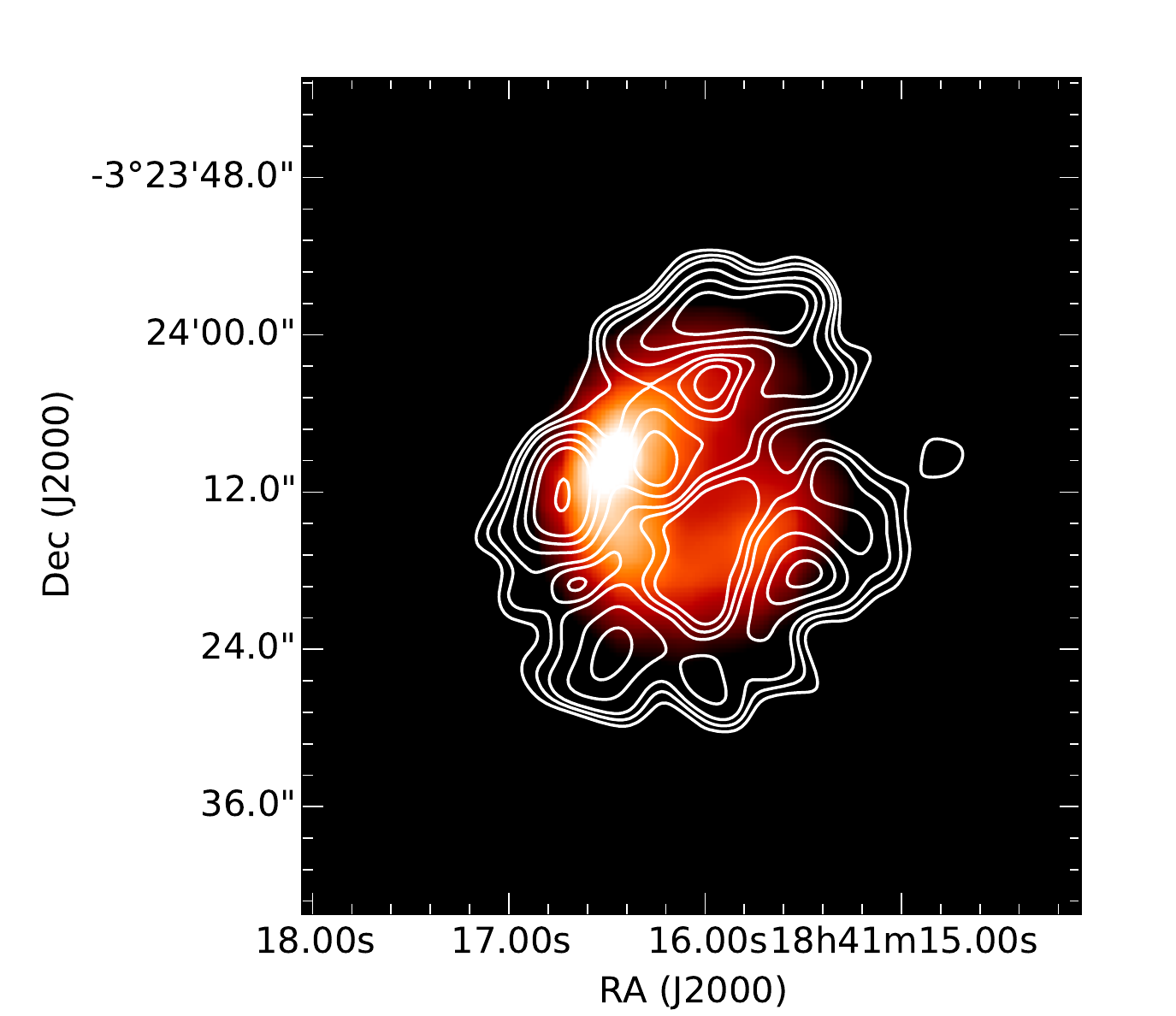}
\caption{Superposition of 24-$\umu$m image and 6-cm contours of MGE 028.7440+00.7076.}
\label{fig:3333}
\end{figure}

This bubble has a very peculiar radio morphology, being composed of two different ring-like structures and an overall `8' shape. The nebula is not detected in any IRAC band, also because the field is particularly crowded with stars and diffuse emission. However, the relatively large extension of the radio nebula ($33\um{arcsec}$) allows a significant comparison with the 24-$\umu$m image (Figure \ref{fig:3333}).
From the superposition it emerges that the 24-$\umu$m emission is less extended than radio, and radio itself seems to wrap the former. In Paper I we had reported a positive spectral index ($\alpha=0.39\pm0.26$) between 1.4 and $5\um{GHz}$, suggesting a thermal emission. This bubble is barely detected in H$\alpha$ images from the SHS, appearing slightly more extended than the radio nebula. It is possible that this bubble is a ionization-stratified PN (with a central [O \textsc{iv}]/He \textsc{ii} Str\"omgren zone). The PN nature is supported by the lack of central star candidates in IRAC images and at shorter wavelengths, excluding therefore the presence of a massive star. The peculiar morphology makes this source an interesting target for future studies.

\subsection{MGE 032.4982+00.1615} 
This bubble is somehow puzzling. In Paper I we had reported that it was probably characterized by a non-thermal spectral index $\alpha=-0.30\pm0.19$ calculated between 1.4 and $5\um{GHz}$. The high resolution radio image reveals that it is a compact source approximately $10\times7\um{arcsec^2}$. \citet{Flagey2014} suggested that its central star could be an O5-6V, based on near-IR spectroscopy, but they however warned that this classification is uncertain. They estimate a distance of $11\um{kpc}$ based on near-IR colours.

\begin{figure}
\begin{center}
\includegraphics[width=\columnwidth]{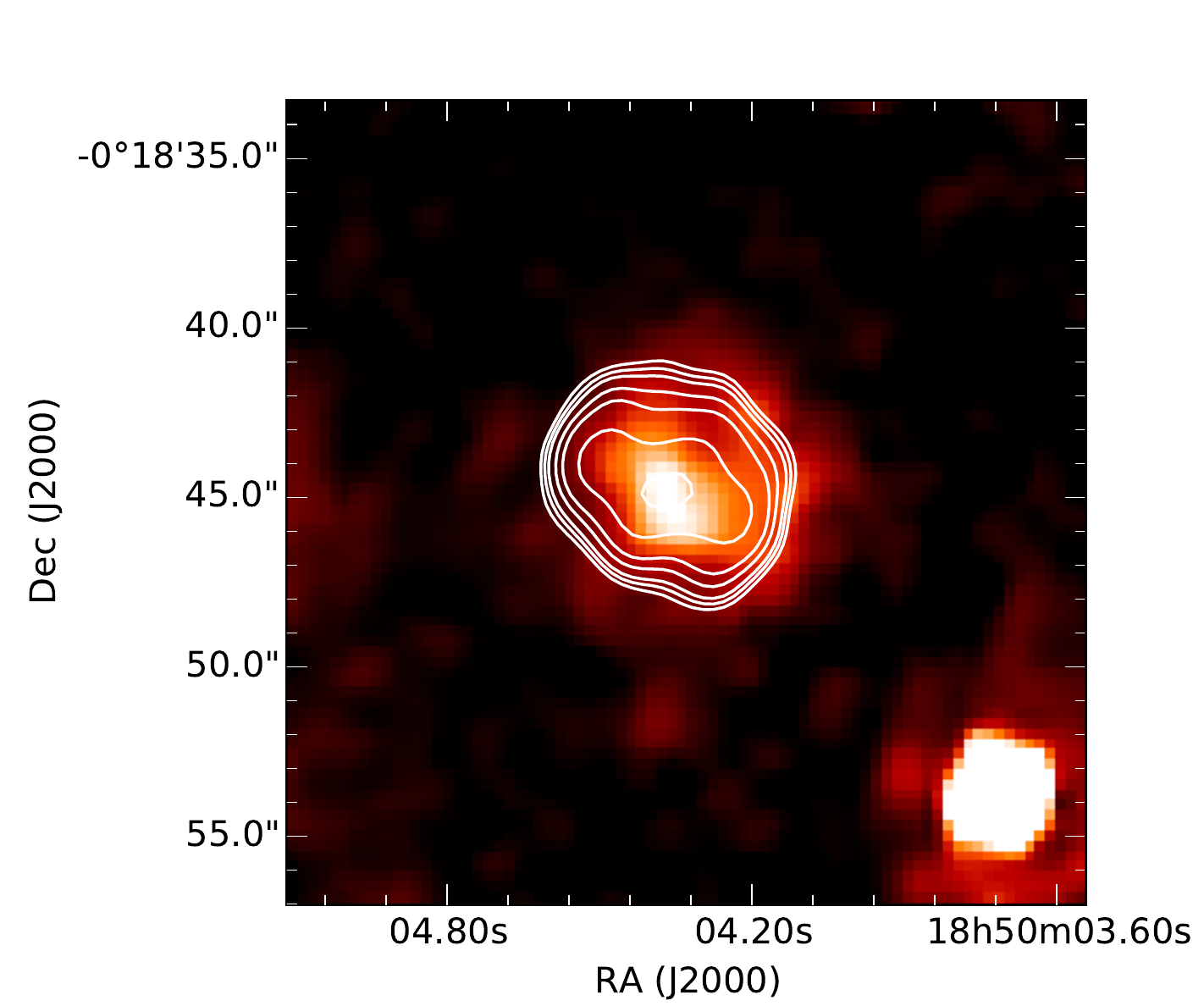}
\caption{Superposition of 8-$\umu$m image and 6-cm contours of MGE 032.4982+00.1615.}
\label{fig:3367_sup}
\end{center}
\end{figure}

From these new data at $5\um{GHz}$ we found that the spectral index of the nebula is $\alpha=-0.8$, even steeper than the one reported in Paper I. For the same source \citet{Bihr2016} found a spectral index $\alpha=-1.04\pm0.39$ between 1 and $2\um{GHz}$. Searching for flux density variability we found that the flux density at $5\um{GHz}$ of the 2015 data is $5.5\pm0.3\um{mJy}$, in comparison with the 2010 value of $4.7\pm0.9\um{mJy}$. Though still within the error, the two measurements suggest that an increase in flux density may have taken place. The time variability over year-scale as well as a non-thermal radio spectrum have been recently associated with proto-PNe (e.g. \citealt{Bains2009,Suarez2015}; Cerrigone et al. \textit{submitted}). In this scenario the post-AGB star at the centre of the nebula is evolving from cool spectral types ($T_\mathrm{eff}\sim\!5000\um{K}$) toward hotter temperature ($T_\mathrm{eff}>\!30\,000\um{K}$). During its evolution the star can therefore appear as spectral type O \citep{Davis2005}. The distance calculated by \citet{Flagey2014} is however too large to be compatible with this hypothesis. We suggest that the distance may have been over-estimated in the hypothesis of a post-AGB star. In fact the distance would be considerably less in this case, because post-AGB OB stars have similar $T_\mathrm{eff}$ and $\log g$ to main-sequence OB stars, but are less luminous. In IRAC images in fact the nebula appear extremely compact and co-spatial with radio (Figure \ref{fig:3367_sup}). Therefore our hypothesis of a proto-PN cannot be ruled out and further investigation are needed.

\subsection{MGE 356.7168-01.7246} 
The radio morphology of this bubble is strongly bipolar. A hint of bipolarity is observed also at $24\mic{m}$, while at $8\mic{m}$ the source, though partially outshined by a field star, has roughly the same morphology as in radio. The source is not unambiguously visible in any other of IRAC bands. In Paper I we proposed this bubble as a PN candidate, based on IRAS colours. A spectral index analysis confirms the value reported in Paper I (compatible with a free-free emission) and shows no variation within the nebula. Collimated bipolar outflows are sometimes observed in young PNe \citep{Redman2000} as well as in forming stars \citep{Gusdorf2015}. If the IRAS colour analysis is correct the latter option should be dropped in favour of the first. The nebula is likely detected in the H$\alpha$ images from the SHS, where it shows an elliptical shape with the same orientation of the radio nebula. The H$\alpha$ nebula is too faint to identify a bipolar structure similar to the radio nebula. Doubts on the classification of this bubble remains and apart from having revealed the radio bipolar morphology, the present observations cannot give us further hints on its nature.

\section{Discussion on Planetary Nebulae}
\label{sec:PNe}
\subsection{Physical properties}
As stated in Section \ref{sec:ell}, the morphological analysis of the elliptical bubbles suggests that, along with the known PN and the PN candidate, they can be proposed as PN candidates. In this section we assume this hypothesis valid and we derive some of their physical properties. In a previous work \citep{Ingallinera2014b} we showed that the 6-cm map alone is a gold mine for deriving important physical characteristics of a PN, provided that the nebula is resolved (since the method we are using is based on the nebula brightness). As a first step we need to estimate the distance of these bubbles, since no values are available in literature. A possible distance estimation is the statistical distance, calculated following the method by \citet{vandeSteene1995} which assumes that the nebula is optically thin. The distance is derived from the flux density and angular extension of the nebula. This method, though very easy, is affected by large uncertainties (\citealt{vandeSteene1995} declare an accuracy of 40 per cent) and supplies a reasonable value only if the nebula is truly a PN.

To derive the mean electron density and the ionized mass we followed the same procedure reported in \citet{Ingallinera2014b}. For each bubble we first create an optical depth map according to
\begin{equation}
B=B_\mathrm{bb}(T_e)\tau_\nu,
\end{equation}
where $B$ is the brightness, $T_e$ is the electron temperature, $B_\mathrm{bb}(T_e)$ is the brightness of a black body at a temperature $T_e$ and $\tau_\nu$ is the optical depth at the frequency $\nu$. Since for these bubbles $T_e$ is not known, we assumed for it a typical value of $10^4\um{K}$ \citep{Ingallinera2014b}. From the brightness map it is then possible to derive an optical-depth map. The optical depth is related to the emission measure $\mathrm{EM}$ through the relation
\begin{equation}
\tau_\nu=8.235\times10^{-2}T_e^{-1.35}\left(\frac{\nu}{\mathrm{GHz}}\right)^{-2.1}\left(\frac{\mathrm{EM}}{\mathrm{pc\times cm^{-6}}}\right),
\end{equation}
hence we can derive the electron density $n_e$ from the relation
\begin{equation}
\mathrm{EM}=\int_0^s\!\!\!n_e^2ds,
\end{equation}
where $s$ is the line-of-sight coordinate. It is finally possible to derive the PN ionized mass as a volume integral:
\begin{equation}
M_\mathrm{ion}\approx\int_Vn_em_pdV,
\label{eq:mion}
\end{equation}
where $m_p$ is the proton mass. For the actual calculation we assumed that the bubble is simply a uniform sphere. This assumption has no effect on the total ionized mass but we warn the reader that the electron density is just an average value over the entire volume.

\begin{figure}
\includegraphics[width=\columnwidth]{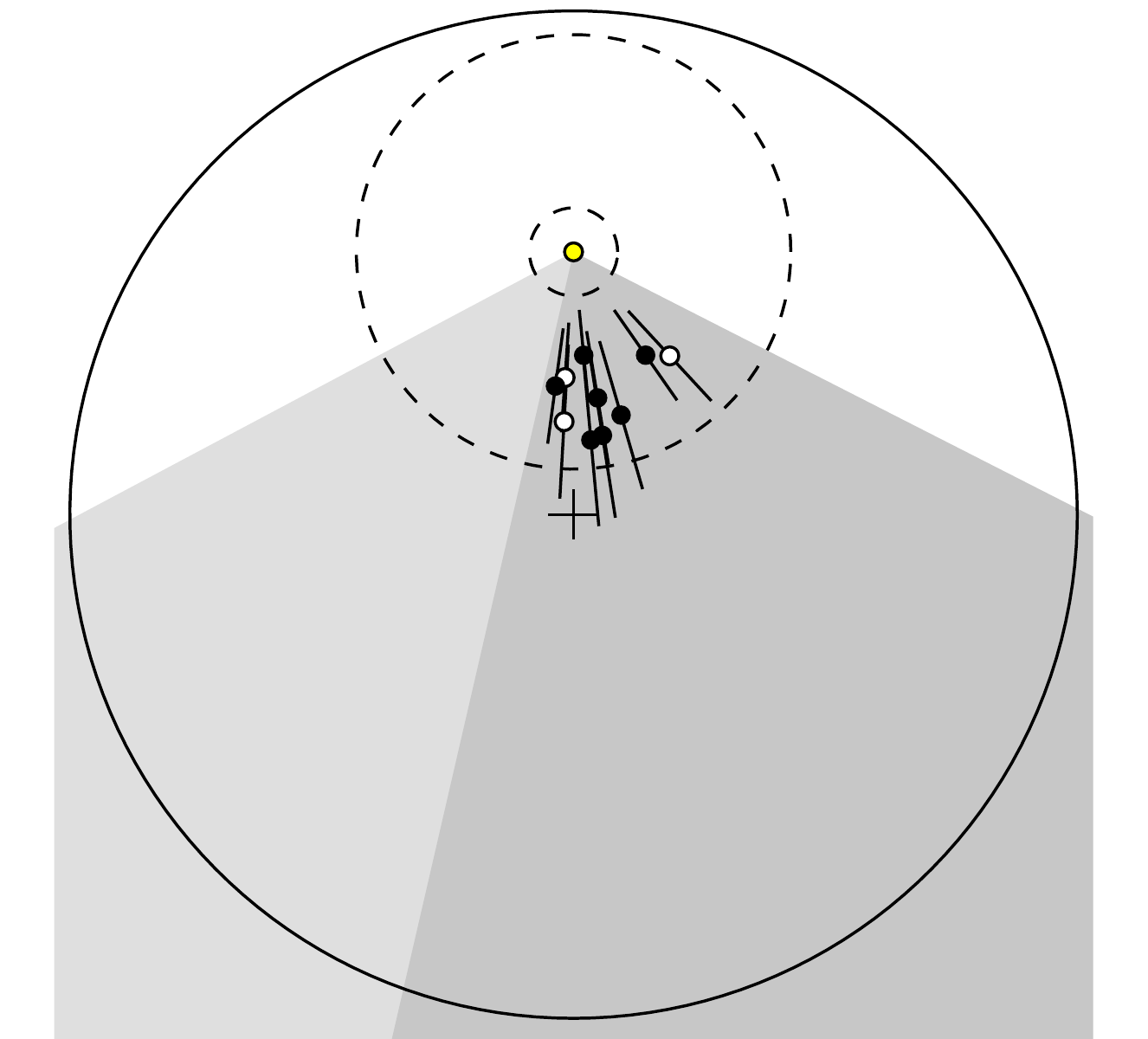}
\caption{Calculated position of the elliptical bubbles (solid circles and error bars) and filled elliptical bubbles (open circles and error bars) onto the Galactic plane. The Sun is represented with a yellow circle, the Galactic centre with a cross and the Galaxy outer limit is assumed to be $16\um{kpc}$ from the centre (solid circumference). The entire shaded area represents the portion of the Galactic plane covered by MIPSGAL. Its darker part satisfies $\delta>-40^\circ$, the lower declination limit of VLA. The two dashed circumferences represent the distance from the Sun at which a nebula with a diameter of $0.4\um{pc}$ is seen under an angle of $12\um{arcsec}$ (outer circumference) and $60\um{arcsec}$ (inner).}
\label{fig:gal}
\end{figure}

In Table \ref{tab:PNe} we report the distance, diameter, mean electron density and ionized mass, derived as discussed above, for the seven elliptical bubbles. Please notice that since $M_\mathrm{ion}\propto D^{5/2}$, $D$ being the distance, the error on the ionized mass is of order of 100 per cent and those values must be regarded as indicative just of the order of magnitude.
\begin{table}
\caption{Physical parameters of PNe and PN candidates.}
\begin{tabular}{ccccc}\hline
[MGE] & Distance & Diam. & $\langle n_\mathrm{e}\rangle$ & $M_\mathrm{ion}$\\
& (kpc) & (pc) & ($\mathrm{cm}^{-3}$) & ($\mathrm{M}_{\sun}$)\\\hline
005.2641+00.3775 & $6.0\pm2.7$ & 0.41 & $\phantom{1}52$ & 0.04\\ 
005.6102-01.1516 & $3.3\pm1.4$ & 0.29 & 110 & 0.03\\ 
008.9409+00.2532 & $5.9\pm2.6$ & 0.46 & $\phantom{1}31$ & 0.07\\ 
009.3523+00.4733 & $4.7\pm2.1$ & 0.32 & $\phantom{1}74$ & 0.05\\ 
016.2280-00.3680 & $5.4\pm2.4$ & 0.45 & $\phantom{1}60$ & 0.12\\ 
034.8961+00.3018 & $4.0\pm1.7$ & 0.47 & $\phantom{1}27$ & 0.06\\ 
352.3117-00.9711 & $4.3\pm1.8$ & 0.46 & $\phantom{1}56$ & 0.12\\ 
\hline
\end{tabular}
\label{tab:PNe}
\end{table}
Though the statistical distance method is intrinsically affected by relevant errors, it is possible to notice how all these bubbles have a similar distance. This can be ascribed to the homogeneity of the original 24-$\umu$m sample, consisting of objects with a similar angular extension and shape. We recall that \citet{Nowak2014} pointed out that almost all bubbles classified as PNe show a disk or ring morphology. PNe with very different 24-$\umu$m shape were excluded \textit{a priori} by \citet{Mizuno2010}. Furthermore, more distant PNe are likely not resolved at $24\mic{m}$, and so even in this case they would have been excluded from the bubble catalogue. In Figure \ref{fig:gal} we report a sketch of the Galactic plane and we locate onto it the seven elliptical bubbles, highlighting the uncertainties associated with their distances.

The morphological uniformity of the MIPSGAL bubbles translates also into a homogeneity of the physical characteristics of these objects. \citet{Kwok1985} presented a plot showing the relationship between brightness temperature of a PN and the mean electron density. For our bubbles we have mean brightness temperature ranging from about $2\um{K}$ to about $10\um{K}$, which, according to that relationship, would translate to mean electron densities of order of $10^1-10^2\um{cm^{-3}}$, in perfect agreement with the values we derived. Therefore the relative homogeneity in the mean electron density can be ascribed to a uniformity in the bubble dimension (original sample selection bias) and in the radio flux densities (limited by the sensitivity of present instruments). \citet{vandeSteene1995} report that the PNe with those values of brightness temperature and diameter are usually evolved PNe, with an age after post-AGB phase end much greater than $10^3$ years. As shown by \citet{Volk1985}, at these ages the central star of the PN starts its contraction toward the white dwarf phase. The contraction implies that the total flux of UV (ionizing) photons decreases. The PN enters a phase of recombination where the radius of the ionized nebula is approximately constant, but its ionised mass decreases because of the gas dispersion (considering also the neutral gas, the nebula is still expanding). This can justify why five bubbles show ionized mass values significantly below $0.1\um{M_{\sun}}$, in the sense that this could be a hint of aged PNe. Indeed young PNe are characterized by these values of ionized mass, however their mean electron density is higher and their radius is smaller \citep{Cerrigone2008}. We are aware that for young PNe the statistical distance method may not be applied if the nebula is still optically thick at $5\um{GHz}$, but the spectral indices reported in Paper I and the spectral index maps obtained from these new data confirm that these seven bubbles are all optically thin at $5\um{GHz}$. The ionized masses found for these PNe are smaller than the median value found by \citet{Frew2010}, namely $0.5\sqrt{\epsilon}\ \mathrm{M}_{\sun}$ where $\epsilon$ is the filling factor. It is possible that the MIPSGAL sample is biased towards low-mass, high-excitation PNe (with the 24-$\umu$m emission dominated by [O \textsc{iv}] line; e.g. \citealt{Frew2016}). In this case very aged PNe become unobservable with MIPS as the nebular excitation (and consequently the [O \textsc{iv}] integrated flux) decreases when the central star becomes a white dwarf \citep{Chu2009}.

In Section \ref{sec:fill} we discussed the possibility that also the filled elliptical bubbles could be PNe. Now we apply to them the same procedure described above for elliptical bubbles to derive their distance and then the other related physical parameters. We collect the result in Table \ref{tab:fill}.
\begin{table}
\caption{Physical parameters for filled elliptical bubbles if they are considered PNe.}
\begin{tabular}{ccccc}\hline
[MGE] & Distance & Diam. & $\langle n_\mathrm{e}\rangle$ & $M_\mathrm{ion}$\\
& (kpc) & (pc) & ($\mathrm{cm}^{-3}$) & ($\mathrm{M}_{\sun}$)\\\hline
042.7665+00.8222 & $4.5\pm1.9$ & 0.44 & 52 & 0.10\\ 
356.1447+00.0550 & $4.0\pm1.7$ & 0.58 & 36 & 0.16\\ 
356.8155-00.3843 & $5.4\pm2.4$ & 0.34 & 70 & 0.06\\ 
\hline
\end{tabular}
\label{tab:fill}
\end{table}
The distance calculation is not performed on MGE 002.2128-01.6131 since, as reported in Paper I, it is not optically thin at $5\um{GHz}$. Again we underline that the values reported in Table \ref{tab:fill} are meaningful only if the sources are truly PNe and, even if those values are in agreement to what should be expected from PNe, they cannot be used to demonstrate the PN nature of these bubbles. However, if the PN hypothesis is true, we can notice how the values in Table \ref{tab:fill} are very similar to those in Table \ref{tab:PNe}. We report the filled elliptical bubbles in Figure \ref{fig:gal} as open circles.

In \citet{Ingallinera2014b} we found that the distance of the four PNe studied in that work ranged between 1.6 and $3.2\um{kpc}$. Those bubbles were selected because they were resolved also in D-configuration observations. From their angular extension we estimate diameter values between 0.28 and $0.41\um{pc}$, in accordance to those derived in this paper. The different angular dimension of the bubbles in \citet{Ingallinera2014b} should then been ascribed mostly to their smaller distance rather than to physical differences.


\subsection{Completeness of the bubbles catalogue with respect to planetary nebulae}
\label{sec:compl}
One of the most promising aspect of the bubbles catalogue was that it appeared as the perfect place where to search for `missing PNe'. It is expected that the number of Galactic PNe is around $20\,000$ (e.g. \citealt{Zijlstra1991}), but only about 3300 of them are known \citep{Sabin2014}. When \citet{Mizuno2010} published the bubbles catalogue, about 15 per cent of them were associated with already classified object, and 90 per cent of them were PNe. The idea was that the PNe represented the dominant kind of object among the MIPSGAL bubbles and, in turn, the catalogue could contain a large quantity of previously unknown PNe. \citet{Mizuno2010} divided the bubbles into six categories, based on their morphology. The categories most populated are `ring' and `disk' nebulae, accounting for 339 objects out of 428.\footnote{The complete catalogue with the morphology types is available at: http://vizier.u-strasbg.fr/viz-bin/VizieR?-source=J/AJ/139/1542} As previously discussed, \citet{Nowak2014} found that more than 80 per cent of classified `ring' and `disk' bubbles are PNe, and that few other PNe are found in other categories. It is possible to estimate that the entire bubbles sample should contain about 300 PNe.

It is important now to establish whether the bubbles catalogue is complete with respect to round and elliptical PNe, that is we are going to check if MIPSGAL successfully detected all the round and elliptical PNe resolved as bubbles lying in its field of view. A round or elliptical PN is listed in the bubbles catalogue only if it is a resolved object. Since MIPSGAL resolution is $6\um{arcsec}$ we can state that the minimum angular extension for a PN in the bubbles sample must be at least around $12\um{arcsec}$. In fact this is also the lower limit of the angular extension for bubbles (except only for two). In the previous section we found out that a typical value for an elliptical bubble diameter is around $0.4\um{pc}$, a value in accordance with many other estimates of the dimensions of PNe. A comparable mean value is derived for filled elliptical bubbles, in the hypothesis that they are PNe, and for PNe in \citet{Ingallinera2014b}. Similar values are found by \citet{Bojicic2011} in their study of the radio emission of 26 PNe from the Macquarie/AAO/Strasbourg H$\alpha$ (MASH) catalogue \citep{Parker2006}. A PN with a diameter of $0.4\um{pc}$ is seen under an angle of $12\um{arcsec}$ if it is located at about $6.9\um{kpc}$ from the observer. In Figure \ref{fig:gal} we represent this limit with the outer dashed circumference. Beyond this distance MIPSGAL is not able to fully resolve an object extended less than $0.4\um{pc}$, therefore any other PN beyond this limit unlikely is part of the bubbles sample. On the other hand the largest ring bubble is $63\um{arcsec}$ wide (and only 18 other bubbles are larger than it), the largest bubble classified as PN is $50\um{arcsec}$ and the largest disk bubble is $42\um{arcsec}$. We can set then the approximate upper limit of $60\um{arcsec}$ for the PNe in the bubbles sample. A PN with a diameter of $0.4\um{pc}$ appears 60-arcsec wide when it is located at a distance of about $1.4\um{kpc}$. This distance (from the observer) is represented in Figure \ref{fig:gal} with the inner dashed circumference. These two distances and the MIPSGAL field of view (shaded region in Figure \ref{fig:gal}) encompass an area of $49.8\um{kpc^2}$. Now the outer limit of the Milky Way is somehow arbitrary, here we assume a radius of $16\um{kpc}$, a boundary beyond which the star density drops rapidly (for a discussion on the Galactic dimensions and outer rings see \citealt{Xu2015}). This radius encompasses a total area of about $800\um{kpc^2}$.
Taking into account an exponential radial distribution with a scale-length of $3.5\um{kpc}$ \citep{Zijlstra1991}, we found that about 3000 PNe fall in the area where they can be observed as MIPSGAL bubbles.
Morphological studies of PNe show that about 2/3 of them are elliptical or round, with the remaining showing bipolar or irregular shapes \citep{Sabin2014}. 
As previously said, in some bipolar PNe the emission at $8\mic{m}$, mainly tracing dust, is concentrated at its center. The 24-$\umu$m emission is due to both thermal dust continuum and gas lines, with the former decreasing as the PN ages and expands. The morphologies at $8\mic{m}$, $24\mic{m}$ and radio can therefore be very different, depending on the evolutionary age of the PN \citep{Parker2012}.
We can therefore safely state that the number of PNe that can be seen as a bubble by MIPS is approximately around 2000.

The distribution of the PNe heights from the Galactic disk was investigated by \citet{Zijlstra1991}, who found a height scale value of $250\um{pc}$ for an exponential disk model. Limiting our search between $D_\mathrm{min}=1.4\um{kpc}$ and $D_\mathrm{max}=6.9\um{kpc}$ the average distance of an object in this area is
\begin{equation}
\langle D\rangle=\sqrt{\frac{D_\mathrm{min}^2+D_\mathrm{max}^2}{2}}=5.0\um{kpc}.
\label{eq:mion}
\end{equation}
At this distance, $250\um{pc}$ correspond to an angle of $2.9^\circ$. The MIPSGAL latitude coverage is $|b|<1^\circ$, except for $-5^\circ<l<7^\circ$ where $|b|<3^\circ$. This means that the MIPSGAL latitude coverage allows us to observe only about the 30 per cent of PNe. This percentage gives us a total of about 600 PNe that should appear as a bubble in MIPSGAL. The fact that we likely observe only about 300 PNe as MIPSGAL bubbles can then be ascribed to their aging, as discussed in previous Section.
This in turn means that the bubbles catalogue should contain approximately half of the elliptical and round PNe located between $1.4\um{kpc}$ and $6.9\um{kpc}$ from us falling into the MIPSGAL field of view.

If future studies on the bubbles will confirm our estimation then this result will have two important implications on the search for Galactic PNe. First, going backwards through what discussed above, it will pose a constraint on the total number of Galactic PNe. Since this constraint is mainly derived from a statistical base it could be used as an independent test for the estimates derived from stellar evolution models. Second, it will turn out that the IR band around $24\mic{m}$ is probably the best band where next-generation IR telescope shall search for PNe. In particular the \textit{James Webb Space Telescope}, improving seven-fold the \textit{Spitzer} resolution, should be able to observe and resolve all the Galactic PNe as bubbles (possibly with the only exception of most evolved ones), even if we assume that its sensitivity with respect to \textit{Spitzer} will just scale up as the ratio of their areas.


\section{Summary and conclusions}
\label{sec:sumcon}
In this paper we reported radio observations at $5\um{GHz}$ of 18 MIPSGAL bubbles, performed with the VLA in configuration B and BnA. The images show that the bubbles can be grouped in five different categories according to their radio morphology. Three bubbles show a clear evidence of a central point-like object; seven are characterized by an elliptical ring shape and the total lack of a compact central object or diffuse emission toward their centre; for four bubbles the radio morphology is intermediate between the two previous categories, with diffuse emission toward the centre but without a clear central object; one has a bipolar appearance; the remaining three bubbles have a peculiar shape.

The bubbles with a central object in radio images (MGE 027.3839-00.3031, MGE 030.1503+00.1237, MGE 042.0787+00.5084) were associated with massive evolved stars, in particular LBVs. This classification is compatible with the spectroscopic identification of the central star of two of them as B/B[e]/LBV stars. The radio emission from the central object is likely due to stellar winds. The CSE of MGE 042.0787+00.5084 shows relevant variations of spectral index, on the contrary the CSE of MGE 030.1503+00.1237 has a more homogeneous spectral index. The 24-$\umu$m images of these bubbles show a morphological similarity with radio only in two cases, while at IRAC wavelengths only the central object is detected.

The elliptical bubbles were associated with PNe. One of them is already classified as PN in literature. For the others morphological considerations suggest that can be all proposed as PN candidates. If this hypothesis holds, we derive for them the distance, the diameter, the mean electron density and the ionized mass. We found that the elliptical bubbles are very similar also from the physical point of view. We ascribe this similarity to the selection bias of the bubbles catalogue.

Four bubbles presented a morphology intermediate between the first two, with the lack of a central object but with diffuse emission toward their centre. On average they appear fainter at $8\mic{m}$ with respect to elliptical bubbles. Nevertheless the radio and the 24-$\umu$m morphology suggest that they can also be proposed as PN candidates.

The remaining bubbles show different radio morphologies. One of them, MGE 356.7168-01.7246, has a clear bipolar shape. The 8-$\umu$m emission of MGE 031.7290+00.6993 is located outside the radio nebula and is proposed as an H \textsc{ii} region. MGE 032.4982+00.1615 is only barely resolved in our radio images. Its flux density probably has varied with respect to 2010 data and its radio spectral index suggests that its radio emission is non-thermal. We propose that it could be a proto-PN. Finally MGE 028.7440+00.7076 is a very peculiar object with the radio emission placed outside the $24\mic{m}$.

As discussed in Section \ref{sec:compl} the knowledge achieved so far in the discovery of PNe among the MIPSGAL bubbles allows us to outline some general assertions about Galactic PNe. The number of known Galactic PNe is about the 15 per cent of what expected. The explanation is that, lying in proximity of the Galactic disk, many of them are heavily obscured in visible light. IR surveys, like MIPSGAL, are able to observe through the Galactic clouds and potentially discover a large quantity of previously unknown PNe. In this sense, preliminary results indicated that the bubbles catalogue could harbour many PNe. Given the typical diameter of a PN, the field of view and the resolution of MIPSGAL, we infer that the bubbles catalogue may contain about half of the Galactic PNe that would be resolved in MIPSGAL. It is likely that the most evolved PNe are not detected in MIPSGAL since as the central star becomes white dwarf, the nebular excitation decreases and consequently their flux at $24\mic{m}$. If this hypothesis will be confirmed, beside the discovery of 250 unknown PNe, it will pose an important constraint on the actual total number of Galactic PNe, being a significant statistical search on a relevant portion of our Galaxy. Furthermore it will give the precise indication that the 24-$\umu$m is the best band where to search for the remaining Galactic PNe when next-generation IR telescopes will provide better resolution and sensitivity than \textit{Spitzer}.

The main goal of this work was to show the great potential of high-resolution high-sensitivity radio images. We showed that different kinds of evolved stars have strikingly different radio morphologies. We showed also how to use radio images to infer several physical properties of these Galactic objects. Even for the bubbles that we were not able to classify the discovery of their peculiar radio morphology could lead to further investigations and to the discovery a new shape prototypes. We want to remark that the imaging performance achieved in these targeted observations are about to be achieved also by several ongoing Galactic surveys used as pathfinders for next generation radio telescopes (e.g. \citealt{Bihr2016} with VLA or \citealt{Umana2015} with the ATCA\footnote{Australia Telescope Compact Array} and ASKAP\footnote{Australian SKA Pathfinder}). This work shows therefore what kind of information we can expect to derive on Galactic sources from future radio observations. Recalling the fundamental help of comparing radio maps with IR images, we highlight also the importance of the synergy with future IR telescopes, above all the \textit{James Webb Space Telescope}, that will provide us with images of unprecedented sensitivity and angular resolution comparable to radio interferometers.



\section*{Acknowledgements}
This work is based on observations made with the Very Large Array of the National Radio Astronomy Observatory, a facility of the National Science Foundation operated under cooperative agreement by Associated Universities Inc.. Archive search made use of the SIMBAD database and the VizieR catalogue access tool, operated by the Centre de Donn\'ees astronomique de Strasbourg, and of the NASA/IPAC Infrared Science Archive, which is operated by the Jet Propulsion Laboratory, California Institute of Technology, under contract with the National Aeronautics and Space Administration. This publication makes use of data products from the Two Micron All Sky Survey, which is a joint project of the University of Massachusetts and the Infrared Processing and Analysis Center/California Institute of Technology, funded by the National Aeronautics and Space Administration and the National Science Foundation. F.B. acknowledges support from the VIALACTEA Project, a Collaborative Project under Framework Programme 7 of the European Union, funded under Contract \# 607380. CA acknowledges support from FONDECYT grant No. 3150463.

\end{document}